\documentclass[a4paper,twoside,titlepage,11pt]{article}
\usepackage{hyperref}

\usepackage{amssymb,amsmath,amsfonts} \usepackage{epsfig}

\newcommand{\be}{\begin{equation}} \newcommand{\ee}{\end{equation}}
\newcommand{\bea}{\begin{eqnarray}} \newcommand{\eea}{\end{eqnarray}}
\newcommand{\bino}[2]{\left( \begin{array}{c} #1 \\ #2 \end{array}
\right)} \newcommand{\CW}{\mathcal{W}} \newcommand{\CG}{\mathcal{G}}
\newcommand{\CO}{\mathcal{O}} \newcommand{\CN}{\mathcal{N}}
\newcommand{\CI}{\mathcal{I}}

\newcommand{\id}{\hbox{1\kern-.27em l}}
\newcommand{\sid}{\hbox{\scriptsize1\kern-.27em l}}

\newcommand{\we}{\kern-.1em\wedge\kern-.1em}
\newcommand{\scal}{\kern-.13em\cdot\kern-.13em}

\newcommand{\II}{I\kern-.09em I}

\newcommand{\Ga}{\Gamma} 
 \newcommand{\de}{\delta}
 \newcommand{\si}{\sigma}
 
\newcommand{\om}{\omega}

\newcommand{\Z}{\mathbb{Z}} \newcommand{\C}{\mathbb{C}}
\newcommand{\R}{\mathbb{R}}

\newcommand{\ads}{{\rm AdS}}

\newcommand{\nn}{\nonumber} \newcommand{\spa}{\ \ ,\ \ \ \ }
 \newcommand{\tr}{\mathop{{\rm
Tr}}} 
\newcommand{\gym}{g_{\mathrm{YM}}} 

\newcommand{\vecto}[2]{\left( \begin{array}{c} #1 \\ #2 \end{array}
\right) } \newcommand{\matrto}[4]{\left( \begin{array}{cc} #1 & #2 \\
#3 & #4 \end{array} \right) }

\newcommand{\Hlc}{H_{\rm lc}} \newcommand{\Hlct}{\tilde{H}_{\rm lc}}
\newcommand{\gqgt}{g_{\mathrm{QGT}}}

\newcommand{\map}{\cong} \newcommand{\mapsim}{\sim}

\numberwithin{equation}{section}
\begin{document}

\begin{titlepage}

\rightline{\vbox{\small\hbox{HUTP-02/A048} \hbox{NORDITA-2002/61 HE}
 \hbox{ITFA-2002-37} \hbox{DFTT 28/2002}\vskip.5ex  \hbox{\tt
 hep-th/0209201}}}
\vskip 1.3cm

\centerline{\LARGE \bf Gauge theory description of compactified
pp-waves}
\vskip 1.2cm  \centerline{{\Large   M. Bertolini${}^a$, J. de
Boer${}^b$, T. Harmark${}^c$, E. Imeroni${}^d$, N. A. Obers${}^e$} }
%\centerline{{\bf People}}
\vskip 0.3cm
\begin{center}
{\sl ${}^a$NORDITA, Blegdamsvej 17, DK-2100 Copenhagen \O,
Denmark\\[.75ex] ${}^b$ Institute for Theoretical Physics, University
of Amsterdam \\[-.3ex] Valckenierstraat 65, 1018 XE Amsterdam, The
Netherlands \\[.75ex] ${}^c$Jefferson Physical Laboratory, Harvard
University, Cambridge, MA 02138, USA \\[.75ex] $^d$Dipartimento di
Fisica Teorica, Universit\`a di Torino \\[-.3ex] and I.N.F.N., Sezione
di Torino, Via P. Giuria 1, I-10125 Torino, Italy \\[.75ex] $^e$The
Niels Bohr Institute, Blegdamsvej 17, DK-2100 Copenhagen \O, Denmark}
\vskip .1in  {\small \sffamily teobert@nbi.dk, jdeboer@science.uva.nl,
harmark@bose.harvard.edu, \\imeroni@to.infn.it, obers@nbi.dk}
\end{center}
\vskip 0.8cm  \centerline{\bf \large Abstract} \vskip 0.1cm
\noindent We find a new Penrose limit of $\ads_5 \times S^5$ that
gives the maximally symmetric pp-wave background of Type IIB string
theory in a coordinate system that has a manifest space-like
isometry. This induces a new pp-wave/gauge-theory duality which on the
gauge theory side involves a novel scaling limit of $\CN=4$  SYM
theory.  The new Penrose limit, when applied to $\ads_5 \times S^5 /
\Z_M$, yields a pp-wave with a space-like circle. The dual gauge
theory description involves a triple scaling limit of an $\CN=2$
quiver gauge theory. We present in detail the map between gauge theory
operators and string theory states including winding states, and
verify agreement between the energy eigenvalues  obtained from string
theory and those computed in gauge theory, at least to one-loop order
in the planar limit.  We furthermore consider other related new
Penrose limits and explain how these limits can be understood as part
of a more general framework.

\end{titlepage}

%-----------------------------------
\tableofcontents  \setcounter{page}{1}

\newpage
%--------------------------------------------

\section{Introduction}

Since the discovery of the AdS/CFT correspondence there has been
extensive work on finding new dualities between large $N$ gauge
theories and string theory on various backgrounds. Recently
\cite{Berenstein:2002jq} it was considered what happens in the AdS/CFT
correspondence for $N \rightarrow \infty$. Since the curvature of
$\ads_5 \times S^5$ is proportional to $N^{-1/2}$ one has flat space
in this limit. This is a difficult limit to control.  In fact, the
novelty of Ref.~\cite{Berenstein:2002jq} has been to take a double
scaling limit of $\CN=4$ Super Yang-Mills (SYM) theory where $N
\rightarrow \infty$ is taken together with the limit  of large
R-charge $J$ so that $J^2 /N$ and $\gym^2$ are fixed.  This gives on
the gauge theory side a well-defined double scaling limit
\cite{Kristjansen:2002bb,Constable:2002hw}, while on the geometric
side it becomes a Penrose limit that gives the recently discovered
maximally symmetric pp-wave background of type IIB string theory of
Ref.~\cite{Blau:2001ne}.

The pp-wave/gauge theory duality of Ref.~\cite{Berenstein:2002jq} thus
provides a tractable step toward obtaining string theory in flat space
from large $N$ gauge theory. Inspired by this, we propose in this
paper a new pp-wave/gauge theory duality between string theory on a
pp-wave background which has a space-like circle and a certain triple
scaling limit of an $\CN=2$ quiver gauge theory. That the pp-wave
background has a space-like circle means that it is geometrically
``close'' to $\R^{1,8} \times S^1$.  We are thus addressing how to get
string theory on $\R^{1,8} \times S^1$ from large $N$ gauge theory.

One cannot directly compactify the maximally symmetric pp-wave
background  of type IIB string theory in the coordinate system used in
Ref.s~\cite{Blau:2001ne,Berenstein:2002jq} since it does not display
manifest space-like isometries.  Our first step is therefore to put
forward a new Penrose limit of $\ads_5 \times S^5$ that results in a
pp-wave background with a manifest space-like isometry and show that
it induces a new pp-wave/gauge-theory duality between type IIB string
theory on the pp-wave background and a certain scaling limit of
$\CN=4$ SYM theory. It turns out that on the gauge theory side this
duality looks quite different from the one of
Ref.~\cite{Berenstein:2002jq}.  The resulting pp-wave background has
previously been considered by Michelson \cite{Michelson:2002wa} where
it was obtained by a coordinate transformation from the maximally
symmetric type IIB pp-wave background considered in
Ref.s~\cite{Blau:2001ne,Berenstein:2002jq}. Here, by deriving it
directly from a Penrose limit, we are able to make manifest what the
corresponding scaling limit of the dual ${\cal N}=4$ gauge theory
should be, and hence construct the appropriate dual gauge theory
operators.

It is interesting to note that, although our scaling limit and that of
Ref.~\cite{Berenstein:2002jq} look different, they are required to be
physically equivalent since the two corresponding pp-wave backgrounds
are related by a coordinate transformation. In fact, the existence of
different Penrose limits is related to inequivalent ways in which we
can choose the neighborhood of null geodesics. As defined in
Ref.~\cite{Penrose:1976} (see e.g. also Ref.s~\cite{Gueven:2000ru} and
\cite{Blau:2002mw}), Penrose limits involve a very specific choice of
coordinates in the neighborhood of the null geodesic, but one of the
points of the present work is that there are other choices that lead
to well-defined but inequivalent scaling limits.  We address and
explain this issue  after having presented our pp-wave/gauge-theory
duality for the compactified pp-wave in detail.

To get a pp-wave with a space-like circle we implement our new Penrose
limit on $\ads_5 \times S^5 / \Z_M$ by taking $M \rightarrow \infty$
in such a way that we get a circle with a finite radius  along the
direction of the space-like isometry.  This induces a duality between
type IIB string theory on a pp-wave background with a space-like
circle and  a (triple) scaling limit of the superconformal $\CN=2$
quiver gauge theory (QGT) which is dual \cite{Kachru:1998ys} to the
$\ads_5 \times S^5 / \Z_M$ background.\footnote{See
Ref.s~\cite{Itzhaki:2002kh,Gomis:2002km,
Alishahiha:2002ev,Kim:2002fp,Takayanagi:2002hv,Floratos:2002uh,
Mukhi:2002ck,Alishahiha:2002jj,Oh:2002sv,Naculich:2002fh}  for other
works on Penrose limits of orbifold geometries.} To check our proposal
we first discuss the type IIB string theory states on this pp-wave
background and then construct the corresponding dual gauge theory
operators of the $\CN=2$ QGT.  Note that this also includes winding
states corresponding to strings winding on the space-like circle.  We
check the correspondence by computing the leading correction to the
anomalous dimensions of various gauge theory operators and by
comparing these to the energy eigenvalues of the corresponding string
states.

One of the interesting aspects of finding a pp-wave/gauge theory
correspondence for a pp-wave with a space-like circle is that one can
T-dualize this type IIB pp-wave background to a type IIA pp-wave
background. By a subsequent S-duality one further obtains  an M-theory
pp-wave background.  If one then finds a Matrix theory
\cite{Banks:1997vh,Sen:1997we,Seiberg:1997ad} description of this
M-theory background, one can dualize this theory back to obtain a
Matrix String theory \cite{Motl:1997th,Banks:1997my,Dijkgraaf:1997vv}
description of the type IIB pp-wave with a space-like circle.  This
Matrix String theory has been considered in
Ref.~\cite{Gopakumar:2002dq,Sugiyama:2002tf}.%
\footnote{See Ref.s~\cite{Bonelli:2002mb,Verlinde:2002ig} for
approaches to Matrix String theory using directly the pp-wave solution
of Ref.~\cite{Blau:2001ne}.}   Thus, our new pp-wave/gauge theory
correspondence could possibly be enhanced to a correspondence between
gauge theory and Matrix String theory on the pp-wave background with a
space-like circle.  To this end, we also find a new Penrose limit of
$\ads_5 \times S^5 / (\Z_{M_1} \times \Z_{M_2})$ that not only gives
the space-like circle but also a compact null-direction. This means
that our pp-wave can have a DLCQ description, similarly to what has
been found for the maximally symmetric type IIB pp-wave  in
Ref.s~\cite{Mukhi:2002ck,Alishahiha:2002jj}.  We can thus hope to find
a correspondence between Matrix String theory with finite matrices
(with sizes  equal to the quantized momentum along the null-direction)
and a quadruple scaling limit of an $\CN = 1$ quiver gauge theory.%
\footnote{We also explain how to get the DLCQ of the pp-wave
background with a space-like isometry from $\ads_5 \times S^5/ \Z_M$,
which means one could make a duality between Matrix String theory and
a scaling limit of $\CN=2$ quiver gauge theory.}  Apart from being
interesting in itself, such a correspondence could perhaps illuminate
the current attempts of understanding interacting string theory from
the dual gauge theory. This will be pursued in a future publication.

In another direction which lies slightly outside the main focus of our
paper, we consider another class of Penrose limits, now with two
space-like isometries. We give three limits corresponding to zero, one
and two compact space-like directions.  Interestingly, these
backgrounds are time-dependent.  The backgrounds have been considered
previously in Ref.~\cite{Michelson:2002wa} where again the uncompact
one is connected to the maximally symmetric type IIB pp-wave
background of Ref.s~\cite{Blau:2001ne,Berenstein:2002jq} by a
coordinate transformation.  Having the explicit Penrose limits could
perhaps make it possible to find the precise scaling limit of the
gauge theory and find the map between gauge theory operators and
string theory states. This could be important since we then would
obtain time-dependent string theory from a limit of gauge theory and
since the understanding of string theory on time-dependent backgrounds
is still at its infancy.

This paper is organized as follows. In section \ref{secsugra} we
discuss the new Penrose limits that yield pp-wave backgrounds with
manifest space-like isometries. We first find a new Penrose limit of
$\ads_5 \times S^5$ that yields the same pp-wave as the one discussed
by BMN, but in a different coordinate system in which a space-like
isometry is manifest. We then apply this new Penrose limit to $\ads_5
\times S^5 / \Z_M$. The group $\Z_M$ acts along the direction of the
space-like isometry. By a suitable scaling of $M$ this direction is
compactified, and we obtain a pp-wave with a finite space-like
circle. The scaling of $M$ is quite distinct from other scalings that
have appeared in the literature, namely $M$ has to scale as $N^{1/3}$,
where $N$ is the rank of the $U(N)$ gauge group factors that appear in
the  dual ${\cal N}=2$ quiver gauge theory.  The duality between this
quiver gauge theory and the compactified pp-wave is elaborated upon in
sections~\ref{secstring} and~\ref{secgauge}. We then continue in
section~\ref{secsugra} to show how to find a Penrose limit of $\ads_5
\times S^5 / ( \Z_{M_1} \times \Z_{M_2} )$ that gives a pp-wave with
both a space-like circle and a compact null direction and another
Penrose limit of $\ads_5 \times S^5 / \Z_M$ that yields a pp-wave with
a manifest space-like isometry and a compact null direction. For these
two Penrose limits we also find the corresponding scaling of the gauge
theory parameters.

In section \ref{secstring} we discuss the type IIB string theory on
the pp-wave background obtained from our new Penrose limit. We
quantize the theory and derive the spectrum. This will be useful when
comparing with the dual gauge theory operators.

In section \ref{secgauge} we discuss the relevant ${\cal N}=2$ QGT
which is dual to type IIB string theory on $\ads_5 \times S^5 / \Z_M$
and find the gauge theory operators surviving the triple scaling limit
derived in section \ref{secsugra}. These operators are shown to be in
one-to-one correspondence with string theory states, including winding
modes. We compute the anomalous dimension of near-BPS operators, at
one loop in the planar limit, and verify agreement with the
expectations from the string theory side.

In section \ref{secn5} we return to the novel scaling limit of $\CN
=4$ SYM that corresponds to the Penrose limit giving rise to the
pp-wave with manifest space-like isometry. We present the relevant
gauge theory operators and discuss the genus counting parameter
arising in non-planar contributions. By studying the algebra of
killing vectors, we explain the mechanism by which it is possible that
different scaling limits of ${\cal N}=4$ SYM give pp-wave backgrounds
which are simply related by coordinate transformations. In particular
we discuss the relation between our limit and the one discussed in
Ref.~\cite{Berenstein:2002jq}. We also show how one can obtain
different sectors in ${\cal N}=2$ QGT by embedding the orbifold group
$\Z_M$ of $\ads_5 \times S^5 / \Z_M$ in the isometry algebra in
different ways. This translates into a different scaling of the order
of the orbifold group $M$ with the radius $R$ of $\ads_5$ and $S^5$.

Finally, in section \ref{ythings} we consider a different class of
Penrose limits that leads to pp-wave backgrounds with a time-dependent
light-cone Hamiltonian. First we describe a Penrose limit of $\ads_5
\times S^5$ giving a pp-wave background with two manifest space-like
isometries. Then we consider a Penrose limit of  $\ads_5 \times S^5 /
\Z_M$ giving a pp-wave with one space-like circle. Finally, we
consider a Penrose limit of $\ads_5 \times S^5 / (\Z_{M_1} \times
\Z_{M_2})$ that has a space-like two-torus.

Section \ref{secconcl} contains a summary of our findings, a
discussion of open questions and future lines of research.

The reader can find many technical details in the appendices.  In
appendix \ref{appkill} we review the coordinate transformations found
in Ref.~\cite{Michelson:2002wa} which relate the original pp-wave
background of Ref.s~\cite{Blau:2001ne,Berenstein:2002jq} to the
backgrounds discussed in sections \ref{secsugra} and \ref{ythings}. In
appendix \ref{appqgt} we describe the superconformal ${\cal N}=2$
quiver gauge theory which is dual to type IIB string theory on $\ads_5
\times S^5 / \Z_M$. We review how this can be seen as a consistent
truncation of ${\cal N}=4$ SYM and present the structure of the chiral
primaries of ${\cal N}=2$ theory as obtained from ${\cal
N}=4$. Appendix \ref{appn2} contains the derivations of many of the
results presented in section \ref{secgauge} together with a complete
translational dictionary between ${\cal N}=2$ and ${\cal N}=4$
formalism, the latter being the one used throughout the
paper. Finally, in appendix \ref{appqgt1} we describe the ${\cal N}=1$
quiver gauge theory which is dual to type IIB string theory on $\ads_5
\times S^5 / (\Z_{M_1} \times \Z_{M_2})$. This is the superconformal
SYM theory in which to implement the two different quadruple scaling
limits of ${\cal N}=1$ SYM discussed in the end of section
\ref{secsugra} and in section \ref{ythings}. One expects the surviving
operators of the gauge theory to be dual to type IIB string theory on
the pp-wave backgrounds of sections \ref{secDLCQ} and \ref{ythings}
respectively.

%%%%%%%%%%%%%%%%%%%%%%%%%%%%%%%%%%%%%%%%%%%%%%%%%%%%%%%%%%%%%%%%%%%%%
%%%%%%%%%%%%%%%%%%%%%%%%%%%%%%%%%%%%%%%%%%%%%%%%%%%%%%%%%%%%%%%%%%%%%
\section{New Penrose limits and space-like isometry}
\label{secsugra}

In this section we describe the new Penrose limit of $\ads_5 \times
S^5$ that realizes an explicit space-like isometry for the pp-wave
background of Ref.s~\cite{Blau:2001ne,Berenstein:2002jq}. Then we show
how one can use this to generate compact space-like and null
directions for the pp-wave starting from orbifolded backgrounds.
The pp-wave background that we obtain after our new Penrose limit is
in the coordinate system first given by Michelson
\cite{Michelson:2002wa}.  We review in appendix \ref{appkill} how this
background is connected to the maximally symmetricc type IIB pp-wave
background of Ref.s~\cite{Blau:2001ne,Berenstein:2002jq} by a
coordinate transformation.

\subsection{Explicit space-like isometry from Penrose limit
of $\ads_5 \times S^5$}
\label{seciso}
Let us start from the ten dimensional $\ads_5 \times S^5$ solution
written in global coordinates. This solution has metric
\begin{equation}
\label{sol1}
ds^2 = R^2 \Big[ - \cosh^2 \rho\, dt^2 + d\rho^2 + \sinh^2 \rho \,
(d\Omega_3')^2 + (d\Omega_5)^2 \Big]
\end{equation}
and Ramond-Ramond five-form field strength
\begin{equation}
\label{sol2}
F_{(5)} = 2\, R^4 \left( \cosh\rho\, \sinh^3 \rho \,dt \,d\rho \,
d\Omega_3'  + d\Omega_5 \right) \spa
\end{equation}
where $R^4 = 4\pi g_s l_s^4 N$, $g_s$ being the string coupling, $l_s$
the string length and $N$ the flux on the $S^5$.  We embed the
five-sphere in $\C^3$ with coordinates $(a_1,a_2,a_3)$ by
\begin{equation}
\label{aemb}
a_1 = R \cos \theta \cos \psi \, e^{i \chi}  \spa a_2 = R \cos \theta
\sin \psi \, e^{i \phi}  \spa a_3 = R \sin \theta \, e^{i \alpha}\spa
\end{equation}
where $0 \leq \theta, \psi \leq \pi/2$ and $0 \leq \alpha, \phi, \chi
\leq 2\pi$.  The metric of the unit $S^5$ in these coordinates is
\begin{equation}
\label{thes5}
(d\Omega_5)^2 = d\theta^2 + \sin^2 \theta \,(d\alpha)^2 + \cos^2
\theta \, (d\Omega_3)^2\spa
\end{equation}
with the three-sphere part given by
\begin{equation}
\label{thes3}
(d\Omega_3)^2 = d\psi^2 + \sin^2 \psi \,(d\phi)^2 + \cos^2 \psi \,
(d\chi)^2 ~.
\end{equation}
We first consider the subgroup $U(1) \times SO(4)$ of the $SO(6)$
symmetry of the $S^5$, where the $U(1)$ factor corresponds to the
$\alpha$ angle. Define now the new coordinates
\begin{equation}
\label{phiLR}
\phi_L = \frac{1}{2} \left( \chi - \phi \right) \spa \phi_R =
\frac{1}{2} \left( \chi + \phi \right)\spa
\end{equation}
with $\partial/\partial \phi_{L,R} = \partial/\partial \chi  \mp
\partial/\partial \phi$.
Note that in these coordinates we have
\begin{equation}
\label{theLRs3}
(d\Omega_3)^2 = d\psi^2 + (d\phi_L)^2 + (d\phi_R)^2 + 2 \cos ( 2 \psi
) d\phi_L d\phi_R ~.
\end{equation}
The new coordinates \eqref{phiLR} define geometrically the relation
$SO(4) = SU(2)_L \times SU(2)_R$ where $\phi_{L,R}$ is the angle of
the Cartan generator of $SU(2)_{L,R}$.  Finally, define the light-cone
coordinates as
\begin{equation}
\tilde{z}^\pm = \frac{1}{2} ( t \pm \phi_R ) ~.
\end{equation}
The Penrose limit now consists of keeping $\alpha$ fixed while taking
\begin{equation}\label{plim1-2}
\begin{gathered}
R \rightarrow \infty \qquad \mbox{with} \qquad  \tilde{z}^+ = \mu z^+
\spa \tilde{z}^- = \frac{1}{\mu R^2} z^-\spa\\ \phi_L = \frac{z^1}{R}
\spa \psi = \frac{\pi}{4} - \frac{z^2}{R} \spa \rho = \frac{r}{R} \spa
\theta = \frac{\tilde{r}}{R}\spa
\end{gathered}
\end{equation}
and gives the pp-wave solution
\begin{equation}
\label{iso1}
ds^2 = - 4 dz^+ dz^- - \mu^2 z^I z^I (dz^+)^2  + dz^i dz^i + 4 \mu z^2
dz^1 dz^+ \spa
\end{equation}
with Ramond-Ramond five-form field strength
\begin{equation}
\label{iso2}
F_{(5)} = 2 \,\mu \,dz^+ \left(dz^1 dz^2 dz^3 dz^4 + dz^5 dz^6 dz^7
dz^8 \right) \spa
\end{equation}
where $i=1,...,8$ and $I=3,...,8$. Here $z^3, z^4$ are defined by $z^3
+ i z^4 = \tilde{r} e^{i \alpha}$ and $z^5,...,z^8$ are defined by
$r^2 = \sum_{I=5}^8 (z^I)^2$ and $dr^2 + r^2 (d\Omega'_3)^2 =
\sum_{I=5}^8 (dz^I)^2$.
We see that the pp-wave background \eqref{iso1}-\eqref{iso2} has an
explicit space-like isometry along the $z^1$ direction (together  with
the usual null killing isometry, which is the general feature of
pp-wave solutions). This is the background we will elaborate on in the
rest of the paper. We can define now the two currents
\begin{equation}\label{JLJR}
J_L = - \frac{i}{2} \frac{\partial}{\partial \phi_L } \spa J_R = -
\frac{i}{2} \frac{\partial}{\partial \phi_R }\spa
\end{equation}
corresponding to the coordinates \eqref{phiLR} so that   $J_{L,R}$ are
the Cartan generator of $SU(2)_{L,R}$.  Using these currents we can
write the generators
\begin{subequations}\label{stringvsgauge}
\begin{equation}
\label{relhlc}
\Hlc = - \frac{1}{\mu} P_+ = \frac{1}{\mu} i \frac{\partial}{\partial
z^+ } = i \frac{\partial}{\partial t } + i \frac{\partial}{\partial
\phi_R } = \Delta - 2 J_R \spa
\end{equation}
\begin{equation}
\label{pplus}
\mu P^+ = - \frac{\mu}{2} P_- = \frac{\mu}{2} \, i
\frac{\partial}{\partial z^- } = \frac{\Delta + 2 J_R}{2R^2} \spa
\end{equation}
\begin{equation}
\label{p1}
P_1 = -i \frac{\partial}{\partial z^1 } = \frac{1}{R} 2 J_L\spa
\end{equation}
\end{subequations}
corresponding to the light-cone Hamiltonian, light-cone momentum and
momentum along the $z_1$-direction respectively.  We can now use
eq.s~\eqref{stringvsgauge} to argue for a duality between type IIB
string theory on the pp-wave background \eqref{iso1}-\eqref{iso2} and
$\CN=4$ SYM theory.  Note that $\Delta = i \partial / \partial t$
corresponds to the scaling dimension of the gauge theory operators in
$\CN=4$ SYM theory.  Since the left-hand sides of
eq.s~\eqref{stringvsgauge} correspond to string theory quantities we
should demand that the right-hand sides are finite. This means that in
the $R \rightarrow \infty$ limit we need $J_R/R^2$ and $J_L/R$ to be
fixed. Using that $R^4 = 4\pi g_s l_s^4 N$ this means that the type
IIB string theory on the pp-wave background \eqref{iso1}-\eqref{iso2}
should be dual to $\CN=4$ SYM theory in the triple scaling limit
\begin{equation}
\label{limitn4}
N \rightarrow \infty \spa \frac{J_R}{\sqrt{N}} = \mbox{fixed} \spa
\frac{J_L}{N^{1/4}} = \mbox{fixed} \spa \gym^2 = \mbox{fixed}~.
\end{equation}
This scaling limit of $\CN=4$ SYM theory is very different from the
one of Ref.~\cite{Berenstein:2002jq}.  Since we know that on the
string theory side the two pp-wave backgrounds are related by a
coordinate transformation this implies an interesting connection
between coordinate transformations of a string theory background and
different sectors in the dual gauge theory. In  section \ref{secn5} we
will find the operators in $\CN=4$ SYM theory that correspond to the
various  string states in the type IIB string theory.

It might seem strange that we have a triple scaling limit rather than
a double-scaling limit as in Ref.~\cite{Berenstein:2002jq}.  However,
this is because we are looking at eigenstates of the momentum $P_1$,
so we are considering a certain sector of $P_1$.  The extra finite
quantity we get in this limit is then compensated by the fact that we
have one less bosonic zero-mode, as  we shall see in section
\ref{secgauge}. In the pp-wave as obtained in
Ref.~\cite{Berenstein:2002jq} there are also space-like isometries and
corresponding momenta. In principle one can consider what happens when
we keep these momenta fixed, to find an analogue of the triple scaling
limit we described here. The main difference is that these momenta
will no longer commute with the light-cone Hamiltonian, and we cannot
associate quantum numbers to both operators at the same time.

\subsection{A space-like circle from $\ads_5 \times S^5/\Z_M$}
\label{seccirc}

In this section we show that one can obtain a pp-wave background with
a space-like circle by taking a Penrose limit of $\ads_5 \times
S^5/\Z_M$. This gives a duality between the pp-wave background with a
space-like circle and a specific scaling limit of an $\CN=2$ quiver
gauge theory. Consider the orbifold $\C^2/\Z_M \times \C$ defined  by
the identification
\begin{equation}
(a_1,a_2,a_3) \equiv (\theta\, a_1,\theta^{-1} a_2, a_3) \spa \theta =
\exp \left( \frac{2\pi \, i}{M} \right) ~.
\end{equation}
It is well known that by placing $N$ coincident D3-branes  at the
orbifold singularity we get a four-dimensional $\CN=2$ quiver gauge
theory \cite{Douglas:1996sw}. The geometry dual to this is $\ads_5
\times S^5/\Z_M$ with $NM$ units of five-form flux on the $S^5$.

The $\CN=2$ quiver gauge theory (QGT) has gauge group
\begin{equation}
U(N)^{(1)} \times U(N)^{(2)} \times \cdots \times U(N)^{(M)}
\end{equation}
and consists of $M$ vector multiplets, one for each $U(N)$ group, and
$M$ bifundamental hypermultiplets.  The gauge coupling of each of the
$U(N)$ factors is $\gqgt^2 = 4\pi g_s M$ in terms of the string
coupling $g_s$. The $\CN=2$ QGT is described in detail in appendix
\ref{appqgt}.

To realize the dual geometry we consider again the embedding of $S^5$
given by eq.~\eqref{aemb}. With this embedding the $\C^2/\Z_M \times
\C$ orbifold induces an $S^5/\Z_M$ orbifold given by the
identifications
\begin{equation}
\label{idchiphi}
\chi \equiv \chi + \frac{2\pi}{M} \spa \phi \equiv \phi -
\frac{2\pi}{M} ~.
\end{equation}
This defines the $\ads_5 \times S^5/\Z_M$ geometry.  The radius $R$ of
$\ads_5$ and $S^5$ is given by $R^4 = 4 \pi l_s^4 g_s N M$.  Note that
we are working in the covering space of $\ads_5 \times S^5/\Z_M$ where
the background is given by \eqref{sol1}-\eqref{sol2}.
If we write the identifications \eqref{idchiphi} using the coordinates
defined in eq.~\eqref{phiLR} we find
\begin{equation}
\phi_L \equiv \phi_L + \frac{2\pi}{M} \spa \phi_R \equiv \phi_R\spa
\end{equation}
so that the $J_L$ current is quantized in units of $M/2$ while the
$J_R$ current is unaffected.
Consider now the Penrose limit \eqref{plim1-2}: we see that $z^1 = R
\phi_L$, so we clearly have that
\begin{equation}
z^1 \equiv z^1 + 2\pi \frac{R}{M}~.
\end{equation}
This means that if we send $M$ to infinity so that $R/M$ is fixed we
have that $z^1$ is a compact direction. Since we know from above that
$z^1$ is a manifest isometry of the pp-wave space
\eqref{iso1}-\eqref{iso2} we have the result that the Penrose limit
\eqref{plim1-2} of $\ads_5 \times S^5/\Z_M$ with $R/M$ fixed gives a
pp-wave with a space-like circle of radius $R/M$. As a consistency
check, note that since $J_L$ is quantized in units of $M/2$,  $P_1$ as
defined in eq.~\eqref{p1} is quantized in units of $M/R$, just as we
would expect for a quantized momentum on a circle of radius $R/M$.

We can now summarize the new Penrose limit of $\ads_5 \times S^5/\Z_M$
giving a pp-wave with a space-like circle of radius $R/M$. In terms of
$\CN=2$ QGT quantities the limit is a triple scaling limit
\begin{equation}
\label{qgtlim}
N \rightarrow \infty \spa \frac{M^3}{N} = \mbox{fixed} \spa
\frac{J_R}{M^2} = \mbox{fixed} \spa \frac{\gqgt^2}{M} = \mbox{fixed}~.
\end{equation}
This follows from having $R/M$, $\Hlc$ and $\mu P^+$ finite, as well
as having $g_s$ and $l_s$ finite and recalling that $R^4 = 4 \pi g_s
l_s^4 N M$.  In this limit it is implicitly understood that we
consider finite values of $J_L/M$. To see this, note that $J_L/M$ is
quantized in units of $1/2$, i.e. we are looking at finite values of
this quantum number.

In Ref.~\cite{Michelson:2002wa} it was shown that the pp-wave
background \eqref{iso1}-\eqref{iso2} with $z^1$ being compact has 24
supersymmetries.  The reason that putting a circle can break the
supersymmetry from 32 to 24 supercharges  is  that the Killing spinors
in the pp-wave background are non-trivial functions of the
coordinates, and they therefore need not be periodic along the circle.
We thus see that the number of preserved supersymmetries of the
$\CN=2$ QGT is enhanced from 16 to 24 in the limit \eqref{qgtlim}.

In the literature, several other Penrose limits of orbifold geometries
have been studied
\cite{Itzhaki:2002kh,Gomis:2002km,Alishahiha:2002ev,Kim:2002fp,Takayanagi:2002hv,Floratos:2002uh,Mukhi:2002ck,Alishahiha:2002jj,Oh:2002sv,Naculich:2002fh}.
Many of these also have supersymmetry enhancement, just as in our case.

The case that perhaps comes closest to our Penrose limit is the one of
Ref.s~\cite{Mukhi:2002ck,Alishahiha:2002jj}.  In these papers a
Penrose limit is taken of $\ads_5 \times S^5/\Z_M$ which instead gives
a null circle, thus providing a DLCQ version of the maximally
symmetric type IIB pp-wave of Ref.~\cite{Berenstein:2002jq}. The
Penrose limit of Ref.s~\cite{Mukhi:2002ck,Alishahiha:2002jj}
translates in taking $N \rightarrow \infty$ with $N \sim M \sim R^2
\sim J_R$. Comparing with our limit \eqref{qgtlim}, we thus see that
for the case of a null circle a completely different sector of the
(same) $\CN=2$ QGT is obtained. This is of course consistent with the
fact that whereas in our case supersymmetry is enhanced to 24
supercharges, in Ref.s~\cite{Mukhi:2002ck,Alishahiha:2002jj} there is
a supersymmetry enhancement to 32 supercharges. In the following
section we show how to obtain a compact null circle along with the
space-like circle by a similar construction as that of
Ref.s~\cite{Mukhi:2002ck,Alishahiha:2002jj}.

\subsection{Space-like circle and DLCQ from
$\ads_5 \times S^5/(\Z_{M_1} \times \Z_{M_2})$}
\label{secDLCQ}

In this section we show that one can obtain a pp-wave background with
a space-like circle which in addition has $z^-$ compact, by taking a
Penrose limit of $\ads_5 \times S^5/(\Z_{M_1} \times \Z_{M_2})$. This
provides a DLCQ version of the pp-wave background with a space-like
circle that we considered above. Also, it implies a duality between a
pp-wave background with a space-like circle and $z^-$ compact and a
special scaling limit of an $\CN = 1$ QGT, the SYM theory which is
dual to the $\ads_5 \times S^5/(\Z_{M_1} \times \Z_{M_2})$ background.

As explained in the introduction, having a Penrose limit that gives a
DLCQ version of the pp-wave with a space-like circle is the first step
towards finding a pp-wave/gauge-theory duality between matrix string
theory and gauge theory.  We furthermore find in section
\ref{secDLCQ2} the Penrose limit corresponding to the space-like
isometry being non-compact but the null one compact. Although we do
not discuss these two correspondences further  in the rest of the
paper, we find in the following  the appropriate scaling limits of the
gauge theories, providing the starting point of a more detailed
investigation.

Consider the orbifold $\C^3 / (\Z_{M_1} \times \Z_{M_2})$ defined by
the identifications
\begin{equation}\label{defn1}
\begin{gathered}
(a_1,a_2,a_3) \equiv (\theta_1 a_1,\theta_1^{-1} a_2, a_3) \spa
(a_1,a_2,a_3) \equiv (a_1,\theta_2^{-1} a_2, \theta_2 a_3)\spa\\
\theta_1 = \exp \left( \frac{2\pi \, i}{M_1} \right) \spa \theta_2 =
\exp \left( \frac{2\pi \, i}{M_2} \right) ~.
\end{gathered}
\end{equation}
Placing $N$ coincident D3-branes at the orbifold singularity we  get a
four-dimensional $\CN=1$ QGT theory on the branes
\cite{Douglas:1997de,Lawrence:1998ja}.  The dual geometry of this is
$\ads_5 \times S^5 /(\Z_{M_1} \times \Z_{M_2})$ with $N M_1 M_2$ units
of five-form flux on the $S^5$.
The $\CN=1$ gauge theory has gauge group
\begin{equation}
U(N)^{(1)} \times U(N)^{(2)} \times \cdots \times U(N)^{(M_1 M_2)}
\end{equation}
and consists of $M_1M_2$ vector multiplets and $3 M_1 M_2$
bifundamental chiral multiplets. The gauge coupling is the same for
all group factors, $\gqgt^2 = 4\pi g_s M_1M_2$ in terms of the string
coupling. This ${\mathcal N}=1$ QGT is described in appendix
\ref{appqgt1}.

Using the embedding of $S^5$ in $\C^3$ given by \eqref{aemb} we see
that the orbifold $\C^3 / (\Z_{M_1} \times \Z_{M_2})$ defined above
induces an $S^5 /(\Z_{M_1} \times \Z_{M_2})$ orbifold given by the
identifications
\begin{equation}
\label{iden3}
\chi \equiv \chi + \frac{2\pi}{M_1} n_1 \spa \phi \equiv \phi -
\frac{2\pi}{M_1} n_1 - \frac{2\pi}{M_2} n_2 \spa \alpha \equiv \alpha
+ \frac{2\pi}{M_2} n_2\spa
\end{equation}
for any $n_1,n_2 \in \Z$. The radius $R$ of both $\ads_5$ and $S^5$ is
given by $R^4 = 4\pi g_s l_s^4 N M_1 M_2$. Define now the coordinates
\begin{equation}
\gamma_1 = \frac{1}{2} ( \chi - \phi - \alpha ) \spa \gamma_2 =
\frac{1}{2} ( \chi + \phi - \alpha ) \spa \gamma_3 = \frac{1}{2} (
\chi + \phi + \alpha )\spa
\end{equation}
where $\partial/\partial \gamma_1 = \partial/\partial \chi -
\partial/\partial \phi \,,\, \partial/\partial \gamma_2 =
\partial/\partial \phi - \partial/\partial \alpha$ and
$\partial/\partial  \gamma_3 = \partial/\partial \chi +
\partial/\partial \alpha$.  We can then rewrite the identifications
\eqref{iden3} as
\begin{equation}
\label{idengam}
\gamma_1 \equiv \gamma_1 + \frac{2\pi}{M_1} n_1 \spa \gamma_2 \equiv
\gamma_2 + \frac{2\pi}{M_2} n_2 \spa \gamma_3 \equiv \gamma_3\spa
\end{equation}
for any $n_1,n_2 \in \Z$. In the Penrose limit \eqref{plim1-2} this
means that we should identify
\begin{equation}
\begin{gathered}
z^+ \equiv z^+ - \frac{\pi}{2 \mu M_2} n_2 \spa z^- \equiv z^- +
\frac{\mu R^2 \pi}{2 M_2} n_2\spa\\ z^1 \equiv z^1 + \frac{2\pi
R}{M_1} n_1 + \frac{\pi R}{M_2} n_2 \spa \alpha \equiv \alpha +
\frac{2\pi}{M_2} n_2\spa
\end{gathered}
\end{equation}
for any $n_1,n_2 \in \Z$.  We now see that if we let $M_1/R$ and
$M_2/R^2$ be fixed when $R \rightarrow \infty$ we find the
identifications
\begin{equation}
z^+ \equiv z^+ \spa z^- \equiv z^- + \frac{\mu R^2 \pi}{2 M_2} n_2
\spa z^1 \equiv z^1 + \frac{2\pi R}{M_1} n_1 \spa \alpha \equiv
\alpha\spa
\end{equation}
for any $n_1,n_2 \in \Z$ in the $R \rightarrow \infty$ limit.  We have
thus found a Penrose limit of $\ads_5 \times S^5 /(\Z_{M_1} \times
\Z_{M_2})$ that gives a circle in the $z^1$ direction of radius
$R/M_1$ and a circle in the $z^-$ direction of radius  $R_{-} \equiv
\mu R^2 / (4 M_2)$.

Turning to the currents, we define the three currents  $J_{(k)} = -
\frac{1}{2} i \, \partial/\partial \gamma_k$ ($k=1,2,3$).  {}From
\eqref{idengam} we then see that $J_{(1)}$ is quantized in units of
$M_1/2$, $J_{(2)}$ is quantized in units of $M_2/2$ and $J_{(3)}$ is
quantized in units of $1/2$.  Using \eqref{stringvsgauge} together
with $J_L = J_{(1)}$ and $J_R = J_{(2)} + J_{(3)} $ we have
\begin{equation}
\Hlc = \Delta - 2J_{(2)} - 2J_{(3)} \spa \mu P^+ = \frac{\Delta +
2J_{(2)} + 2J_{(3)}}{2R^2} \spa P_1 = \frac{2 J_{(1)}}{R}~.
\end{equation}
We now see that the Penrose limit in terms of $\CN=1$ QGT quantities
is the quadruple scaling limit
\begin{equation}
\label{quadlim}
N \rightarrow \infty  \spa \frac{M_1}{N} = \mbox{fixed} \spa
\frac{M_2}{N^2} = \mbox{fixed} \spa \frac{J_{(3)}}{M_2} = \mbox{fixed}
\spa  \frac{\gqgt^2}{M_1 M_2} = \mbox{fixed}~.
\end{equation}
A few remarks are in order here.  First, it is implicitly understood
in this limit that we consider finite quantum numbers of $J_{(1)}/M_1$
and $J_{(2)}/M_2$.  Moreover, we also consider finite quantum numbers
of $J_{\alpha} = - i \partial / \partial \alpha$. Clearly $P_1$ is
correctly quantized in units of $M_1/R$.  To see that \eqref{quadlim}
implies that $P^+$ is quantized we write
\begin{equation}
\mu P^+ = \frac{\Hlc + 8J_{(2)} + 4( J_{(3)} - J_{(2)} )   }{2R^2} =
\frac{\Hlc + 8J_{(2)} + 4( J_{(1)} + J_\alpha )}{2R^2} ~.
\end{equation}
Observing that $\Hlc/R^2$, $J_{(1)}/R^2$ and $J_\alpha/R^2$ all go to
zero, we find that in the limit
\begin{equation}
P^+ = \frac{4J_{(2)}}{\mu R^2} = \frac{2J_{(2)}}{M_2} \frac{1}{2
R_{-}}~.
\end{equation}
Thus we get the right quantization of $P^+$ since the fact that $z^-
\equiv z^- + R_{-}$ means, using $g_{+-} = -2$, that $P^+$ should be
quantized in units of $1/(2R_{-})$. Finally, we note that the above
method to obtain the null circle is similar to the way it was
constructed in Ref.s~\cite{Mukhi:2002ck,Alishahiha:2002jj}, although
many of the details  are obviously different.

%%%%%%%%%%%%%%%%%%%%%%%%%%%%%%%%%%%%%%%%%%%%%%%%%%%%%%%%%%%%%%%%%%
\subsection{Space-like isometry and DLCQ from
$\ads_5 \times S^5/ \Z_{M}$}
\label{secDLCQ2}

As anticipated in this section, we note that one can in fact also
construct a Penrose limit with a non-compact space-like isometry and a
compact null direction  for the pp-wave background. This could be
useful for constructing a  duality between matrix string theory on the
pp-wave and gauge theory since this background still admits a matrix
string description while we have an $\CN=2$ QGT on the gauge side,
which makes the duality simpler  than if we consider the one of the
previous section.

Since most of what we do will be simple repetitions of the previous
sections we will be brief. In accordance with previous notation, we
are now  considering the orbifold $\C \times \C^2 / \Z_{M}$ defined by
the identification
\begin{equation}
(a_1,a_2,a_3) \equiv (a_1,\theta^{-1} a_2, \theta a_3) \spa \theta =
\exp \left( \frac{2\pi \, i}{M} \right)~.
\end{equation}
Thus, in terms of angles on $S^5$ the $S^5 / \Z_M$ orbifold is defined
by the identification
\begin{equation}
\chi \equiv \chi \spa \phi \equiv \phi - \frac{2\pi}{M} \spa \alpha
\equiv \alpha + \frac{2\pi}{M}~.
\end{equation}
Repeating the steps of section \ref{secDLCQ} this means that
\begin{equation}
z^+ \equiv z^+ - \frac{\pi}{2 \mu M}  \spa z^- \equiv z^- + \frac{\mu
R^2 \pi}{2 M}  \spa z^1 \equiv z^1 + \frac{\pi R}{M}   \spa \alpha
\equiv \alpha + \frac{2\pi}{M}~.
\end{equation}
If we now keep $M/R^2$ fixed when $R\rightarrow \infty$ we get
\begin{equation}
z^+ \equiv z^+  \spa  z^- \equiv z^- + \frac{\mu R^2 \pi}{2 M}  \spa
z^1 \equiv z^1 \spa \alpha \equiv \alpha~.
\end{equation}
Thus, we have found a Penrose limit of $\ads_5 \times S^5 / \Z_M$ that
gives a circle in the $z^-$ direction of radius  $R_{-} \equiv \mu R^2
/ 4 M$.
The left and right currents of the above $\C^2 / \Z_M$ orbifold are
defined by
\begin{equation}
\tilde{J}_L = \frac{1}{2} ( J_\phi - J_\alpha ) \spa \tilde{J}_R =
\frac{1}{2} ( J_\phi + J_\alpha )
\end{equation}
where $J_\alpha = - i \partial/ \partial \alpha$ and $J_\phi = - i
\partial/ \partial \phi$.  As usual $\tilde{J}_L$ is quantized in
units of $M/2$ and $\tilde{J}_R$ in units of $1/2$.  We can now write
\begin{equation}
\Hlc = \Delta - 2J_R \spa \mu P^+ = \frac{\Delta + 2J_R}{2R^2} \spa
P_1 = \frac{2 J_L}{R}
\end{equation}
with
\begin{equation}
J_L = \frac{1}{2} ( J_\chi - \tilde{J}_R - \tilde{J}_L ) \spa J_R =
\frac{1}{2} ( J_\chi + \tilde{J}_R + \tilde{J}_L )
\end{equation}
where $J_\chi = - i \partial/ \partial \chi$.  The scaling limit of
$\CN=2$ QGT theory that we need to take is thus
\begin{equation}
N \rightarrow \infty \spa \frac{M}{N} = \mbox{fixed}  \spa
\frac{J_R}{M} = \mbox{fixed} \spa \frac{J_L}{\sqrt{M}} = \mbox{fixed}
\spa \frac{\gqgt^2}{M} = \mbox{fixed}~.
\end{equation}
Some remarks are needed.  First of all we also have $\tilde{J}_R/M$
and $J_\chi/M$ both fixed in the limit. Moreover we are considering
finite quantum numbers of $\tilde{J}_L / M$ and $J_\alpha$, this being
consistent with the fact that  $\tilde{J}_R$ scales like $M$. We also
see that $P^+ = (2 \tilde{J}_L /M) \times (1/(2R_-))$ and hence $P^+$
is correctly quantized.  Finally, if we compare with the limit of
section \ref{secDLCQ}, we see that the present limit  precisely
corresponds to a ``continuum version'' of the previous one, i.e. the
gauge theory operators that one would use would be the same, only
without the quantization condition  for the momentum $P_1$ along the
space-like direction.

\section{String Theory on pp-wave with space-like circle}
\label{secstring}

We now turn to the study of string theory living in the pp-wave
background with one compact space-like dimension, obtained from the
new Penrose limit presented in sections~\ref{seciso} and
\ref{seccirc}. The spectrum of the type IIB superstring in this
background has already been studied in Ref.~\cite{Michelson:2002wa}
(see  Ref.~\cite{Alishahiha:2002nf} for the T-dual type IIA case), but
we rederive it here in some detail, in view of the comparison with the
gauge theory operators that we present in section \ref{secgauge}.

\subsection{Bosonic sector}
\label{secstract}
Let us first consider the bosonic sector of the superstring, which
consists of the coordinate fields $Z^i$, $i=1,\ldots,8$. The bosonic
$\sigma$-model action describing a string living in the
background~\eqref{iso1} is given in conformal gauge by
\begin{multline}
    S^{\text{B}}=-\frac{1}{4\pi l_s^2}\int d\tau d\sigma \bigg[ -4
        \partial^\alpha Z^+ \partial_\alpha Z^- - \mu^2 Z^I Z^I
        \partial^\alpha Z^+ \partial_\alpha Z^+\\ + \partial^\alpha
        Z^i \partial_\alpha Z^i + 4 \mu Z^2 \partial^\alpha Z^+
        \partial_\alpha Z^1 \bigg]\spa
\end{multline}
where $I=3,\ldots,8$. The equation of motion that follows from varying
with respect to $Z^-$ is $\partial^\alpha \partial_\alpha Z^+ =
0\,$. This fact allows us to choose the light-cone gauge $Z^+ = l_s^2\
p^+ \tau\,$. The effective dynamics of the fields $Z^i$ is then
described by the following light-cone action
\begin{equation}\label{lcactz}
    S_{\text{lc}}^{\text{B}}=-\frac{1}{4\pi l_s^2}\int d\tau d\sigma
        \left[\partial^\alpha Z^i \partial_\alpha Z^i + f^2 Z^I Z^I -4
        f Z^2 \dot{Z}^1 \right]\spa
\end{equation}
where we have defined $f = l_s^2\ p^+ \mu\,$. Here and in the
following, dots and primes will denote respectively derivatives with
respect to the worldsheet coordinates $\tau$ and $\sigma$.

The equations of motion one gets from the light-cone
action~\eqref{lcactz} are
\begin{subequations}
\begin{align}
    (\partial^\alpha \partial_\alpha - f^2) Z^I &= 0\spa\qquad
    I=3,\ldots,8\spa \label{eomzI}\\ \partial^\alpha \partial_\alpha
    Z^1 + 2 f \dot{Z}^2 &= 0\spa\label{eomz1}\\ \partial^\alpha
    \partial_\alpha Z^2 - 2 f \dot{Z}^1 &= 0~. \label{eomz2}
\end{align}
\end{subequations}
Notice that the equations satisfied by the fields $Z^I$ are the same
as in the case of the standard maximally supersymmetric
pp-wave. Imposing on these the usual closed string boundary conditions
$Z^I(\tau,\sigma+2\pi)=Z^I(\tau,\sigma)\,$, one finds the solution
\begin{equation}\label{modezI}
    Z^I = i \frac{l_s}{\sqrt{2}} \sum_{n=-\infty}^{+\infty}
        \frac{1}{\sqrt{\omega_n}} \left(a_n^I
        e^{-i(\omega_n\tau-n\sigma)} - (a^\dagger_n)^I
        e^{i(\omega_n\tau-n\sigma)}\right)\spa
\end{equation}
where
\begin{equation}\label{omegan}
    \omega_n =  \sqrt{n^2+f^2}\spa\qquad \forall n\in\mathbb{Z}~.
\end{equation}
We use here a notation analogous to BMN, in which left-moving and
right-moving modes are respectively labeled by positive and negative
values of $n$. However, the form of the mode expansion differs from
the one usually given (compare e.g. Ref.~\cite{Metsaev:2002re}) in the
zero mode part, since we use an oscillator notation also for the zero
modes instead of center of mass position and momentum. The connection
between the two equivalent formalisms can be clarified by defining
\begin{equation}
    a_0^I = \frac{l_s}{\sqrt{2f}} \left( p^I - i\frac{f}{l_s^2} x^I
        \right) \qquad  (a^\dagger_0)^I = \frac{l_s}{\sqrt{2f}} \left(
        p^I + i\frac{f}{l_s^2} x^I \right)\spa
\end{equation}
so that the mode expansion~\eqref{modezI} of $Z^I$ gets modified as
\begin{equation}
    Z^I = \cos(f\tau) z^I + \frac{\sin(f\tau)}{f} l_s^2 p^I  + i
       \frac{l_s}{\sqrt{2}}\sum_{n\neq0} \frac{1}{\sqrt{\omega_n}}
       \left(a_n^I e^{-i(\omega_n\tau-n\sigma)} - (a^\dagger_n)^I
       e^{i(\omega_n\tau-n\sigma)}\right)\spa
\end{equation}
which is the usual expression given in the literature.

To solve the eq.s~\eqref{eomz1}-\eqref{eomz2}, it is useful to
decouple them by introducing a complex field
\begin{equation}\label{zcomplexfield}
    Z = Z^1 + i Z^2\spa
\end{equation}
in terms of which the above equations read as follows
\begin{subequations}
\begin{align}
    \partial^\alpha \partial_\alpha Z - 2 i f \dot{Z} &=
    0\spa\label{eomz}\\ \partial^\alpha \partial_\alpha \bar{Z} + 2 i
    f \dot{\bar{Z}} &= 0~.
\end{align}
\end{subequations}
One can see that a solution of the form $Z = e^{-i f \tau} Y$ solves
\eqref{eomz} if $Y$ satisfies the same equation as the fields $Z^I$,
that is $(\partial^\alpha \partial_\alpha - f^2) Y = 0\,$. We also
have to specify the boundary conditions for $Z^1$ and $Z^2$. We have
seen in section~\ref{seccirc} that in the case under consideration the
coordinate $Z^1$ is compact with radius $R_T=R/M$. We then want to
allow for winding modes around the compact direction $Z^1$, by
implementing the boundary conditions
$Z(\tau,\sigma+2\pi)-Z(\tau,\sigma)=2\pi R_T m$ (and the same for
$\bar{Z}$). Therefore the solutions of the equations of motion can be
written in the following form
\begin{equation}
    Z = e^{-i f \tau} Y + mR_T\sigma \spa \qquad \bar{Z} = e^{i f
    \tau} \bar{Y} + mR_T\sigma\spa
\end{equation}
where $Y$ and $\bar{Y}$ have the following mode expansions
\begin{subequations}\label{modey}
\begin{align}
    Y &= i\ {l_s} \sum_{n=-\infty}^{+\infty} \frac{1}{\sqrt{\omega_n}}
        \left(a_n e^{-i(\omega_n\tau-n\sigma)} - \tilde{a}^\dagger_n
        e^{i(\omega_n\tau-n\sigma)}\right)\spa\\  \bar{Y} &= i\ {l_s}
        \sum_{n=-\infty}^{+\infty} \frac{1}{\sqrt{\omega_n}}
        \left(\tilde{a}_n e^{-i(\omega_n\tau-n\sigma)} - {a}^\dagger_n
        e^{i(\omega_n\tau-n\sigma)}\right)~.
\end{align}
\end{subequations}
{} From the expression of the action \eqref{lcactz}, one can also
 compute the conjugate momenta
\begin{equation}\label{piz}
    \Pi_I = \frac{\dot{Z}^I}{2\pi l_s^2}\spa\qquad \Pi_1 =
    \frac{\dot{Z}^1-2fZ^2}{2\pi l_s^2}\spa\qquad \Pi_2 =
    \frac{\dot{Z}^2}{2\pi l_s^2}\spa
\end{equation}
so that the classical bosonic hamiltonian $\Hlct^{\text{B}} =\int
d\sigma ( \Pi_i \dot{Z}^i - \mathcal{L} )$ is given by
\begin{equation}\label{HZ}
    \Hlct^{\text{B}} = \frac{1}{4\pi l_s^2} \int_0^{2\pi} d\sigma
        \left[ \dot{Z}^i \dot{Z}^i + (Z^i)' (Z^i)' + f^2 Z^I
        Z^I\right]~.
\end{equation}

\subsection{Fermionic sector}

Let us now consider the fermionic sector of the theory. The field
content consists in  real anticommuting fields $\theta^{Aa}$, where
$A=1,2$ and the index $a$ runs from 1 to 16 as appropriate for a
Majorana-Weyl spinor in ten dimensions. As discussed in
Ref.~\cite{Metsaev:2002re}, in the light-cone gauge
\begin{equation}\label{fermlcc}
    Z^+ = l_s^2\ p^+ \tau\spa\qquad \Gamma^+ \theta^A=0\spa
\end{equation}
the Green-Schwarz fermionic action is given by
\begin{equation}\label{fermact}
    S^{\text{F}}_{\text{lc}} = \frac{i}{4\pi l_s^2} \int d\tau d\sigma
        \left[ \left( \eta^{\alpha\beta} \delta_{AB} -
        \epsilon^{\alpha\beta}(\sigma_ 3)_{AB} \right) \partial_\alpha
        Z^+ \bar{\theta}^A \Gamma_+ ( \mathcal{D}_\beta \theta )^B
        \right]\spa
\end{equation}
where here and in the following the $\sigma_k$'s are the Pauli
matrices. In the case at hand, where only the Ramond-Ramond five-form
field strength is turned on, the generalized covariant derivative
appearing in the above action assumes the form
\begin{equation}\label{fermD}
    \mathcal{D}_\alpha = \partial_\alpha +\frac{1}{4} \partial_\alpha
        Z^+ \left( \omega_{\rho\sigma +} \Gamma^{\rho\sigma}
        -\frac{1}{2\cdot 5!} F_{\lambda\nu\rho\sigma\kappa}
        \Gamma^{\lambda\nu\rho\sigma\kappa} i\sigma_2 \Gamma_+
        \right)\spa
\end{equation}
where $\omega$ is the spin connection of the metric. We recall from
eq.~\eqref{iso2} that the five-form $F_{(5)}$ has the components
$F_{+1234}=F_{+5678}=2\mu$ turned on, while one can compute from the
metric \eqref{iso1} the only relevant component $\omega_{12+}=-\mu$ of
the spin connection.

We want to express the action in terms of the canonical
eight-component GS spinors $S^{Ab}$, $b=1,\ldots,8$, which are related
to the fields $\theta^a$ through the relation $\Gamma^{+-}\theta^{Aa}
= 2^{-3/4}\sqrt{p^+} S^{Aa}$. We will be using a chiral representation
for the $32\times32$ gamma-matrices of $SO(1,9)$, such that
\mbox{$\Gamma^i = \begin{pmatrix} 0&\hat{\gamma}^i \\ \hat{\gamma}^i
&0\end{pmatrix}\,$,} $i=1,\ldots8$. We also define $8\times8$ matrices
$\gamma^i$ as $(\gamma^{i_1\dotsm i_n})^{ab}=(\hat{\gamma}^{i_1\dotsm
i_n})^{ab}$, $a,b=1,\ldots,8$. Notice that we can work with
eight-component spinors because of the light-cone condition
\eqref{fermlcc}. With some manipulations, we can then rewrite the
fermionic light-cone action as
\begin{equation}\label{fermactS}
    S^{\text{F}}_{\text{lc}} = \frac{i}{2\pi} \int d\tau d\sigma
        \left[ S^1 \left( \partial_+ - \frac{f}{2} \gamma^{12}
        \right)S^1 + S^2 \left( \partial_- - \frac{f}{2} \gamma^{12}
        \right)S^2 - 2f S^1 \Pi S^2\right]\spa
\end{equation}
where $\partial_{\pm} = \partial_\tau \pm \partial_{\sigma}\,$,
$\Pi=\gamma^{1234}$ and we recall that $f=l_s^2\ p^+\mu\,$.

{} From the action \eqref{fermactS} we can derive the equations of
motion
\begin{subequations}\label{fermeom}
\begin{align}
    \left( \partial_+ - \frac{f}{2}\gamma^{12} \right) S^1 - f \Pi S^2
    &= 0\spa   \\ \left( \partial_- - \frac{f}{2}\gamma^{12} \right)
    S^1 + f \Pi S^1 &= 0~.
\end{align}
\end{subequations}
It is useful to observe that a field of the form $S^A =
e^{\frac{f}{2}\gamma^{12}\tau}\Sigma^A$ satisfies the above equations
if the fields $\Sigma^A$ obey the equations of motion of the fermionic
fields in the usual pp-wave background \cite{Metsaev:2002re}:
\begin{subequations}
\begin{align}
    \partial_+ \Sigma^1 - f \Pi \Sigma^2 &= 0\spa   \\ \partial_-
    \Sigma^1 + f \Pi \Sigma^1 &= 0~.
\end{align}
\end{subequations}
Therefore, the mode expansions of the fields $S^A$ which solve the
eq.s~\eqref{fermeom} can be easily written as
\begin{subequations}\label{fermmodes}
\begin{align}
    S^1 &= e^{\frac{f}{2}\gamma^{12}\tau} \Big\{ \Big[ c_0\
        e^{-if\tau} S_0  \nn\\ &\qquad\qquad - \sum_{n>0} c_n\
        e^{-i\omega_n\tau} \Big( S_n e^{i n\sigma} +
        \tfrac{\omega_n-n}{f} S_{-n} e^{-i n\sigma} \Big) \Big] +
        \text{h.c.}\Big\}\spa \\ S^2 &= e^{\frac{f}{2}\gamma^{12}\tau}
        \Big\{ \Big[ - c_0\ e^{-if\tau} i\Pi S_0 \nn\\ &\qquad\qquad -
        i\Pi \sum_{n>0} c_n\ e^{-i\omega_n\tau} \Big( S_{-n} e^{- i
        n\sigma} - \tfrac{\omega_n-n}{f} S_n e^{i n\sigma} \Big) \Big]
        + \text{h.c.}\Big\}\spa
\end{align}
\end{subequations}
where, for all values of $n$, $\omega_n$ is defined as in
eq.~\eqref{omegan} as $\omega_n = \sqrt{n^2 + f^2}$, while $c_n=
\tfrac{1}{\sqrt{2}}\big[1+\big(\tfrac{\omega_n-n}{f}\big)^2\big]^{-1/2}$. We
have imposed the usual closed string boundary conditions $S^A
(\tau,\sigma+2\pi) = S^A (\tau,\sigma)$, which are left unchanged by
the compactification along $Z^1$.

{} From the action \eqref{fermactS} one can also compute the fermionic
conjugate momenta $\lambda_A = i S^A  / 2 \pi$ which are needed for
obtaining the fermionic part of the classical hamiltonian
\begin{align}\label{fermHS}
    \Hlct^{\text{F}} &= \frac{i}{2\pi} \int_{0}^{2\pi} d\sigma \left[
        S^1(S^1)'- S^2 (S^2)' -\frac{f}{2} (S^1\gamma^{12}S^1+
        S^2\gamma^{12}S^2) -2f S^1\Pi S^2\right]\nn\\ &=
        \frac{i}{2\pi} \int_{0}^{2\pi} d\sigma \left( S^1\dot{S}^1+
        S^2\dot{S}^2\right)\spa
\end{align}
where in the last equality we have made use of the equations of motion
\eqref{fermeom}.

Finally, one must also analyze the constraint coming from the
vanishing of the worldsheet energy-momentum tensor that is given by
the condition $\int d\sigma [\Pi_i (Z^i)' + \lambda_A (S^A)' ] = 0\,$,
which in terms of the fields $Z^i$ and $S^A$ becomes
\begin{equation}\label{constrZ}
    \int_0^{2\pi} d\sigma \left[ \dot{Z}^i (Z^i)' - 2 f Z^2 (Z^1)' + i
         l_s^2 S^A (S^A)' \right] = 0~.
\end{equation}

\subsection{The String spectrum}\label{secquant}

Let us now turn to the spectrum of the theory introduced in the
previous sections. To quantize the theory, one has to impose the
canonical equal time (anti)commutation relations\footnote{For the
fermionic fields one has to take into account the second class
constraints coming from the expression of the fermionic momenta
$\lambda_A$, which must be treated following the Dirac quantization
procedure (see for instance Ref.~\cite{Metsaev:2001bj} for its
application to the standard pp-wave background).}
\begin{equation}
\begin{split}
        [Z^i (\tau,\sigma), \Pi_j (\tau,\sigma')] &= i \delta^{ij}
        \delta(\sigma - \sigma')\spa\\
        \{S^{Aa}(\tau,\sigma),S^{Bb}(\tau,\sigma')\} &= \frac{1}{2}
        \delta^{AB} \delta^{ab}\delta(\sigma-\sigma')\spa
\end{split}
\end{equation}
which imply the following relations for the oscillators
\begin{gather}\label{oscalgebra}
     [a_n^I, (a^J_m)^\dagger ]  = \delta^{IJ} \delta_{nm}\spa\qquad
            [a_n, {a}^\dagger_m]  = [\tilde{a}_n, \tilde{a}^\dagger_m]
            = \delta_{nm}\spa\nn\\ \{S^a_n,(S^b_m)^\dagger \} =
            \delta^{ab}\delta_{nm}~.
\end{gather}

The spectrum of the theory is then obtained by acting with the raising
operators $a^\dagger_n\,$, $\tilde{a}^\dagger_n\,$, $(a^I_n)^\dagger$
and $(S^\dagger_{n})^b$ on the vacuum defined by
\begin{equation}\label{stringvac}
    a_n\lvert 0 \rangle = \tilde{a}_n\lvert 0 \rangle = a_n^I\lvert 0
    \rangle = S_{n}^b\lvert 0 \rangle = 0 \qquad \forall
    n\in\mathbb{Z}~.
\end{equation}

We can now express the light-cone hamiltonian and energy-momentum
constraint  in terms of the oscillators. For this purpose, let us
introduce the following number operators
\begin{gather}
    N_n = a^\dagger_n a_n\spa \qquad \widetilde{N}_n =
    \tilde{a}^\dagger_n \tilde{a}_n\spa \nn\\  N^{(I)}_n =
    (a^I_n)^\dagger a_n^I\spa \qquad F^{(b)}_n = (S^b_{n})^\dagger
    S^{b}_n~.\label{numbops}
\end{gather}
After some algebra, one gets the following expression for the
hamiltonian
\begin{equation}\label{Hlc}
    \Hlct = \Hlct^{\text{B}} + \Hlct^{\text{F}}\spa
\end{equation}
where the bosonic and fermionic parts, coming respectively from
eq.s~\eqref{HZ} and \eqref{fermHS}, can be expressed as follows
\begin{subequations}
\begin{align}
    \Hlct^{\text{B}} &= 2f N_0 + \frac{m^2R_T^2}{2l_s^2} +
        \sum_{n\neq0} \left[ ( \omega_n + f )  N_n + ( \omega_n - f )
        \widetilde{N}_n \right] + \sum_{n=-\infty}^{+\infty}
        \sum_{I=3}^8 \omega_n N^{(I)}_n \spa\label{HlcB}\\
        \Hlct^{\text{F}} &= \sum_{n=-\infty}^{+\infty} S^\dagger_n
        \left(\omega_n + i\frac{f}{2}\gamma^{12}\right)
        S_n~.\label{HlcF}
\end{align}
\end{subequations}
Notice that with our choice of vacuum \eqref{stringvac} the
hamiltonian comes with vanishing zero-point energy.

Since the eigenvalues of $i \gamma^{12}$ are $\pm 1$, each with
multiplicity four, we see from the expression \eqref{HlcF} that a
``splitting'' of energy eigenvalues is realized at each level of
excitation of the fermionic oscillators.  With a suitable choice of
basis, we can then rewrite the fermionic part of the hamiltonian as
\begin{equation}
    \Hlct^{\text{F}} = \sum_{n=-\infty}^{+\infty}\left[ \sum_{b=1}^4
        \left( \omega_n - \tfrac{f}{2} \right) F_n^{(b)} +
        \sum_{b=5}^8 \left( \omega_n + \tfrac{f}{2} \right)
        F_n^{(b)}\right]~.
\end{equation}
In view of the comparison with gauge theory states, we are
particularly interested in the states of the spectrum which are
obtained by acting on the vacuum with a single bosonic or fermionic
zero-mode creation operator. The spectrum of such states is summarized
in Table~\ref{tableZM}.

\begin{table}
\begin{center}
\begin{tabular}{|ccc|c|}
\hline \multicolumn{3}{|c|}{State} & $\Hlc\equiv\Hlct/f$\\ \hline \ \
$a_0^\dagger \lvert 0 \rangle$ &&& 2\\ \hline $(a_0^I)^\dagger \lvert
0 \rangle$ & for & $I =3,\ldots,8$ & 1\\ \hline $(S_0^b)^\dagger
\lvert 0 \rangle$ & for & $b =1,2,3,4$ & 1/2\\ \hline $(S_0^b)^\dagger
\lvert 0 \rangle$ & for & $b =5,6,7,8$ & 3/2\\ \hline
\end{tabular}
\end{center}
\caption{The lowest-lying zero-mode states of the superstring on the
compactified pp-wave.\label{tableZM}}
\end{table}

Finally, the constraint \eqref{constrZ} coming from the
energy-momentum tensor assumes the following form
\begin{equation}\label{prelevmat}
    \sum_{n\neq0} n \left[ N_n + \widetilde{N}_n + \sum_{I=3}^8
        N^{(I)}_n + \sum_{b=1}^8 F^{(b)}_n \right] =
        \frac{\sqrt{f}}{l_s} mR_T ( \tilde{a}^\dagger_0 + \tilde{a}_0
        ) ~.
\end{equation}
In order to understand better the structure of the above constraint,
we have to take into account the quantization of momentum along the
direction $Z^1\,$.  This amounts to the condition $\int d\sigma
\Pi_1=k/R_T\,$, which by means of eq.s~\eqref{zcomplexfield} and
\eqref{piz} can be expressed as follows
\begin{equation}\label{Pi1calc}
    \frac{1}{4\pi l_s^2} \int_0^{2\pi} d\sigma \left[
    \dot{Z}+\dot{\bar{Z}}+2if \left( Z-\bar{Z} \right)\right] =
    \frac{\sqrt{f}}{l_s} ( \tilde{a}^\dagger_0 + \tilde{a}_0 ) =
    \frac{k}{R_T}\spa
\end{equation}
where in the second step we have used the mode expansions of the
fields $Z$, $\bar{Z}$.  Therefore, from eq.~\eqref{prelevmat} we can
now obtain the following final expression for the level-matching
condition:
\begin{equation}\label{levmat}
    \sum_{n\neq0} n \left[ N_n + \widetilde{N}_n +  \sum_{I=3}^8
N^{(I)}_n + \sum_{b=1}^8 F^{(b)}_n \right] = k m ~.
\end{equation}
A similar computation to the one performed in eq.~\eqref{Pi1calc}
shows that the oscillators $a_0$, $a_0^\dagger$ (and consequently the
state $a_0^\dagger \lvert 0 \rangle$) are associated only to the
uncompactified direction $Z^2$.

It is important to realize that we wrote everything in terms of
oscillators for convenience, but that the quantization of $
\tilde{a}^\dagger_0 + \tilde{a}_0 $ in units of $l_s/R_T \sqrt{f}$
implies that the Hilbert space associated to the modes
$\tilde{a}^\dagger_0\,$, $\tilde{a}_0$ is not the usual harmonic
oscillator Hilbert space. The latter is isomorphic to $L^2(\R)$,
whereas the Hilbert space we need is $L^2(S^1)$. Any element of
$L^2(S^1)$ has infinite norm in $L^2(\R)$, since they correspond to
periodic functions on the real line. One can indeed verify that any
harmonic oscillator state that is an eigenstate of $
\tilde{a}^\dagger_0 + \tilde{a}_0 $ has infinite norm.

\section{Space-like circle and $\CN =2$  quiver gauge theory}
\label{secgauge}

In this section we put forward our proposal for the gauge theory
operators of the ${\cal N}=2$ QGT surviving the scaling limit
\eqref{qgtlim} that are dual to the states of type IIB string theory
on the pp-wave background with a space-like circle described so far.

\begin{figure}[ht]
\begin{center}
{\scalebox{1}{\includegraphics{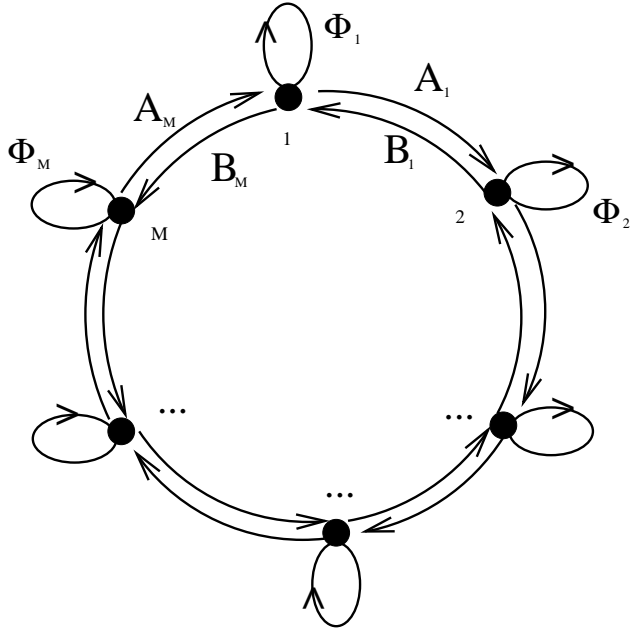}}} \caption{\small The quiver
diagram for the $\CN=2$ QGT that we consider.  Each dot represents a
$U(N)$ gauge factor. For the matter fields, we  use a ${\mathcal N}=1$
notation: each line between dots correspond to a complex scalar of
either type ${A}_I$ or ${B}_I$ ($I=1,...,M$), the two making-up the
corresponding bifundamental hypermultiplet.  Arrows go from
fundamental to anti-fundamental representations of the corresponding
gauge groups. The fermionic partner of each bosonic field is implicit
in the figure.} \label{quiv1}
\end{center}
\end{figure}

A detailed analysis of the relevant ${\cal N}=2$ QGT is presented in
appendix~\ref{appqgt}, with all definitions and notations that will be
used in the present section. For the sake of simplicity, let us just
briefly recall here the field content of the theory.  We have a
product gauge group with $M$ gauge factors $U(N)$ and $M$
corresponding vector multiplets whose complex scalars we denote by
$\Phi_I$ ($I=1,...,M$). Each of these scalars transforms in the
adjoint representation of the corresponding $I$-th gauge group factor.
Using ${\cal N}=1$ notation we define $\psi_I$ as the fermionic
partner of the gauge field $A_{\mu I}$ and $\psi_{\Phi I}$ as the
fermionic partner of the complex scalar $\Phi_I$. The matter content
is given by $M$ bifundamental hypermultiplets whose quaternionic
scalars we define in ${\cal N}=1$ notation to be $A_I,B_I$ where $A_I$
transforms in the $({ N}_I,{ \bar N}_{I+1})$ and $B_I$ transforms in
the $({ \bar N}_I,{ N}_{I+1})$, where ${ N}_{I}$ (${\bar N}_{I}$)
represents the fundamental (anti-fundamental) representation  of the
gauge group $U(N)^{(I)}$. The fermionic fields $\chi_{A,I} ,
\chi_{B,I}$ are the superpartners of the complex scalars $A_I$ and
$B_I$ respectively. The field content of the gauge theory can be
conveniently summarized by a quiver diagram, see figure \ref{quiv1}.

When writing down gauge theory operators in the following we use
mostly the ${\cal N}=4$ notation of the $U(NM)$ parent gauge theory,
where the orbifold projection that determines the QGT restricts the
form of the fields to those given in appendix~\ref{appqgt}. It is thus
understood that, for example, whenever the field $A$, $B$ or $\Phi$
appears we mean the specific expressions in
\eqref{AmPAB}. Analogously, by $\psi$, $\psi_\Phi$, $\chi_A$ and
$\chi_B$ we mean the specific expressions \eqref{psichis}. In
appendix~\ref{appn2} we present the resulting forms in ${\cal N}=2$
notation and provide many of the calculational details for the results
of this section.  As the essential building blocks for these operators
are the fields $A$, $B$, $\Phi$ and their conjugates (and similarly
for the fermions), we have listed in Table~\ref{tableAB} their
relevant quantum numbers, being the conformal dimension $\Delta$ and
the eigenvalues under the Cartan currents $J_{L,R}$ of the
$SU(2)_{L,R}$,   where $SU(2)_L \times SU(2)_R \simeq SO(4)$ is a
subgroup of the  original $SO(6)$ R-symmetry group of the parent gauge
theory.   The  $J_{L,R}$ eigenvalues for the bosons are obtained by
respectively identifying  $A$, $B$ and $\Phi$ with $a_1$, $a_2$, $a_3$
defined in eq.~\eqref{aemb}, and then using eq.s~\eqref{phiLR} and
\eqref{JLJR}. The eigenvalues of the fermions follow similarly using
supersymmetry and the fact they are in the spinor representation
$(\mathbf{4} + \mathbf{\bar{4}})$ of $SO(6)$. Note that the R-symmetry
of the $\CN=2$ QGT is $SU(2) \times U(1)$ where the $SU(2)$ is
identified with $SU(2)_R$ and the $U(1)$ corresponds to the angle
$\alpha$ in the direction $a_3$ which is inert under the action of the
orbifold.

\begin{table}
\begin{center}
\begin{tabular}{|c|c|c|c|c|c|c|c|c|c|c|c|c|c|c|}
\hline  & $A$ & $B$ & $\bar{A}$ & $\bar{B}$ & $\Phi$ & $\bar{\Phi}$ &
$\chi_A$ & $\bar{\chi}_B$ & $\chi_B$ & $\bar{\chi}_A$ &
$\bar{\psi}_\Phi$ &  $\bar{\psi}$ & $\psi_\Phi$ & $\psi$  \\ \hline
$\Delta$ & $1$ & $1$ & $1$ & $1$ & $1$ & $1$ & $\frac{3}{2}$
&$\frac{3}{2}$ &$\frac{3}{2}$ &$\frac{3}{2}$ &$\frac{3}{2}$
&$\frac{3}{2}$ &$\frac{3}{2}$ &$\frac{3}{2}$ \\ \hline  $J_L$ &
$\frac{1}{2}$ &$-\frac{1}{2}$ & $-\frac{1}{2}$ & $\frac{1}{2}$  & $0$
& $0$ & $\frac{1}{2}$ &$\frac{1}{2}$ & $-\frac{1}{2}$ & $-\frac{1}{2}$
&  $0$ & $0$ & $0$ & $0$ \\ \hline  $J_R$ & $\frac{1}{2}$
&$\frac{1}{2}$ & $-\frac{1}{2}$ & $-\frac{1}{2}$  & $0$ & $0$ & $0$ &
$0$ & $0$ & $0$ & $\frac{1}{2}$ &$\frac{1}{2}$ & $-\frac{1}{2}$ &
$-\frac{1}{2}$ \\ \hline
\end{tabular}
\end{center}
\caption{$\Delta$, $J_L$ and $J_R$ eigenvalues for bosonic and
fermionic operators \label{tableAB}}
\end{table}

\subsection{Predictions from string theory
\label{secgenrem}}

We start with some general remarks. It is first worth to point out
that the dictionary between the string theory and the gauge theory is
given by eq.s~\eqref{stringvsgauge}, which we repeat here
\begin{equation}
\label{dict} \Hlc = \Delta - 2 J_R \spa \mu P^+ = \frac{\Delta + 2
J_R}{2R^2} \spa k \equiv R_T P_1 = \frac{2}{M} J_L\spa
\end{equation}
where
\begin{equation}
 \label{dict2} R^4 = 4\pi g_s l_s^4 N M \spa g_{\rm QGT}^2 = 4 \pi g_s
M ~.
\end{equation}
String theory on the pp-wave solution \eqref{iso1}-\eqref{iso2} with
$z^1$ being compact with radius $R_T$ now predicts the light-cone
Hamiltonian \eqref{Hlc}.  Using that $\mu P^+ = 2 J_R / \sqrt{ \gqgt^2
N}$, which follows from eq.~\eqref{dict} using eq.~\eqref{dict2}, we
can write the prediction of the energy eigenvalues from free string
theory as
\begin{subequations}\label{strpredbf1}
\begin{multline}
\Hlc^B = m^2 \frac{\gqgt^2 N}{4 M^2 J_R} + 2N_0 + \sum_{n \neq 0}
    \left[ 1  + \sqrt{ 1 + n^2 \frac{\gqgt^2 N}{4J_R^2} } \right]
    N_n \\ + \sum_{n \neq 0} \left[ -1  + \sqrt{ 1 + n^2
    \frac{\gqgt^2 N}{4J_R^2} } \right] \widetilde{N}_n +
    \sum_{n=-\infty}^\infty \sqrt{ 1 + n^2 \frac{\gqgt^2
    N}{4J_R^2} } \sum_{I=3}^8 N_n^{(I)} \spa \label{strpred1}
\end{multline}
\begin{equation}
\Hlc^F = \sum_{n=-\infty}^\infty \left[-\frac{1}{2}+ \sqrt{ 1 + n^2
        \frac{\gqgt^2 N}{4J_R^2} } \right] \sum_{b=1}^4 F_n^{(b)} +
        \sum_{n=-\infty}^\infty \left[ \frac{1}{2} + \sqrt{ 1 + n^2
        \frac{\gqgt^2 N}{4J_R^2} }  \right] \sum_{b=5}^8 F_n^{(b)}
        \spa\label{strpredf1}
\end{equation}
\end{subequations}
where we have defined
\begin{equation}
\Hlc = \Hlc^B + \Hlc^F~.
\end{equation}
It is important to notice that here we use the rescaled light-cone
Hamiltonian $\Hlc = \Hlct / f$, where $\Hlct$ is the Hamiltonian
discussed in section \ref{secstring}, since this is more natural when
we compute the energy eigenvalues from gauge theory.  We further
remind the reader that $m$ is the winding number and that the number
counting operators $N_n$, $\widetilde{N}_n$, $N^{(I)}_n$ and
$F^{(b)}_n$ have been defined in eq.~\eqref{numbops}.  Note also that
we used the fact that the radius of the space-like circle is $R_T =
R/M$.

To make the comparison between string theory and gauge theory even
more clear, we note that if we consider eq.s~\eqref{strpredbf1} to
first order in $\gqgt^2 N / J_R^2\,$, we get
\begin{subequations}\label{strpredbf2}
\begin{eqnarray}
\Hlc^B &=& m^2 \frac{\gqgt^2 N}{4 M^2 J_R} +2 N_0 + \sum_{n \neq 0}
        \left[ 2  + n^2 \frac{\gqgt^2 N}{8J_R^2} \right] N_n + \sum_{n
        \neq 0} n^2 \frac{\gqgt^2 N}{8J_R^2} \widetilde{N}_n \nn \\ &&
        + \sum_{n=-\infty}^\infty \left[ 1 + n^2 \frac{\gqgt^2
        N}{8J_R^2} \right] \sum_{I=3}^8 N_n^{(I)}\spa \label{strpred2}
\end{eqnarray}
\begin{equation}
\Hlc^F = \sum_{n=-\infty}^\infty \left[ \frac{1}{2} + n^2
        \frac{\gqgt^2 N}{4J_R^2} \right] \sum_{b=1}^4 F_n^{(b)} +
        \sum_{n=-\infty}^\infty \left[ \frac{3}{2} + n^2 \frac{\gqgt^2
        N}{4J_R^2} \right] \sum_{b=5}^8 F_n^{(b)}~.\label{strpredf2}
\end{equation}
\end{subequations}
In the following, after having identified the various string states as
gauge theory operators, we reproduce the free string  spectrum
\eqref{strpredbf2} from gauge theory along with the level-matching
condition \eqref{levmat}.

\subsection{Bosonic gauge theory operators without anomalous dimensions
\label{secgt1}}

Let us first consider the bosonic part of the spectrum (the fermionic
part will be discussed in section \ref{fermoper}). We start by
considering operators which are  chiral primaries in the gauge
theory. The chiral primary gauge theory operators corresponds to the
BPS states of the superconformal algebra. This means that they do not
receive quantum corrections, e.g. that they do not have any anomalous
dimension.  Just as in the general AdS/CFT correspondence, the chiral
primaries of the gauge theory are dual to the supergravity states of
the given background.

{} From eq.~\eqref{strpred1} we see that we should reproduce the
spectrum
\begin{equation}
\label{sugrapred}
 \Hlc = 2 N_0 + \sum_{I=3}^8 N_0^{(I)}~.
\end{equation}
This we know from the fact that the supergravity modes corresponds to
the zero-modes of the string theory or the  first quantized string
theory vacua.

\subsubsection*{Ground states}

We first consider the ground states%
\footnote{Note that we do not consider states with non-zero winding
$m$ as ground states since they have $\Hlc \neq 0$.}, i.e. the states
which have $\Hlc = 0$.  Consider first the ground state $| k = 0 , m =
0 \rangle$ with zero momentum along the circle, which has $\Hlc = 0$
and $k=0$.  Our dictionary \eqref{dict} tells us that we are looking
for a chiral primary with $\Delta = 2J_R$ and $J_L=0$.  From table
\ref{tableAB} and appendix \ref{appqgt} we see that it corresponds to
the single-trace operator
\begin{equation}
\label{grst1}
\mbox{Tr} \left[ \mbox{sym} ( A^{J_R} B^{J_R} ) \right]\spa
\end{equation}
where with ``sym'' we mean that we symmetrize over all possible
combinations of the A's and B's. Note that the trace is here over the
A's and B's as $NM \times NM$ matrices, as defined in \eqref{AmPAB}.
Clearly the state \eqref{grst1} is gauge invariant.

In parallel with \eqref{grst1}, we see that the ground state $| k , m
= 0 \rangle$  with non-zero momentum $k = 2 J_L / M$ corresponds to
the gauge theory operator
\begin{equation}
\label{grst2}
\mbox{Tr} \left[ \mbox{sym} ( A^{J_R+J_L} B^{J_R-J_L} ) \right]~.
\end{equation}
Thus, we have identified the general ground state with momentum $k$.

Before going on to more advanced operators we first establish some
useful notation.  An efficient way of writing down the symmetrization
of A's and B's in \eqref{grst1} and \eqref{grst2} is to use a
generating function.  Consider the function
\begin{equation}
\label{genfun}
{\cal{G}} (x,y) = (x A + y B)^{P}~.
\end{equation}
This is a generating function of symmetrized sums of all possible
``words'' that can be formed with a total of $P$ ``letters'' $A$ or
$B$. In quantum mechanics, this way of ordering operators goes under
the name of Weyl ordering. We define a word of type $(K,L)$ to contain
$K$ letters $A$ and $L$ letters $B$. The generating function
\eqref{genfun} then enables us to select specific types of words by
differentiation, e.g.
\begin{equation}
\label{genfun1}
{\cal{G}}_{K,L}(A,B)  = \frac{1}{\sqrt{(K+L)! K! L!}}  \partial_x^K
\partial_y^L (x A + yB)^{K+L} \big\rvert_{x=y=0} =
\frac{1}{\sqrt{w_{K,L}}}\sum_{\si \in \si (K,L)} {\cal{W}}_\si\spa
\end{equation}
where
\begin{equation}
\label{wdef}
w_{K,L} \equiv \bino{K + L}{K}~.
\end{equation}
$\CG_{K,L}$ is now the symmetrized sum of all possible words of type
$(K,L)$. Moreover, $w_{K,L}$ is the number of words of $(K,L)$-type.
The last expression in eq.~\eqref{genfun1} is a short-hand expression
for the sum over words of $(K,L)$-type. In further detail, we write a
given word of length $P$ as
\begin{equation}
\label{genword}
 {\cal{W}}_\si \equiv U_{\si (1)} U_{\si (2)} \cdots U_{\si (P)} \spa
\si (i) = \pm 1 \spa U_1 \equiv A  \spa U_{-1} \equiv B
\end{equation}
and define the surplus of $A$'s versus $B$'s as
\begin{equation}
\label{index}
{\cal{I}}(\si) \equiv \sum_{i=1}^P \si (i) ~.
\end{equation}
which we call the index of the word.  Then it follows that
\begin{equation}
\label{wordsum}
 \sum_{\si \in \si (K,L)} {\cal{W}}_\si = \sum_{i=1}^{K+L} \sum_{\si
 (i) = \pm 1} \de_{{\cal{I}}(\si), K - L}  U_{\si (1)} U_{\si (2)}
 \cdots U_{\si (K+L)}
\end{equation}
is an alternate way of parameterizing the sum over all $(K,L)$-type
words. We will use both representations: the generating function is
most useful to compute free properties such as level matching, while
the more explicit form in terms of sums over words turns out to be
superior when computing anomalous dimensions.

In terms of \eqref{genfun1}, the ground state \eqref{grst1} then takes
the form
\begin{equation}
\label{gs}
|k = 0, m=0 \rangle \map \frac{1}{N^{J_R} \sqrt{2 J_R M}} {\rm Tr}
 \left[ {\cal{G}}_{J_R,J_R} (A,B) \right]\spa
\end{equation}
containing a sum over all possible words with $J_R$ $A$'s and $B$'s.
Note that we are now working with the right normalization factors, and
we use here and in the following the notation $\map$ to denote this
exact map (including normalization) between string theory states and
gauge theory operators.  Note also that we have suppressed here the
$P^+$-dependence of the string theory ground state, which, for
simplicity of notation, is implicitly assumed here and in all of the
following operators. Moreover, in all operators the normalization
factors are computed in the planar limit only and we refer to
appendix~\ref{appn2} for their derivation.

Using the generating function \eqref{genfun1}, the operators
\eqref{grst2} with non-zero momentum $k$ along the compactified
direction are also easily constructed
\begin{equation}
\label{gsk}
 |k, m=0 \rangle \map \frac{1}{N^{J_R} \sqrt{2 J_R M}} {\rm
  Tr}[{\cal{G}}_{J_R+J_L,J_R-J_L} (A,B) ] \spa J_L = \frac{kM}{2}\spa
\end{equation}
since for $k > 0$ ($k < 0$) these contain the words that have an
excess (deficit) of $2J_L $ $A$'s with respect to the $B$'s.
Consequently, the state has $\Delta = 2 J_R$, while using $J_L = kM/2$
in eq.~\eqref{p1} yields the quantized momentum $P_1 = k / R_T$ with
the compact radius $R_T = R/M$ finite in the Penrose limit. As
explicitly shown in appendix~\ref{appn2}, the quantization of $J_L = k
M/2$ guarantees that these operators are indeed well defined in the
$\CN=2$ theory. It is also shown in that appendix that the operators
\eqref{gsk} are orthonormal.

\subsubsection*{Zero modes}

We turn now to the gauge theory operators corresponding to the zero
modes on the string. Because one direction is compact, we need seven
bosonic zero modes, which are the ones summarized in
Table~\ref{tableZM} of section~\ref{secquant}. These are given by
\begin{equation}
\label{1osc}
(a^i_0)^{\dagger} |k , m=0 \rangle \map \frac{1}{N^{J_R+\frac{1}{2}}
\sqrt{M}} {\rm Tr} [ \Psi_i\ {\cal{G}}_{J_R+J_L,J_R-J_L} (A,B) ]\spa
\end{equation}
where $J_L = k M / 2$ and the form of the  impurity $\Psi_i$ depends
on the direction in the transverse space. For six of these seven zero
modes we have
\begin{equation}
\label{transsp}
\Psi_i = \left\{
\begin{array}{ll}
\frac{1}{\sqrt{2}} ( \Phi + \bar{\Phi} ) &  \mbox{ for } i = 3 \\
\frac{1}{\sqrt{2}i} ( \Phi - \bar{\Phi} ) & \mbox{ for } i = 4 \\
D_{i-4} & \mbox{ for } i=5,6,7,8
\end{array} \right.~.
\end{equation}
Note that in \eqref{1osc}  the $D_{i-4}$ is understood to act on the
$A$ or $B$ to the right of it.  The six operators of table
\ref{transsp} have $\Delta = 2 J_R +1 $ so that $\Hlc = 1$, in
agreement with the zero-mode spectrum of \eqref{sugrapred}.  For the
seventh non-compact direction $i = 2$ the analysis is more subtle. The
two simplest impurities that have $J_L = J_R=0$ beyond those in
\eqref{transsp} are $ A \bar A$ and $B \bar B$.  In
appendix~\ref{appqgt}  it is shown by analyzing the ${\cal N}=4$
scalar chiral primaries in $SU(3)$ notation that the only combination
of these two operators (not involving $\Phi \bar \Phi$) is ${\rm
Tr}(A\bar A - B \bar B)$ for the particular case $\Delta=2$.
More generally, we find that for the zero mode in the $z_2$-direction
the corresponding gauge theory state is obtained by acting on the
ground state \eqref{gs} with the differential operator
\begin{equation} \label{apq}
{a}_0^\dagger \qquad \rightarrow \qquad  \mathcal{D}_{2} \sim  \bar A
\ \partial_B - \bar B \ \partial_A
\end{equation}
which is a particular element of the R-symmetry group $SU(2) \times
U(1)$. This corresponds to effectively interspersing the operator
$\Psi_{2}\sim A\bar{A}-B\bar{B}$ in a totally symmetric way in the
ground state. As an important check, one verifies that the resulting
state has $\Delta= 2 J_R + 2$, which is in perfect agreement with the
energy $\Hlc=2$ of the zero mode $a_0^\dagger | 0 \rangle$ that
follows from \eqref{sugrapred}.
It is moreover important to note that we have accounted for all of the
scalar chiral primaries in this particular scaling limit of $\CN = 2$
QGT, so we really have a one-to-one correspondence between the bosonic
supergravity modes on the string side and the scalar chiral primaries
on the gauge theory side. Note in this connection also that it is in
fact the $SU(2) \times U(1) $ R-symmetry that provides the precise
definition of the bosonic zero modes. The same is true for the
fermionic zero modes that we will discuss in section \ref{fermoper}.

We can also consider bosonic zero modes with more than one zero-mode
excitation. In order to do this, we introduce the notation%
\footnote{Note that we can only write this for $l_1 \leq l_2 \leq
\cdots \leq l_q$.  If for example $l_1 > l_2$ when $q=2$ we resolve
this by defining ${\cal{G}}_{K,L; i_1 (l_1) , i_2(l_2) } \equiv
{\cal{G}}_{K,L; i_2 (l_2) , i_1(l_1) } $.}
\begin{equation}\label{genfunA}
\begin{split}
    {\cal{G}}&_{K,L; i_1(l_1) \cdots i_q(l_q)} (A,B) \\ &=
        \frac{1}{\sqrt{(K+L)! K! L!}}  \partial_x^K \partial_y^L ( x A
        + yB)^{l_1} \Psi_{i_1} (xA + yB)^{l_2-l_1} \Psi_{i_2} \dotsm
        \\ &\qquad\qquad  \dotsm \Psi_{i_q} (xA+yB)^{K+L -
        \sum_{s=1}^q l_s} \vert_{x=y=0} \\ &= \frac{1}{\sqrt{w_{K,L}}}
        \sum_{\si \in \si (K,L)} {\cal{W}}_{\si;i_1(l_1) \cdots i_q
        (l_q)}\spa
\end{split}
\end{equation}
in terms of which we have the identification%
\begin{equation}
\label{genzero}
 \prod_{s=1}^q (a_{0}^{i_s})^\dagger | k,m=0 \rangle \mapsim
\sum_{s=1}^q \sum_{l_s} {\rm Tr}[ {\cal{G}}_{J_R +
J_L,J_R-J_L;i_1(l_1) \cdots i_q(l_q)} (A,B) ]\spa
\end{equation}
which indeed reduces to \eqref{1osc} for $q=1$. Notice that, in order
to get exact chiral primaries, we have to define the range of the sum
in~\eqref{genzero} to be $0\leq l_s \leq 2J_R$ in the case of $\Phi$,
$\bar\Phi$ insertions and $0\leq l_s \leq 2J_R$ for $D_{i_s-4}$
insertions. In addition no two consecutive insertions of $D_{i_s-4}$
are allowed in eq.~\eqref{genfunA}, since one is in fact inserting
$D_{i-4} A$ or $D_{i-4} B$ rather than $D_{i-4}$.

In eq.~\eqref{genzero} we have used the symbol $\mapsim$ to indicate
that we have omitted for brevity the normalization factor. We will
generally omit such factors when writing down general states. The
appropriate planar normalization factors of the gauge theory operators
are however easily determined by the following rule. If the operator
is a sum of terms with phases and this sum is cyclically invariant,
the factor is $(N^{w/2} M w)^{-1/2}$, with $w$ the number of letters
in each word. Here $N^{w/2}$ comes from the planar contractions, $M$
arises from the sum over nodes, and the extra factor of $w$ is due to
cyclicity. When there is no cyclical invariance the latter factor is
not present.

\subsection{Bosonic gauge theory operators with anomalous dimensions
\label{secgt2}}

Next, we turn to those non-BPS operators that are nearly BPS, so that
their anomalous dimensions are non-zero but finite in the Penrose
limit.

\subsubsection*{Oscillators and no winding}

Here the first class of operators that we consider are the operators
corresponding to higher oscillator modes on the string theory side.

A first thing to notice is that, while there is no zero-mode for the
compact direction $z_1$, there are of course massive modes associated
to the corresponding world-sheet boson. Therefore, in the same spirit
as in section~\ref{secgt1}, one must figure out what kind of insertion
is needed to describe the corresponding operators in the gauge theory.

However, let us first consider the case of the insertions $\Psi_i$
defined in eq.~\eqref{transsp}, which correspond to states obtained by
acting on the string ground state with the non-zero mode oscillators
$a_n^\dagger$ and $(a_{n}^I)^\dagger$, $I=3,\ldots,8$.  Using the
notation defined in \eqref{genfunA} we can write the general map for
the insertion of these kinds of impurities as
\begin{multline}
\label{genmas}
\prod_{s=1}^q (a_{n_s}^{i_s})^\dagger | k,m=0 \rangle  \\ \mapsim
\sum_{l_1,\ldots,l_q} \beta^{\sum_{s=1}^q n_s l_s} {\rm Tr}[
{\cal{G}}_{J_R + J_L,J_R-J_L;i_1(l_1) \cdots i_q(l_q)} (A,B) ]\spa
\end{multline}
with
\begin{equation}
\label{beta}
\beta \equiv \exp \left( \frac{2\pi i }{2J_R} \right)~.
\end{equation}
The Fourier transformation with phase $\beta$ implies that
\begin{equation}
\label{ABinter}
\Psi A \rightarrow \beta A \Psi \spa \Psi B \rightarrow \beta B \Psi
\end{equation}
when we interchange an impurity $\Psi$ with either $A$ or $B$.

Using cyclicity of trace, it is easy to see that \eqref{genmas} is
zero%
\footnote{It is either simply zero or proportional to $1/J_R$ which
goes to zero in the scaling limit.}  unless
\begin{equation}
\sum_{s=1}^q n_s = 0\spa
\end{equation}
which precisely fits with the level-matching rule \eqref{levmat} of
the dual string theory.

Now we turn to the modes of the string along the direction $z^1$,
which are obtained by acting on the ground state with the oscillators
$\tilde{a}_n^\dagger$ for $n$ different from zero.  As can be seen
from the string spectrum \eqref{strpredbf2}, these massive modes have
$\Hlc=0$ to zeroth order in $\gqgt^2N / J_R^2\,$, so they should be a
small deformation of the ground state chiral primary
\eqref{gsk}. Moreover, we expect that they should be related to the
translation operator along the compact direction. In fact, one finds
that the gauge theory operators corresponding to the string modes
along $z^1$ can be expressed using the same general
formula~\eqref{genmas}, where now the insertions are instead given by
\begin{equation}\label{JLinsertion}
    \Psi_1 \equiv J_L \sim \frac{1}{2}
(A\partial_A - \bar{A} \partial_{\bar{A}} - B\partial_B + \bar{B}
\partial_{\bar{B}})
    \spa
\end{equation}
two consecutive insertions are allowed and and the range of the sum
over $l_s$ goes from 0 to $2J_R-1$.

The effect of the insertion is to modify the ground state
generating function~\eqref{genfun1} by changing the $(xA+yB)$
factor in the $l$-th spot of the product into $(xA-yB)$. Thus, one
easily sees that the ``bare'' quantum numbers are the correct
ones, namely $\Delta-2J_R=0$. In order to study their properties
in more detail, we can also write down the expression of these
operators by implementing the word notation introduced in
eq.~\eqref{genword} in the following way
\begin{equation}\label{wordz1}
    \prod_{s=1}^q \tilde{a}_{n_s}^\dagger | k,m=0 \rangle \mapsim
     \sum_{\sigma \in \sigma(J_R+J_L,J_R-J_L)} d_\sigma {\rm Tr}\ \mathcal{W}_\sigma\spa
\end{equation}
where the coefficients $d_\sigma$ are given by
\begin{equation}
    d_\sigma = \frac{1}{2^q}\sum_{l_1,\ldots,l_q=0}^{2J_R-1}
        \beta^{\sum_{r=1}^q n_r l_r} \sigma(l_1+1)\dotsm
        \sigma(l_q+1)
\end{equation}
In this notation it is first of all clear that, as expected, the
states~\eqref{wordz1} are small deformations of the ground state
obtained by assigning suitable weights to the different words of
the latter. In addition, one can see that when $n=0$ one has
$d_\sigma= ({\cal{I}}(\sigma)q/2)^q$ so that the states reduce to
the ground state~\eqref{grst2} multiplied by their $J_L$
eigenvalue to the $q$th power. This is yet another manifestation
of the fact that the zero-mode for string excitations along $z^1$
are given by the momentum operator which is proportional to $J_L$
in the gauge theory, so that for these modes we are considering
the ground state corresponding to the given $J_L$ eigenvalue.

For all of the considered states, the first non-trivial massive modes
appear for two insertions. After explicitly performing the $l_1$-sum
in \eqref{genmas} one obtains
\begin{equation}
\label{twoins}
(a^{i}_{n})^\dagger (a^{j}_{-n'})^\dagger |k, m=0 \rangle \map
\frac{\de_{n\, n'}}{N^{J_R+1} \sqrt{ 2 J_R M} } \sum_{l} {\rm
Tr}[{\cal{G}}_{J_R+J_L,J_R-J_L;i(0),j(l)}(A,B)] \beta^{n l}\spa
\end{equation}
where $\de_{n\, n'}$ correctly incorporates the level matching. These
operators, which have $J_L=0$ and $\Delta - 2 J_R = 0,1,2 $
(respectively in the case of two $J_L$ insertions, one $J_L$ and one
$\Psi_i$ defined in eq.~\eqref{transsp}, or two $\Psi_i$), are the
analogues of the simplest near-BPS operators of BMN with two
insertions.  As an important check, we verify  in the next section
that, with $\beta$ as in eq.~\eqref{beta}, a one-loop computation in
the planar limit of the gauge theory gives an anomalous dimension to
these operators that is in agreement with the string theory spectrum
\eqref{strpred1} through first order in $\gqgt^2 N / J_R^2$.

\subsubsection*{Winding and no oscillators}

We now turn to another new feature of our correspondence, namely the
operators that correspond to non-zero winding $m$ on the string theory
side. We first focus on zero momentum $k=0$ in the compact direction,
and introduce a generalization of \eqref{genfun1} that assigns a
weight to the ordering of the letters $A$ and $B$ in the words,
\begin{equation}\label{genfun4}
\begin{split}
{\cal{G}}_{K,L} (A,B;\omega) & =  \frac{1}{\sqrt{(K+L)! K! L!}}
\partial_x^K \partial_y^L \prod_{r=0}^{K+L-1} (\om^{-r/2} x A +
\om^{r/2} yB) \vert_{x=y=0}  \\ &  =  \frac{1}{\sqrt{w_{K,L}}}
\sum_{\si \in \si (K,L)} \om^{{\cal{N}}(\si)} {\cal{W}}_\si\spa
\end{split}
\end{equation}
where the phase factor is
\begin{equation}
\label{omega}
\om = \exp ( 2 \pi i/(M J_R))
\end{equation}
and
\begin{equation}
\label{weight}
{\cal{N}}(\si) \equiv -\frac{1}{2} \sum_{i=1}^{K+L} (i-1) \si (i)
\end{equation}
is called the weight of word ${\cal{W}}_\si$.  In parallel with
 \eqref{ABinter} this construction now incorporates the phase shift
\begin{equation}
B A \rightarrow \omega A B
\end{equation}
for the interchange of $A$ and $B$ in a word. However, our
construction also needs the ${\cal N}=2$ twist matrix
\begin{equation}
\label{twistm}
S = \theta\; {\rm diag} (1,\theta,\theta^2, \ldots , \theta^{M-1})
\spa \theta = \exp( 2 \pi i/M) ~.
\end{equation}
Then, in terms eq.~\eqref{genfun4} and $S$, we may write our proposal
for the operators with zero momentum and winding $m$ as
\begin{equation}
\label{wingau}
|k= 0, m \rangle \map \frac{1}{N^{J_R} \sqrt{2J_R M}} {\rm Tr}[ S^m
{\cal{G}}_{J_R,J_R} (A,B;\om^m) ]\spa
\end{equation}
which have $\Delta = 2J_R $, $J_L=0$.

We pause here for a number of important observations.  All our
operators so far were directly inherited from the parent ${\cal N}=4$
theory (see also section~\ref{secbn4}). However, because of the
appearance of the twist matrix $S$, the winding states \eqref{wingau}
involve the twisted sectors of ${\cal N}=2$ QGT. Secondly, the quantum
numbers of the state coincide with those of the ground state
\eqref{gs} suggesting an infinite degeneracy of the string theory
ground state.  However, this is only true in the free QGT, since, as
we will see in the next section, this degeneracy gets lifted once
interactions are turned on in the gauge theory. In particular, we will
show that the one-loop correction (in the planar limit) reproduces the
exact string theory result. This implies a non-renormalization theorem
beyond one-loop for the gauge theory operators \eqref{wingau}. It
would be interesting to prove this, for example using supersymmetry,
directly in the gauge theory.%
\footnote{See Ref.~\cite{Gross:2002su} for a planar two-loop check and
Ref.~\cite{Santambrogio:2002sb} for an all-order check  of anomalous
dimensions in the original BMN setup.  See also Ref.~\cite{Oz:2002wy}
for an all-order check of anomalous dimensions in the orbifolds
discussed in
Ref.s~\cite{Alishahiha:2002ev,Kim:2002fp,Takayanagi:2002hv}.}

We leave a detailed proof of the winding operators above and those in
the sequel (including properties such as level-matching and
orthonormality) to appendix~\ref{appn2}, but present here a quick
consistency check on the state \eqref{wingau}. First we note that for
a given word $\CW_\si$ in ${\cal N}=4$ notation, after substitution of
\eqref{AmPAB} one obtains
\begin{equation}
\label{n4ton2}
\CW_\sigma = \mbox{diag} ( \CW_{\sigma,I} )_{I=1}^M\spa
\end{equation}
where the components $\CW_{\sigma,I}$ are  words in ${\cal N}=2$
notation. Consider now the  specific class of words
\begin{equation}
\label{wordq}
\CW_{\sigma_q} = A^q B^{J_R} A^{J_R-q}\spa
\end{equation}
for which  the ${\cal N}=2$ components can be computed to be
\begin{equation}
\CW_{\sigma_q,I} = A_I \cdots A_{I+q-1} B_{ I+q-1} \cdots B_{I+q-J_R}
A_{I+q-J_R} \cdots A_{I-1}~.
\end{equation}
Using \eqref{weight}, the weight of word \eqref{wordq} is
\begin{equation}
\label{qweight}
{\cal{N}} ( \sigma_q ) = qJ_R - \frac{1}{2} J_R^2\spa
\end{equation}
so that, according to eq.s~\eqref{wingau}-\eqref{n4ton2} for each word
$\CW_{\sigma_q,I}$ we have a phase
\begin{equation}
\label{qphase}
\omega^{{\cal{N}}(\sigma_q)} \theta^I\spa
\end{equation}
where we have taken $m=1$ for simplicity.  Now, under cyclicity of the
trace in \eqref{wingau} we find that $A_I$ in the beginning of
$\CW_{\sigma_q,I}$ is moved to the end of the word, one obtains the
${\cal N}=2$ word $\CW_{\sigma_{q-1},I+1}$ rather than
$\CW_{\sigma_{q-1},I}$. This implies that under cyclicity we have the
equivalence relation
\begin{equation}
\CW_{\sigma_q,I} \equiv \CW_{\sigma_{q-1},I+1}\spa
\end{equation}
which, using \eqref{qphase} means that we need
\begin{equation}
\omega^{{\cal{N}}(\sigma_q)} \theta^I =
\omega^{{\cal{N}}(\sigma_{q-1})} \theta^{I+1}
\end{equation}
and hence, using eq.~\eqref{qweight}, that
\begin{equation}
\omega^{J_R} = \theta~.
\end{equation}
Thus, given $\theta$ in the twist matrix \eqref{twistm}, this
determines $\omega$ as in eq.~\eqref{omega}.  This achieves that all
words are preserved under cyclicity and the phase is constructed so
that it is independent of which letter one starts with in a given
$\CN=2$ word.

\subsubsection*{Winding and oscillators}

Now that we have seen how to introduce i) compact momentum ii) massive
string modes and iii) winding, we can combine all three.  The simplest
non-trivial state that combines these features has non-zero compact
momentum and winding, accompanied by one impurity insertion.  To this
end define
\begin{equation}\label{genfun5}
\begin{split}
    {\cal{G}}&_{K,L;i(l)} (A,B;\omega) \\ & = \frac{1}{\sqrt{(K+L)! K!
    L!}}  \partial_x^K \partial_y^L \prod_{r_1=0}^{l-1} (\om^{-r_1/2}
    x A + \om^{r_1/2} yB) \Psi_i  \\ & \qquad \qquad  \times
    \prod_{r_2=l}^{K+L-1}(\om^{-r_2/2} x A  + \om^{r_2/2} yB)
    \vert_{x=y=0} \\ & = \frac{1}{\sqrt{w_{K,L}}} \sum_{\si \in \si
    (K,L)} \om^{{\cal{N}}(\si)} {\cal{W}}_{\si;i(l)}\spa
\end{split}
\end{equation}
which combines \eqref{genfunA} with \eqref{genfun4}, and comprises a
weighted sum over $(K,L)$-type words with an insertion of $\Psi_i$
after the $l$-th position in the word.  Then the appropriate gauge
theory operator is
\begin{equation}
\label{mow}
(a_n^i)^\dagger | k,m \rangle \map \frac{1}{N^{J_R + \frac{1}{2}}
\sqrt{2J_R M}} \sum_{l} {\rm Tr}[ S^m {\cal{G}}_{J_R +
J_L,J_R-J_L;i(l)} (A,B;\om^m) ] \beta^{n l}\spa
\end{equation}
which has $\Delta = 2J_R +1$, $J_L = kM/2$. An important check on the
state \eqref{mow} is that vanishes unless it satisfies the level
matching condition
\begin{equation}
\label{levmat1}
 n = k m   ~.
\end{equation}
This is verified in appendix~\ref{appn2} using cyclicity of the trace
and the explicit forms of $\beta$, $\omega$ and $\theta$.

We conclude by presenting the most general state, for which we need
\begin{equation}\label{genfun6}
\begin{split}
    {\cal{G}}&_{K,L; i_1(l_1) \cdots i_q(l_q)} (A,B;\omega) \\ &=
        \partial_x^K \partial_y^L \prod_{r_1=0}^{l_1-1} (\om^{-r_1/2}
        x A + \om^{r_1/2} yB) \Psi_{i_1} \cdots  \\ & \qquad \qquad
        \cdots \Psi_{i_q} \prod_{r_{q+1}= \sum_{i=1}^q l_i }^{K+L-1}
        (\om^{-r_{q+1}/2} x A + \om^{r_{q+1}/2} yB) \vert_{x=y=0}  \\
        &= \sum_{\si \in \si (K,L)} \om^{{\cal{N}}(\si)}
        {\cal{W}}_{\si;i_1(l_1) \cdots i_q (l_q)}\spa
\end{split}
\end{equation}
in terms of which we have the identification
\begin{equation}
\label{mowgen}
\prod_{s=1}^q (a_{n_s}^{i_s})^\dagger | k,m \rangle \mapsim
\sum_{s=1}^q \sum_{l_1,\ldots,l_q} {\rm Tr}[ {\cal{G}}_{J_R +
kM/2,J_R-kM/2;i_1(l_1) \cdots i_q(l_q)} (A,B;\om^m) ] \beta^{
\sum_{r=1}^q n_r l_r}~.
\end{equation}
As a check on this general expression, one may verify again that the
general form of the level matching condition
\begin{equation}
 \sum_{s=1}^q n_s = k m
\end{equation}
is satisfied when using the generating function in \eqref{genfun6} and
cyclicity of the trace.

\subsection{Anomalous dimensions and comparison with string theory
\label{andsec}}

Following the techniques of Ref.~\cite{Berenstein:2002jq}, we now
compute the anomalous dimensions of the gauge theory operators in the
previous section, and verify that they are indeed reproduced by the
string theory results.  We restrict to the leading (one-loop)
correction in the planar limit.

As remarked above, the operators in section \ref{secgt1} are all
chiral primaries and hence do not receive corrections, while those of
section \ref{secgt2} are expected to possess non-zero anomalous
dimensions.  We limit our discussion here to the three simplest
operators of this type. First, we consider gauge theory operators of the type of eq.~\eqref{twoins}, of which we consider only the cases of two $\Phi$ and two $J_L$ insertions%
\footnote{As was done for the BMN operators in
Ref.~\cite{Gursoy:2002yy}, it would be also interesting to
consider other impurities.}
\begin{subequations}\label{Oogeneral}
\begin{gather}
    \label{Oo} (a_n^\Phi)^\dagger ( a_{-n}^{\Phi})^\dagger |k=0,m=0\rangle
        \mapsim {\cal{O}}_{n}^{(\rm o)} \spa
        {\cal{O}}_{n}^{(\rm o)} \equiv \sum_{l=0}^{2J_R}
        {\rm Tr}[{\cal{G}}_{J_R,J_R; \Phi(0) \Phi(l)} (A,B) ] \beta^{nl}\spa\\
    \label{OoJL} (\tilde a_n)^\dagger ( \tilde a_{-n})^\dagger |k=0,m=0\rangle
        \mapsim {\cal{O}}_{n}^{(\tilde{\rm o})} \spa
        {\cal{O}}_{n}^{(\tilde{\rm o})} \equiv \sum_{l=0}^{2J_R-1}
        {\rm Tr}[{\cal{G}}_{J_R,J_R; J_L(0) J_L(l)} (A,B) ] \beta^{nl}~.
\end{gather}
\end{subequations}
The third kind of states are the pure winding ones
\begin{equation}
\label{Ow}
| k=0, m \rangle \mapsim {\cal{O}}_{m}^{(\rm w)} \spa
{\cal{O}}_{m}^{(\rm w)} \equiv \mbox{Tr} \left[ S^m
\CG_{J_R,J_R}(A,B;\omega^m) \right]~.
\end{equation}
To compute the one-loop anomalous dimension of these operators we need
to calculate the ratio
\begin{equation}
\label{ratio}
R({\cal{O}})  = \frac{\langle {\cal{O}}(x) \bar{\cal{O}}(0)
\rangle_{\rm (1-loop)}} {\langle {\cal{O}}(x) \bar{\cal{O}}(0)
\rangle_{\rm (free)}}
\end{equation}
between the one-loop contribution to the two-point function and the free part.

\begin{figure}[ht]
\begin{center}
{\scalebox{1}{\includegraphics{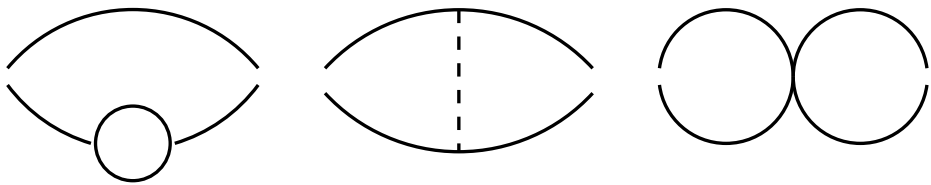}}}
\caption{\small The three relevant graphs contributing to the
radiative correction of two point scalar field trace operator  ${\cal
O}_{(n)}$.}
\label{00}
\end{center}
\end{figure}

For simplicity we present here only the main points of the
derivation, referring to appendix \ref{appn2} for details.  It
turns out that, besides some important differences that will
become clear below, the computation for the operators
\eqref{Oogeneral} and \eqref{Ow} has some analogous features, so
that we may, at least partly, present them in parallel.  For all
these operators the relevant graphs contributing to the leading
radiative correction of two point scalar field trace operators are
the self-energy, the gluon exchange and the four-point
interaction, as depicted in the figure \ref{00}.
For the oscillator operators \eqref{Oo}, it is immediately obvious
that we may invoke the same observation as in the original BMN
computation, namely that the only type of diagrams among those in
figure \ref{00} that have a momentum $n$-dependent contribution
are those arising from F-terms of the four-point interaction.
Interestingly, the same property holds for the  states
\eqref{OoJL} as well as the winding states \eqref{Ow}, in the
latter case because only the F-terms will give a winding
$m$-dependent contribution.

This means that for \eqref{Oo} we need those diagrams that
exchange an $A$ or a $B$ with a $\Phi$ field.  Moreover, for the
$n=0$ operators (i.e. two zero modes), the overall contribution to
the anomalous dimension coming from {\it all} diagrams should
cancel since in that case the operator, ${\cal O}_{0}^{(\rm o)}$,
is BPS and protected.  Thus the $n=0$ contribution from the
F-terms should exactly cancel the overall contribution coming from
all other diagrams. Likewise, for \eqref{OoJL} and \eqref{Ow} we
only need those diagrams that exchange an  $A$ with a $B$. Then
the same argument as above can  be used, since the $n=0$ operator
${\cal O}_{0}^{(\tilde{\rm o})}$ and the $m=0$ operator ${\cal
O}_{0}^{(\rm w)}$ reduce both to the ground state, which obviously
does not receive any corrections.  As a consequence, in computing
the anomalous dimension of the operators ${\cal O}_{n}^{(\rm o)}$,
${\cal O}_{n}^{(\tilde{\rm o})}$ or ${\cal O}_{m}^{(\rm w)}$, one
only needs to compute the F-term contribution from diagrams as the
third one in figure \ref{00} and subtract  from the result the
$n=0$ or $m=0$ contribution. This subtraction will automatically
take into account the effective contribution from all other
diagrams.

\begin{figure}[ht]
\begin{center}
{\scalebox{1}{\includegraphics{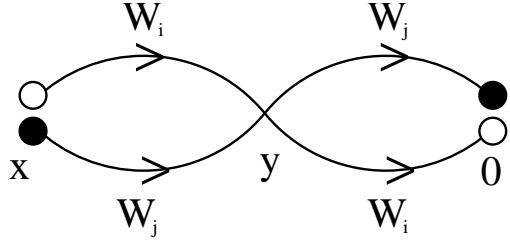}}}
\caption{\small One-loop F-term contribution. This diagram includes
both  $\Phi$ interchange with $A$ or $B$ as well as $A$ and $B$
interchange.}
\label{01}
\end{center}
\end{figure}

After some algebra, it can be shown that for all three operators
\eqref{Oo},\eqref{OoJL} and~\eqref{Ow} the F-term contribution to
the ratio \eqref{ratio} takes the form
\begin{equation}
\label{ratioF}
R({\cal{O} }_p)_{\rm (F)} = \mathfrak{m} \cos(p \alpha ) \frac{
 \langle W_i (x) W_j(x) \bar W_i (0) \bar W_j(0) \rangle}{ \langle W_i
 (x)  \bar W_i (0) \rangle \langle W_j (x)  \bar W_j (0) \rangle} \spa
 W_i \in \{  A, B, \Phi\}~.
\end{equation}
Here, the correlator in the numerator is the one-loop diagram, in
the $U(N)$ gauge theory, depicted in figure~\ref{01}, while the
denominator involves the product of the corresponding scalar
two-point functions. The integer $p$ is the integer $n$ or $m$
that specifies the gauge theory operator.
 For the operators \eqref{Oo} and
\eqref{Ow} $\alpha$ is the phase corresponding to the interchange
$W_i W_j = e^{\alpha} W_j W_i$ and the {\it multiplicity factor}
$\mathfrak{m}$ is given by the number of nearest neighbor pairs of
the form $W_i W_j$. For the operator \eqref{OoJL} the assignment
of the phase  $\alpha$ and factor $\mathfrak{m}$ is less direct,
but, as shown in appendix \ref{appn2}, the final result can still
be cast in the general form \eqref{ratioF}.

The main ingredients that enter the derivation of \eqref{ratioF} for
the oscillator and winding states, is that in the planar limit we can
show for either of these operators that:
\begin{itemize}
\item  in both the free and one-loop two-point functions each word
${\cal{W}}_\si$ in the operator  $\sum_\si {\cal{W}}_\si (x)$ has a
non-zero contraction with only one word in $\sum_{\si'} \bar
{\cal{W}}_{\si'} (0)$,
\item  in both  the free and one-loop two-point functions the
contractions are diagonal in the $U(N)^M$ product space,
\item only nearest neighbor interchanges of fields contribute at the
one-loop level.
\end{itemize}
Moreover, the specific form of the sum over words that makes up
the operator ${{\cal O}_p}$ then determines the multiplicity
factor. We note that the origin of the cosine factor is different
in nature for the three cases \eqref{Oo}, \eqref{OoJL} and
\eqref{Ow}, and refer again to appendix \ref{appn2} for the
details.

In further detail, for the oscillator operators \eqref{Oo} we have
\begin{equation}
\label{oscv}
W_i = \Phi \spa W_j = A\;\, \mbox{or} \;\, B \spa p=n \spa
\alpha = {\rm arg}(\beta) \spa \mathfrak{m} =4\spa
\end{equation}
which is quite analogous to the original BMN computation.  Here
the multiplicity of 4 is easy to understand as there are two
$\Phi$ fields that can undergo an interchange with their nearest
neighbor. For the oscillator operators \eqref{OoJL} we have
\begin{equation}
\label{oscv1} W_i = A \spa W_j = B \spa p=n \spa \alpha = {\rm
arg}(\beta) \spa \mathfrak{m} =4\spa
\end{equation}
where now the multiplicity factor arises due to the fact that for
the F-term contribution only for two occurrences  of $A$ or $B$ is
there a relevant nearest neighbor interchange of the fields $A$
and $B$. Finally, for the winding states \eqref{Ow} we find
\begin{equation}
\label{winv}
W_i = A \spa W_j = B \spa p = m \spa \alpha = {\rm arg}( \omega)  \spa
\mathfrak{m} = J_R~.
\end{equation}
In this case the multiplicity factor is obtained by computing the
number $N_r$ of words that have $r$ nearest neighbors $AB$ and $BA$
and then the average
\begin{equation}
\mathfrak{m}= 2 \frac{\sum_r r N_r}{\sum_r N_r}\spa
\end{equation}
the details of which can be found in appendix \ref{appn2}.

We now make use of the one-loop integral \eqref{onelform} to
compute
\begin{equation}
\label{1loop}
\frac{ \langle W_i (x) W_j(x) \bar W_i (0) \bar W_j(0) \rangle}{
\langle  W_i (x)  \bar W_i (0) \rangle \langle W_j (x)  \bar W_j (0)
\rangle} = \frac{g_{\rm QGT}^2 N}{8 \pi^2} \ln ( |x| \Lambda)^2\spa
\end{equation}
where the factor $g_{\rm QGT}^2 N$ originates from the coupling at the
vertex along with an extra closed loop in $U(N)$ color space in the
one-loop diagram.  We then easily find the F-term contribution to the
anomalous dimension by comparing with \eqref{corf0}. Finally,
subtracting the $p=0$ term we find the universal expression
\begin{equation}
\label{andimg}
\delta \Delta = \frac{g_{\rm QGT}^2 N }{8 \pi^2} \mathfrak{m} [ 1-
\cos(p \alpha) ]
\end{equation}
for the one-loop correction to the anomalous dimension for the
three operators.

Using \eqref{andimg} along with the specific substitution
\eqref{oscv} and \eqref{oscv1} for the oscillator operators and
\eqref{winv} for the winding operators, and expanding for large
$J_R$, we obtain the following expressions for the one-loop
corrected anomalous dimensions of the operators
\begin{subequations}\label{an1l}
\begin{align}
{\cal{O}}_n^{(\rm o)} \quad &: \qquad  \Delta - 2J_R = 2 +
\frac{g_{\rm QGT}^2 N n^2}{(2J_R)^2}\spa\\
{\cal{O}}_n^{(\tilde{\rm o})} \quad &: \qquad  \Delta - 2J_R =
\frac{g_{\rm QGT}^2 N n^2}{(2J_R)^2}\spa\\
{\cal{O}}_m^{(\rm w)} \quad &: \qquad  \Delta - 2J_R =
\frac{g_{\rm QGT}^2 N m^2}{ 4 J_R M^2}
\end{align}
\end{subequations}
where we used $\beta$ and $\omega$ in \eqref{beta} and \eqref{omega}
respectively.

%In this case, the computation of the anomalous dimension goes
%along quite different lines with respect to the two cases that we
%just presented.

%The details of the computation are presented in
%appendix~\ref{appn2}, and the final result is that the anomalous
%dimension is again given by an expression similar to
%\eqref{andimg}, namely
%\begin{equation}
%\delta \Delta = \frac{g_{\rm QGT}^2 N }{8 \pi^2} 4 [ 1- \cos(n\
%{\rm arg}(\beta) ) ]\spa
%\end{equation}
%which determines the one-loop corrected dimension of the operators
%in eq.~\eqref{OoJL} to be
%\begin{equation}
%    {\cal{O}}_n^{(1)} \quad : \qquad  \Delta^{(1)} - 2J_R =
%    \frac{g_{\rm QGT}^2 N n^2}{(2J_R)^2}~.
%\end{equation}
Comparing eq.s~\eqref{an1l} with the string theory predictions in
\eqref{strpred2} we observe that for all considered kinds of
operators we find exact agreement. That this works out for the
oscillator operators with $\Phi$ insertions is a good check.
Though it seemingly works out in a way that is analogous to the
BMN computation, there are some ingredients required that
specifically relate to the ${\cal N}=2$ structure. A much more
stringent and non-trivial check of our proposed correspondence is
given by the other two cases, because the lines of the relevant
computations are quite different from the BMN ones. For instance,
for the new winding operators the phase factor $\omega$, together
with the multiplicity factor $\mathfrak{m}$, elegantly conspire to
reproduce the string theory prediction.

\subsection{Fermionic gauge theory operators}
\label{fermoper}

We finally turn to the fermionic part of the spectrum. We find in the
following the gauge theory operators corresponding to the string states
\begin{equation}
(S_n^b)^\dagger | k,m \rangle \ \ \mbox{with} \ \ b = 1,...,8\spa \ \
n \in \Z\spa
\end{equation}
which correspond to turning on the fermionic number operators
$F_n^{(b)}$ and the winding term in the free light-cone hamiltonian
\eqref{strpredbf1} or its one-loop approximation
\eqref{strpredbf2}. In what follows we do not consider states of the
form $(S_n^b)^\dagger (S_{\tilde{n}}^{\tilde{b}})^\dagger | k,m
\rangle$ but it should be apparent from our construction how to build
such states.

Consider first the modes with $b=1,2,3,4$.  For $b=1,2$ we propose
that%
\footnote{Here we have made specific assignments to the components of
$b=1,2,3,4$ in relation to the gauge theory spinor component. This we
can do since we still have a certain amount of freedom left with
respect to the basis choice of our spinors, i.e.  with respect to the
choices of Gamma matrices. The same is true for the $b=5,6,7,8$
components.}
\begin{equation}
\label{b12}
(S^b_n)^\dagger |k,m \rangle \mapsim \sum_{l=0}^{2J_R} \beta^{n l}
\sum_{\sigma} \delta_{\CI(\sigma) = kM } \, \omega^{m\CN(\sigma)} \tr
\left( S^m U_{\sigma(1)} \cdots U_{\sigma(l)} \, \bar{\psi} \,
U_{\sigma(l+1)} \cdots U_{\sigma(2 J_R - 1)} \right)\spa
\end{equation}
corresponding to the two components of $\bar{\psi}$. Here we have used
the notation defined in eq.~\eqref{genword} and we recall that $S$ is
the twist matrix in eq.~\eqref{twistm}. We can similarly write down
the map for $b=3,4$ by inserting $\bar{\psi}_\Phi$ instead. We see
that these four states precisely have the right eigenvalues for $\Hlc$
when we consider the zero modes, since in that case the gauge theory
operator in \eqref{b12} reduces to a chiral primary with $\Delta =
2J_R + \frac{1}{2}$ and thus $\Hlc = \frac{1}{2}$, according to table
\ref{tableAB}. More generally, the one-loop part of the spectrum is
given by the first term in eq.~\eqref{strpredf2} together with  the
winding contribution in \eqref{strpred2}.  We expect this result to be
reproducible by the same techniques as used in section \ref{andsec}
and appendix \ref{appn2}.

We consider now the four remaining components $b=5,6,7,8$.  For
$b=5,6$ we propose that
\begin{multline}\label{b56}
(S_n^b)^\dagger |k,m \rangle \mapsim \sum_{l=0}^{2J_R-1} \beta^{n l}
\sum_{\sigma} \delta_{\CI(\sigma) = kM } \, \omega^{m\CN(\sigma)}\\
\times \tr \left( S^m U_{\sigma(1)} \cdots U_{\sigma(l)} \,
\chi_{\sigma(l+1)} \, U_{\sigma(l+2)} \cdots U_{\sigma(2J_R)}
\right)\spa
\end{multline}
where $\chi_1 \equiv \chi_A$ and $\chi_{-1} \equiv \chi_B$.
Furthermore, for $b=7,8$ we propose that
\begin{multline}\label{b78}
(S_n^b)^\dagger |k,m \rangle \mapsim \sum_{l=0}^{2J_R-1} \beta^{n l}
\sum_{\sigma} \delta_{\CI(\sigma) = kM } \, \omega^{m\CN(\sigma)}\\
\times \tr \left( S^m U_{\sigma(1)} \cdots U_{\sigma(l)} \,
\bar{\chi}_{\sigma(l+1)} \, U_{\sigma(l+2)} \cdots U_{\sigma(2J_R)}
\right)\spa
\end{multline}
where $\bar{\chi}_1 \equiv \bar{\chi}_B$ and $\bar{\chi}_{-1} \equiv
\bar{\chi}_A$.

Again, we can check that the four components given by \eqref{b56} and
\eqref{b78} have the right eigenvalues for $\Hlc$ when we consider the
zero modes, since in that case the gauge theory operators reduces to
chiral primaries with $\Delta = 2J_R + \frac{3}{2}$ and thus $\Hlc =
\frac{3}{2}$, again according to table \ref{tableAB}. More generally,
the one-loop part of the spectrum is now given by the second term in
eq.~\eqref{strpredf2} together with  the winding contribution in
\eqref{strpred2}.  Again, we expect this to be reproducible by similar
techniques as used in section \ref{andsec} and appendix \ref{appn2}.

Finally, using similar tricks as in appendix \ref{appn2} we verify for
all eight components the level matching condition
\begin{equation}
n = km\spa
\end{equation}
which indeed is a special case of \eqref{levmat}.

\section{Space-like isometry and $\CN =4$ SYM theory }
\label{secn5}

In this section we return to the novel scaling limit of $\CN =4$ SYM
that corresponds to the Penrose limit giving rise to the pp-wave with
manifest space-like isometry presented in section \ref{seciso}. We
give the relevant gauge theory operators in this limit and discuss the
genus counting parameter arising in non-planar contributions. We also
develop a more general framework in terms of which this new Penrose
limit can be understood and discuss in particular the relation between
our limit and the original BMN limit.

\subsection{$\CN =4$ gauge theory operators \label{secbn4} }

We give here the relevant gauge theory operators in the new scaling
limit \eqref{limitn4} of $\CN =4$ SYM.  Having determined in section
\ref{secgauge} the operators in the $\CN = 2$ QGT it is relatively
simple matter to write down the $\CN =4$ states, by considering the
decompactification limit of the former. In practice, this means that
we may set $M=1$ in these states.  In particular, this has the
consequence that the states of the twisted sector disappear and the
operators \eqref{wingau} corresponding to string winding states
disappear from the spectrum, as expected. In the following the
operators $A,B$ etc.  denote operators in the $U(N)$ theory.

{} From \eqref{gs}, the $\CN =4$ ground state is then
\begin{equation}
\label{gsn4}
|0 \rangle \map \frac{1}{N^{J_R} \sqrt{2 J_R}} {\rm Tr} \left[
 {\cal{G}}_{J_R,J_R} (A,B) \right]\spa
\end{equation}
where we recall that ${\cal{G}}_{J_R,J_R} (A,B)$ generates the
symmetrized sum over all operators with $J_R$ $A$'s and $B$'s.  The
seven zero modes are given by
\begin{equation}
\label{1oscn4}
(a^i_0)^{\dagger} |0 \rangle \map \frac{1}{N^{J_R+\frac{1}{2}}} {\rm
 Tr} [ \Psi_i \, {\cal{G}}_{J_R,J_R} (A,B)]\spa
\end{equation}
where the impurities $\Psi_i$ are as in eq.~\eqref{transsp} A special
feature of this new sector of $\CN =4$ SYM is that the eighth zero
mode, corresponding to the isometric direction, is now in one-to-one
correspondence with the operator $J_L$ in \eqref{p1}. These operators
complete the (bosonic) chiral primaries in the spectrum.

Turning to the near-BPS states, we easily read off from \eqref{genmas}
 the $\CN =4$ SYM states corresponding to string oscillator modes
\begin{equation}
\label{genmasn4}
\prod_{s=1}^q (a_{n_s}^{i_s})^\dagger | 0 \rangle \mapsim \sum_{0\leq
l_1,...,l_q\leq 2J_R} \beta^{\sum_{s=1}^q n_s l_s} {\rm Tr}[
{\cal{G}}_{J_R ,J_R;i_1(l_1) \cdots i_q(l_q)} (A,B) ]\spa
\end{equation}
where $\beta = \exp (2 \pi i /(2J_R))$.  The level matching condition
$\sum_{s=1}^q n_s = 0$ of these states follows easily using cyclicity
of the trace.  Moreover, following the same steps as for the $\CN =2$
case (see appendix \ref{appn2}) it is a simple matter to derive the
one-loop planar correction to the anomalous dimensions of the
two-oscillator state  in the planar limit. The result is
\begin{subequations}\label{andn4}
\begin{align}
(a_n^\Phi)^\dagger (a_{-n}^\Phi)^\dagger | 0\rangle \quad & :&
\quad \Delta - 2J_R = 2 + \frac{g_{\rm YM}^2 N n^2}{(2J_R)^2} \\
(\tilde a_n)^\dagger (\tilde a_{-n})^\dagger | 0\rangle \quad & :&
\quad \Delta - 2J_R = 0 + \frac{g_{\rm YM}^2 N n^2}{(2J_R)^2}
\end{align}
\end{subequations}
in agreement with the string theory spectrum.  For brevity we do not
list the  fermionic states, but these are also easily read off from
the results of section \ref{fermoper}.

We finally remark on the genus counting parameter in this new scaling
limit of $\CN =4$ SYM. Using the same techniques as in
Ref.s~\cite{Kristjansen:2002bb,Constable:2002hw} it is not difficult
to see that the torus contribution to the two-point function of the
gauge theory operators above will carry an extra factor of
\begin{equation}
g_2^2 = \frac{J_R^4}{N^2} \ .
\end{equation}
As a consequence $g_2$ is identified as the genus counting parameter
in this scaling limit. More generally, we expect that using matrix
model techniques it should be possible to compute the exact two-point
functions to all orders in $g_2$.

\subsection{On the new Penrose limit}
\label{explain}

It may be surprising that different Penrose limits exist at the level
of ${\cal N}=4$ SYM theory. Actually, as we already noticed, the
result of the  Penrose limit discussed in section \ref{seciso} is the
same as the usual pp-wave, though written in a different coordinate
system. Thus what is really going on is that the standard BMN pp-wave
can be embedded in many inequivalent ways in ${\cal N}=4$ SYM theory.

To understand this observation in somewhat more detail, we consider
the way the isometries of the pp-wave arise from the isometries of the
original $AdS_5 \times S^5$ solution. This is a well known story, see
e.g.  Ref.~\cite{Blau:2001ne}. The original isometry group of $AdS_5
\times S^5$ is $SO(2,4) \times SO(6)$.  We pick the generator $\Delta$
of $SO(2,4)$ corresponding to conformal weight, and some $U(1)$
generator $J$ of $SO(6)$. We can write $SO(2,4)\times SO(6)={\cal G}_-
\oplus {\cal G}_0 \oplus {\cal G}_+$, where ${\cal G}_0$ is spanned by
$\Delta, J$, together with an $SO(4)\subset SO(2,4)$ that commutes
with $\Delta$, and an $SO(4)\subset SO(6)$ that commutes with $J$. The
other parts ${\cal G}_{\pm}$ are such that $[{\cal G}_0,{\cal
G}_{\pm}]=\pm {\cal G}_{\pm}$, and $[{\cal G}_-,{\cal G}_+]={\cal
G}_0$.

The scaling limit introduced in Ref.~\cite{Berenstein:2002jq} now
tells us that we should keep all generators in ${\cal G}_0$ fixed,
except $\Delta+J$. We should keep $(1/R^2)(\Delta+J)$ fixed instead of
$\Delta+J$, where $R$ is the radius of $AdS_5$ and $S^5$. In addition,
we should keep $1/R$ times all the generators in ${\cal G}_+$ and
${\cal G}_-$ fixed.

This procedure implies that after taking the $R\rightarrow \infty$
limit, $(1/R^2)(\Delta+J)$ will be a central element of the algebra,
and that $(1/R){\cal G}_+$ and $(1/R){\cal G}_-$ will become like
harmonic oscillator creation and annihilation operators. They do in
fact become the zero modes of the string in the pp-wave quantized  in
light-cone gauge. The remaining generators in ${\cal G}_0$ simply
transform those in $(1/R){\cal G}_{\pm}$ into each other, and become
global symmetries of the system.

In this way, the isometry algebra of the original $AdS_5 \times S^5$
is contracted into the isometry algebra of the pp-wave. It is
straightforward to generalize this to include the supersymmetry
generators, but we will not need that here. The killing vectors then
obey the following algebra \cite{Blau:2001ne}
\begin{equation}\label{algk}
\begin{gathered}
    \left[k_{e_-},\text{all}\right] = 0 \spa
    \left[k_{e_+},k_{M_{ij}}\right] = 0\spa \\
    \left[k_{e_+},k_{e_i}\right] = k_{e_i^*} \spa
    \left[k_{e_+},k_{e_i^*}\right] = -4 k_{e_i} \spa
    \left[k_{e_i},k_{e_j^*}\right] = 4 \delta_{ij} k_{e_-}\spa\\
    \left[k_{M_{ij}},k_{e_k}\right] = \delta_{jk} k_{e_i} -
    \delta_{ik} k_{e_j} \spa \left[k_{M_{ij}},k_{e_k^*}\right] =
    \delta_{jk} k_{e_i^*} - \delta_{ik} k_{e_j^*}\spa \\
    \left[k_{M_{ij}},k_{M_{kl}}\right] = \delta_{jk} k_{M_{il}} -
    \delta_{ik} k_{M_{jl}} - \delta_{jl} k_{M_{ik}} + \delta_{il}
    k_{M_{jk}}~.
\end{gathered}
\end{equation}
where $i,j,k,l=1,...,8$ (recall however that $k_{M_{ij}}$ are the
generators of $SO(4)\times SO(4)$ where $i,j=1,...,4$ and
$i,j=5,...,8$ label the two different $SO(4)$ factors).  In the
remainder, we focus on the generator $\Delta$ and the $SO(6)$
generators. The isometries of the pp-wave that one obtains from these
in the pp-wave limit are more explicitly given by
\begin{equation}\label{isomalg}
\begin{split}
    k_{e^-} & =  \frac{1}{R^2}(\Delta+J)  \spa\\ k_{e^+} & =
    (\Delta-J) \\ k_{e_i} & =  \frac{1}{R}(E^i_a T^a)\spa\\
    k_{e_i}^{\ast} & =  \frac{1}{R}(F^i_a T^a)\spa  \\ k_{M_{ij}} & =
    M^{ij}_a T^a~.
\end{split}
\end{equation}
where here the $T^a$ are a set of generators of $SO(6)$, and
$M^{ij}_a$ form a basis of $SO(4)\subset SO(6)$ (this means that now
the indices $i,j$ take value $1,2,3,4$), whereas $E^i_a T^a$ and
$F^i_a T^a$ span ${\cal G}_+$ and ${\cal G}_-$.

Before proceeding, it is worth pointing out that there are different
ways to obtain compact circles in the pp-wave geometry.  Any of the
isometries of the background can be used. For example, if we use
$k_{e^-}$ we obtain the model of
Ref.s~\cite{Mukhi:2002ck,Alishahiha:2002jj}. If we use $k_{M_{ij}}$,
we obtain the models of
Ref.s~\cite{Itzhaki:2002kh,Gomis:2002km}. These two cases are distinct
in that the scaling of the $M$ in ${\bf Z}_M$ is completely
different. If we use $k_{e^-}$, we need to scale $M$ as $R^2$ in order
to get a finite circle in the limit. If we use $k_{M_{ij}}$, no
scaling of $M$ is needed. In this paper we essentially deal with the
third remaining case, where we use $k_{e_i}$. This is a novel scaling
limit, since we now need to scale $M$ as $R$.

Coming back to the different pp-wave limits, in order to find a
different embedding of the pp-wave limit in ${\cal N}=4$ SYM, it is
sufficient to find a different embedding of the isometries of the
pp-wave geometry in the $SO(2,4)\times SO(6)$ isometry group of the
original $AdS_5 \times S^5$ configuration. As long as the right
isometry algebra appears once we take the $R\rightarrow \infty$ limit,
the resulting geometry will still be the same as that of the original
pp-wave limit.

By looking at the algebra \eqref{algk} one can easily convince himself
that the most general modification of \eqref{isomalg} with this
property is
\begin{equation}\label{isomalg2}
\begin{split}
    k'_{e^-} & =  \frac{1}{R^2}(\Delta+J+(e_i E^i_a + f_i F^i_a +
         g_{ij} M^{ij}_a)T^a) \spa \\ k'_{e^+} & =  (\Delta-J)\spa \\
         k'_{e_i} & =  \frac{1}{R}((E^i_a+ m^i_{jk} M^{jk}_a) T^a)
         \spa \\ {k'_{e_i}}^{\ast} & =  \frac{1}{R}((F^i_a+ n^i_{jk}
         M^{jk}_a) T^a) \spa \\ k'_{M_{ij}} & =  M^{ij}_a T^a~.
\end{split}
\end{equation}
for some constants $e_i,f_i,g_{ij},m^i_{jk},n^i_{jk}$. Notice that
$k_{e^+}$ cannot be modified, but $k_{e^-}$ can.

The original pp-wave limit in Ref.~\cite{Berenstein:2002jq} is related
to the one in section \ref{seciso} by a transformation precisely of
this type. In the original pp-wave limit, none of the $k_{e_i}$
commute with $k_{e^+}$. Therefore, to compactify the pp-wave in this
direction is somewhat cumbersome; the isometry does not commute with
the light-cone Hamiltonian.

The new pp-wave limit is of the form \eqref{isomalg2}, where we chose
$k'_{e^-}$ to be equal to $\frac{1}{R^2}(\Delta + 2 J_R)$ rather than
$\frac{1}{R^2}(\Delta +J)$. The light-cone Hamiltonian that naturally
arises in this case is $\Delta -2 J_R$, which is a linear combination
of $k_{e^+}$ and $k_{M_{ij}}$, and which is indeed kept finite. One
immediately verifies that a linear combination of $k_{e_i}$ and
$k_{e_i}^{\ast}$ commutes with $\Delta- 2 J_R$, as expected.

Thus, our new pp-wave limit is just one of a large family of pp-wave
limits, that all have a different origin in the ${\cal N}=4$ SYM
theory but all yield the same pp-wave limit. It would be interesting
to understand this phenomenon directly at the level of the ${\cal
N}=4$ theory, and in particular to understand why correlation
functions are independent on the choice of embedding of the pp-wave in
the ${\cal N}=4$ theory.

\subsection{Relation to the BMN limit}

In this section we would like to explain in a bit more detail to what
extent our Penrose limit is different from the one considered by BMN
in Ref.~\cite{Berenstein:2002jq}. A priori the fact that different
Penrose limits seem to exist may appear strange. After all, Penrose
limits \cite{Penrose:1976,Gueven:2000ru,Blau:2002mw} are based on a
choice of null geodesic, and all geodesics on $S^5$ are related to
each other by the $SO(6)$ global symmetry. Indeed, a geodesic is given
by point on $S^5$ and a unit tangent vector at that point. Using
$SO(6)$ we can move any point to any other point, and the
$SO(5)\subset SO(6)$ that fixes a point can be used to related any
tangent vector to any other tangent vector.  However, the existence of
different Penrose limits is related to inequivalent ways in which we
can choose the neighborhood of null geodesics. Normally
\cite{Penrose:1976}, Penrose limits involve a very specific choice of
coordinates in the neighborhood of the null geodesic, but one of the
points of this paper is that there are other choices that lead to
well-defined but inequivalent scaling limits. Perhaps our use of the
word Penrose limit to describe these other situations is an abuse of
the word Penrose limit, but we use the word to describe any
well-defined scaling limit of the neighborhood of a null geodesic that
gives rise to a plane wave limit.

To illustrate the different choices of neighborhoods of a geodesic on
$S^5$, consider again $S^5$ embedded in $\C^3$ and labeled by three
complex coordinates $(a_1,a_2,a_3)$. A particular choice of geodesic
is the circle $(e^{i\alpha},0,0)$.  We can choose two different
coordinate systems that contain this geodesic once we fix all but one
coordinate. The first choice is $(re^{i\alpha},a_2,a_3)$, the second
choice is $(r_1 e^{i(\alpha+\beta)}, r_2 e^{i(\alpha-\beta)},a_3)$.
Both contain the geodesic if we choose $r=r_1=1$,
$r_2=a_2=a_3=\beta=0$.  But away from the geodesic, $\alpha$ plays a
different role in each coordinate system. In the first system the
circles parameterized by $\alpha$ shrink to zero size for $r_1=0$, but
in the second system they shrink to zero size only if $r_1=r_2=0$. In
the Penrose limit we are only interested in a neighborhood of the
geodesic and we never actually see any of these circles shrink to zero
size.  Nevertheless, the qualitative differences in the choice of
coordinates near the geodesic give rise to different scaling limits.

To illustrate the difference between the Penrose limit of BMN and our
Penrose limit, we write down the Killing vectors $k_{e_-},k_{e_+}$ and
$k_{e_1}+k_{e_2^*}$ for each of the two cases. In terms of the
following $SO(4)$ generators\footnote{Notice that $[J_L,J_R]=0$, and
therefore we can also choose both $J_L,J_R$ block diagonal. In that
representation, $J$ would no longer be block-diagonal.}
\begin{equation}
J = \left( \begin{array}{cccc} 0 & 1 & 0 & 0 \\ -1 & 0 & 0 & 0 \\ 0 &
 0 & 0 & 0 \\ 0 & 0 & 0 & 0 \end{array} \right)\spa\, J_R = \left(
 \begin{array}{cccc} 0 & \frac{1}{2} & 0 & 0 \\ -\frac{1}{2} & 0 & 0 &
 0 \\ 0 & 0 & 0 & \frac{1}{2} \\ 0 & 0 & -\frac{1}{2} & 0 \end{array}
 \right)\spa\, J_L = \left( \begin{array}{cccc} 0 & 0 & 0 &
 \frac{1}{2} \\ 0 & 0 & -\frac{1}{2} & 0 \\ 0 & \frac{1}{2} & 0 & 0 \\
 -\frac{1}{2} & 0 & 0 & 0 \end{array} \right)\spa
\end{equation}
the standard BMN limit is
\begin{equation}\label{set1}
\begin{split}
k_{e_+} & =  (\Delta+J)/R^2 \spa \\ k_{e_-} & =  (\Delta-J) \spa \\
k_{e_1}+k_{e_2^*} & =  2J_L/R \spa
\end{split}
\end{equation}
whereas our limit is
\begin{equation}\label{set2}
\begin{split}
k_{e_+} & =  (\Delta+2J_R)/R^2 \spa \\ k_{e_-} & =  (\Delta-J) \spa \\
k_{e_1}+k_{e_2^*} & =  2J_L/R~.
\end{split}
\end{equation}
The light-cone Hamiltonian in the plane-wave coordinates
\eqref{usualpp1-2} is $\Delta-J$, whereas in the coordinates
\eqref{iso1} with a manifest isometry it is given by
$\Delta-2J_R$. One crucial difference between \eqref{set1} and
\eqref{set2} is that in \eqref{set1} $k_{e_+}$ and $k_{e_1}+k_{e_2^*}$
commute only after we take the large $R$ limit, whereas in
\eqref{set2} they already commute before taking the large $R$
limit. Therefore, at finite $R$, there is no $PSU(2,2|4)$ global
symmetry that maps \eqref{set1} into \eqref{set2}. This is what we
mean when we say that the pp-wave can be embedded in inequivalent ways
in the original $\CN=4$ theory. In fact, as we showed in section~5.1,
there are many-parameter families of inequivalent embeddings of the
pp-wave in $\CN=4$. For each choice of parameters, there is a set of
gauge theory states that are in one-to-one correspondence with the
states of string theory in the pp-wave background. Before taking the
large $R$ limit, the correlation functions of these states will depend
on the choice of embedding of the pp-wave, but in the large $R$ limit
they will all become identical. The precise gauge-theoretic origin of
this phenomenon is not clear, but the large $R$ limit should play a
crucial role.

\section{Penrose limit with isometries and time-dependence}
\label{ythings}

In section \ref{secsugra} we considered a particular Penrose limit of
$\ads_5 \times S^5$ that resulted in a pp-wave background with one
space-like isometry.  In this section we put forward another Penrose
limit of $\ads_5 \times S^5$ that instead gives two space-like
isometries.  Interestingly, this Penrose limit results in a
time-dependent background.  This background has previously been found
in Ref.~\cite{Michelson:2002wa} by making a coordinate transformation
of the pp-wave solution of Ref.s~\cite{Blau:2001ne,Berenstein:2002jq}.
We review this coordinate transformation in appendix \ref{appkill}.
After presenting the Penrose limit of $\ads_5 \times S^5$ we show how
to construct the corresponding Penrose limits for $\ads_5 \times S^5 /
\Z_M$ and  $\ads_5 \times S^5 / (\Z_{M_1} \times \Z_{M_2})$ which
result in one and two compact space-like directions, respectively.

\subsubsection*{New Penrose limit of $\ads_5 \times S^5$ with
two space-like isometries}

Consider again the $\ads_5 \times S^5$ background
\eqref{sol1}-\eqref{sol2} in global coordinates. As in section
\ref{seciso} we embed the $S^5$ in $\C^3$ via \eqref{aemb} and we
write the metric of $S^5$ as \eqref{thes5}, \eqref{thes3} and
\eqref{theLRs3}. Defining the light-cone coordinates
\begin{equation}
\tilde{y}^\pm = \frac{1}{2} ( t \pm \psi )\spa
\end{equation}
we write the Penrose limit as
\begin{equation}\label{ypen1-2}
\begin{gathered}
R \rightarrow \infty \qquad \mbox{with} \qquad \tilde{y}^+ = \mu y^+
\spa \tilde{y}^- = \frac{1}{\mu R^2} y^-\spa\\ \phi_L = \frac{y^1}{R}
\spa \phi_R = \frac{y^2}{R} \spa \rho = \frac{r}{R} \spa \theta =
\frac{\tilde{r}}{R}~.
\end{gathered}
\end{equation}
This gives the pp-wave background
\begin{equation}
\label{ysol1}
ds^2 = - 4 dy^+ dy^- - \mu^2 y^I y^I (dy^+)^2 + dy^i dy^i + 2 \cos (
2\mu y^+ ) dy^1 dy^2\spa
\end{equation}
with Ramond-Ramond five-form field strength
\begin{equation}
\label{ysol2}
F_{(5)} = 2 \,\mu \,dy^+ \left(\sin(2\mu y^+) dy^1 dy^2 dy^3 dy^4 +
dy^5 dy^6 dy^7 dy^8 \right)\spa
\end{equation}
with $i=1,...,8$ and $I = 3,...,8$.  Here $y^3 , y^4$ are defined  by
$y^3 + i y^4 = \tilde{r} e^{i\alpha}$ and $y^5,...,y^8$ are defined by
$r^2 = \sum_{I=5}^8 (y^I)^2 $ and $dr^2 + r^2 (d\Omega_3')^2 =
\sum_{I=5}^8 (dy^I)^2$.  As can be seen from the presence of the
factor $\cos(2\mu y^+ )$ in  the metric and  $\sin(2\mu y^+ )$ in the
Ramond-Ramond field strength, this  pp-wave background is manifestly
time-dependent.  In appendix \ref{appkill} it is explained that  this
solution is in fact  the maximally symmetric type IIB pp-wave
background of Ref.s~\cite{Blau:2001ne,Berenstein:2002jq} in a
different coordinate system than the one used by
Ref.s~\cite{Blau:2001ne,Berenstein:2002jq}. This coordinate system was
found by Michelson \cite{Michelson:2002wa}.  This means that the
physics for the two backgrounds should be equivalent. However, in the
following we compactify one or two of these space-like isometries, and
the string theory is therefore no longer equivalent to string theory
on the pp-wave of  Ref.~\cite{Blau:2001ne,Berenstein:2002jq}. More
precisely, the presence of winding operators in the compact directions
means that we obtain a truly time-dependent string theory on those
backgrounds.  We expect that one should be able to find the gauge
theory operators corresponding to the various string states on these
backgrounds.  If this is true, it could be a new approach to study
string theory in time-dependent backgrounds.

It is also interesting to note that the Penrose limits discussed here
are not in the class considered in section \ref{explain}.  This is
because we have an explicit dependence on the light-cone time in the
coefficients mapping the algebra to the one of the pp-wave in the
coordinate system of Ref.~\cite{Blau:2001ne,Berenstein:2002jq} and
because we do not consider a $U(1)$ in the $SO(6)$ to get the
Hamiltonian. It would be interesting to understand how to extend the
general framework of section \ref{explain} to encompass also the
Penrose limits of this section.

\subsubsection*{Space-like circle from $\ads_5 \times S^5 / \Z_M$}

Let us now see how to find a Penrose limit giving one compact
direction.  We define the orbifolded space $\ads_5 \times S^5 / \Z_M$
as in section  \ref{seccirc}. This obviously gives the identification
\begin{equation}
y^1 \equiv y^1 + 2\pi \frac{R}{M}
\end{equation}
for the $y^1$ coordinate defined in \eqref{ypen1-2},  whereas the
$y^2$ coordinate is not affected by the identification. Thus we should
keep $R/M$ fixed in the Penrose limit \eqref{ypen1-2}.   We thus have
a space-like circle along the $y^1$ direction of radius $R/M$ in this
Penrose limit.  Indeed the momentum along $y^1$ is
\begin{equation}
P_1 = \frac{M}{R} \frac{2J_L}{M}\spa
\end{equation}
which is quantized in units of $M/R$, as it should be.

\subsubsection*{Space-like two-torus from
$\ads_5 \times S^5 / (\Z_{M_1} \times \Z_{M_2})$}

Let us finally describe how to get two space-like compact directions.
We define the orbifolded space $\ads_5 \times S^5 / (\Z_{M_1}  \times
\Z_{M_2})$ as in section \ref{secDLCQ}. In terms of the coordinates
defined in \eqref{ypen1-2} we have the identifications
\begin{equation}
\label{yiden1}
y^1 \equiv y^1 + 2\pi \frac{R}{M_1} n_1 + \pi \frac{R}{M_2} n_2 \spa
y^2 \equiv y^2 - \pi \frac{R}{M_2} n_2 \spa \alpha \equiv \alpha +
\frac{2\pi}{M_2} n_2\spa
\end{equation}
for any $n_1,n_2 \in \Z$. No other coordinates are affected by these
identifications.  We now see that if we choose $M_1$ and $M_2$ so that
\begin{equation}
\label{toruscond}
M_1 = 2  n  M_2 \spa \ n \in \Z\spa
\end{equation}
then \eqref{yiden1} is equivalent to
\begin{equation}
y^1 \equiv y^1 + 2\pi \frac{R}{M_1} n_1 \spa  y^2 \equiv y^2 - \pi
\frac{R}{M_2} n_2 \spa \alpha \equiv \alpha + \frac{2\pi}{M_2} n_2\spa
\end{equation}
for any $n_1,n_2 \in \Z$.  Taking the Penrose limit \eqref{ypen1-2}
keeping $R/M_1$ and $R/M_2$ fixed then gives two space-like compact
directions $y^1$ and $y^2$ with radii $R/M_1$ and $R/M_2$,
respectively.  We see from \eqref{toruscond} that we can only get a
very restricted class of two-tori.  The momenta along $y^1$ and $y^2$
read
\begin{subequations}
\begin{align}
P_1 &= \frac{1}{R}\,2\,J_{L} = \frac{M_1}{R}\frac{2J_{(1)}}{M_1}
\spa\\ P_2 &= \frac{1}{R} \,2\, J_{R} = \frac{1}{R} 2 (J_{(2)} +
J_{(3)}) \simeq  \frac{2 M_2}{R} \frac{J_{(1)} + 2 J_{(2)} }{M_2}
\end{align}
\end{subequations}
and we now see, given the quantization rules for $J_{(1,2)}$, that
these momenta are correctly quantized in units of $M_1/R$ and  $2 M_2
/ R = M_1/ n R$  respectively.

\section{Summary, discussion and conclusions}
\label{secconcl}

In this final section we first summarize  the new Penrose limits that
have been discussed in this paper as compared with previously
discussed ones. Then we present a more general overview of the main
results of this work, along with open directions.

\subsection{Summary of new Penrose limits}

Every Penrose limit can be described by the action of the $U(1)$ that
generates the geodesic on $S^5$, together with the action of the
orbifold groups. Each of them can be described by the action on the
complex coordinates $(a_1,a_2,a_3)$ that parameterize $S^5$ via
$|a_1|^2 +|a_2|^2+|a_3|^2=1$. We have three different geodesics
\begin{subequations}
\begin{align}
G_1 & :  (a_1,a_2,a_3) \rightarrow (i\epsilon a_1, -i\epsilon a_2,a_3)
\spa \\ G_2 & :  (a_1,a_2,a_3) \rightarrow (i\epsilon a_1, a_2,a_3)
\spa \\ G_3 & :  (a_1,a_2,a_3) \rightarrow (-\epsilon a_1 \left|
\frac{a_2}{a_1} \right|, \epsilon a_2 \left| \frac{a_1}{a_2}
\right|,a_3)\spa
\end{align}
\end{subequations}
and two orbifold actions
\begin{subequations}
\begin{align}
H_1 & :  (a_1,a_2,a_3) \rightarrow (\theta a_1, \theta^{-1} a_2,a_3)
\spa \\ H_2 & :  (a_1,a_2,a_3) \rightarrow (a_1,\theta a_2,
\theta^{-1} a_3)\spa
\end{align}
\end{subequations}
for suitable roots of unity $\theta$. The various Penrose limits are
now summarized by
\begin{equation}
\begin{array}{|c|c|c|c|}
\hline & G_1 & G_2 & G_3 \\ \hline - & {\rm isometry} & \mbox{BMN} &
 {\rm isometry,isometry} \\ \hline H_1 & {\rm circle} & \mbox{MRV} &
 {\rm isometry,circle} \\ \hline H_1,H_2 & {\rm circle,DLCQ} & & {\rm
 circle,circle} \\ \hline H_2 & {\rm isometry,DLCQ} & \mbox{IKM,GO} &
 \\ \hline
\end{array} .
\end{equation}
Here, the first column refers to the backgrounds discussed in
section~\ref{secsugra}, and the last column refers to the time
dependent backgrounds of section~\ref{ythings}. The middle column
refers to original Penrose limit of BMN \cite{Berenstein:2002jq}, the
orbifold of MRV \cite{Mukhi:2002ck,Alishahiha:2002jj}, and the
orbifolds of IKM and GO \cite{Itzhaki:2002kh,Gomis:2002km}. We used
the word isometry to indicate a manifest space-like isometry, the word
circle to indicate a space-like circle, and the word DLCQ to indicate
a compact null direction. Though there are probably several other
possible Penrose limits, the table suggests that we have obtained a
reasonably complete picture.

\subsection{Discussion and conclusions}

In this paper we have presented in detail a new correspondence between
string theory on a pp-wave with a space-like circle and $\CN=2$ QGT.
This correspondence includes the detailed map between gauge theory
operators and string theory states and the identification of their
energy eigenvalues, at least to first order in $\gqgt^2 \,N /J_R^2$.
There are several points worth emphasizing:

\begin{itemize}
\item When the space-like isometry is non-compact the  correspondence
is between string theory on the maximally symmetric type IIB pp-wave
in a coordinate system with the space-like isometry manifest and
$\CN=4$ SYM theory in the scaling limit \eqref{limitn4}.  This means
that we have found a new correspondence between theories with 32
supersymmetries. We have put this into a more  general framework in
section \ref{explain}.  This shows an interesting interplay  between
coordinate transformations in string theory backgrounds and selecting
distinct sectors in the dual gauge theory. In particular, we have
shown that the existence of different Penrose limits is related to
inequivalent ways of choosing the neighborhood of null geodesics.

\item When considering the space-like isometry to be compact the
duality  is between string theory on a pp-wave background with a
space-like  circle and the following triple scaling limit of $\CN=2$
$U(N)^M$ QGT
\begin{equation}
\label{seclim}
N \rightarrow \infty \spa \frac{M^3}{N} = \mbox{fixed} \spa
\frac{J_R}{M^2} = \mbox{fixed} \spa \frac{\gqgt^2}{M} = \mbox{fixed}~.
\end{equation}
The number of preserved supersymmetries  of the corresponding subset
of surviving gauge theory operators is enhanced from 16 to 24.

\item The chiral primaries on the gauge theory side precisely give the
spectrum one expects from string theory. This is non-trivial in the
sense that our light-cone Hamiltonian $H=\Delta - 2 J_R$ is such that
we do not have bosonic $H=1$ states in the $z^1$ and $z^2$ directions,
but rather $H=0$ and $H=2$, respectively. This makes this part of the
spectrum very different from that of
Ref.~\cite{Berenstein:2002jq}. Moreover, we do not have zero modes
along the compact direction, $z^1$, and the gauge theory operators
corresponding to oscillators states have vanishing energy in the free
theory. Therefore they correspond to the lowest lying states above the
ground state. The computation of their anomalous dimension works in a
rather different way as compared to the other operators (which
resemble instead the BMN computation), however  we have succeeded in finding
agreement with the string theory predictions, reproducing the result
\begin{equation}
\label{anz1} \Delta - 2 J_R = 0 + \frac{g_{\rm QGT}^2 N n^2}{4
J_R^2}
\end{equation}
at leading order in the planar limit.

\item We have also winding states in the correspondence. All the
operators except the windings are directly inherited from the $\CN=4$
SYM theory since they are in the untwisted sector of $\CN=2$ QGT. The
winding states instead involve the twisted sectors of $\CN=2$ QGT. We have also been able to compute the anomalous dimensions for these states,
obtaining
\begin{equation}
\label{anwin} \Delta - 2 J_R = 0 + \frac{g_{QGT}^2 N m^2}{4 J_R
M^2}~.
\end{equation}
Notice that eq.s~\eqref{anz1} and~\eqref{anwin} map into each
other with the interchange $n/\sqrt{J_R} \leftrightarrow m/M$.
This $m$, $n$ exchange resembles T-duality, which should hold
along $z^1$. Another interesting property of the gauge operators
corresponding to oscillators and winding in the  $z^1$ direction
is that they are degenerate with the ground state in the free
theory while the degeneracy is lifted at one-loop. This is
dictated  by eq.s~\eqref{anz1} and \eqref{anwin}, and the gauge
theory computations perfectly agree with expectations from the
string theory spectrum. We regard this agreement as a non-trivial
check on our proposed correspondence.

\item The presence of the winding states provide other
non-trivial checks on the correspondence. To compare the energy
eigenvalues we used the multiplicity factor counting the average
number of nearest neighbor pairs $AB$ or $BA$ in the totally
symmetrized sum over words ${\rm Tr}[ {\rm sym}(A^{J_R} B^{J_R})]$.
We also used the fact that the phase angle of the winding states is
quantized in units of $2\pi / (J_R M)$ rather than $2\pi/(2J_R)$, as
is the case for insertions of oscillator modes of the string.
Moreover, we have shown that we can derive string theory
level-matching from our general prescription for the operators.
\end{itemize}
There are several further checks and future directions that deserve to
be examined in connection with these new pp-wave/gauge-theory
correspondences. These include:
\begin{itemize}

\item From our analysis it follows that string theory predicts that
the gauge theory operators \eqref{wingau} corresponding to winding
modes are protected in the sense that their anomalous dimension is
not renormalized beyond one loop. It would be interesting to prove
this within a pure field theory computation.

\item We have shown that the genus counting parameter of our new
Penrose limit in the ${\cal N}=4$ theory is given by $g_2^2=J_R^4/N^2$.
By lifting this to the $U(NM)$ ${\cal N}=4$ parent gauge theory of the
${\cal N}=2$ QGT we discussed, one obtains $g_2^2=M^6/N^2$
using the scaling limit \eqref{seclim}. We strongly believe
that this quantity, which indeed is finite in the scaling
limit \eqref{seclim}, is the correct genus counting parameter
of the resulting $\CN =2$ QGT. It would be interesting to
derive this directly from our $\CN =2$ QGT operators.

\item As a related issue, another interesting direction is of course
to study interactions (see
Ref.s~\cite{Kristjansen:2002bb,Berenstein:2002sa,Constable:2002hw,Gopakumar:2002dq,Verlinde:2002ig,Spradlin:2002ar,Kiem:2002xn,Huang:2002wf,Chu:2002pd,Lee:2002rm,Spradlin:2002rv,Chu:2002qj,Klebanov:2002mp,Huang:2002yt,Chu:2002eu,Beisert:2002bb,Gross:2002mh,Zhou:2002mi,Constable:2002vq,Lee:2002vz}
for current work on interactions for the original BMN proposal).
It would be interesting to develop string field theory in backgrounds
with compact directions
and compare in our setting 3-point functions on the gauge theory side
with predictions from  string field theory.

\end{itemize}

In the context of this new correspondence we also found several things
that would be interesting to study further:
\begin{itemize}
\item We found the Penrose limits describing the  space-like isometry
pp-wave background with a compact null direction. This we found both
when the space-like isometry is compact and non-compact.  This makes
it possible to find a correspondence between the DLCQ of string
theory, which should be some kind of matrix string theory, and a
quiver gauge theory.

\item We found a new Penrose limit giving a background with two
space-like isometries. We also found the corresponding Penrose limits
giving one and two compact directions. These backgrounds are
time-dependent, which obviously makes the pp-wave/gauge theory
correspondence very interesting.

\end{itemize}

%%%%%%%%%%%%%%%%%%%%%%%%%%%%%%%%%%%%%%%%%%%%%%%%%%%%%%%%%%%%%%%%%%%
\subsection*{Acknowledgments}

We would like to thank M. Blau, G. Bonelli, P. Di Vecchia,
E. Kiritsis, C.  Kristjansen, J.L. Petersen, J. Plefka, S.-J. Rey,
S. Ross, R.  Russo, K. Savvidis, S. Sciuto, A. Sevrin, N. Toumbas and
S. Vandoren for useful discussions. We would especially like to thank
Assaf Shomer for  pointing out a mistake in a previous version of this
paper. JdB would like to thank the Aspen Institute for Physics for
hospitality while part of this work was being completed. TH thanks the
Niels Bohr Institute for hospitality and support during part of this
work. EI was supported during part of this work by an EC Marie Curie
Training Site Fellowship at Nordita, under contract number
HPMT-CT-2000-00010. MB is supported by an EC Marie Curie Postdoc
Fellowship under contract number HPMF-CT-2000-00847. This work is
partially supported by the European Community's Human Potential
Programme under contract number HPRN-CT-2000-00131.

%%%%%%%%%%%%%%%%%%%%%%%%%%%%%%%%%%%%%%%%%%%%%%%%%%%%%%%%%%%%%%%%%%%%%%
%%%%%%%%%%%%%%%%%%%%%%%%%%%%%%%%%%%%%%%%%%%%%%%%%%%%%%%%%%%%%%%%%%%%%%
\begin{appendix}

\section{Killing vectors and coordinate transformations of the pp-wave}
\label{appkill}

In this appendix we briefly review the connection between the pp-wave
solution of Ref.s~\cite{Blau:2001ne,Berenstein:2002jq} and the
solutions \eqref{iso1}-\eqref{iso2} and
\eqref{ysol1}-\eqref{ysol2}. Note that this appendix is essentially a
brief review of certain results and methods of
Ref.~\cite{Michelson:2002wa}. However, we do have some remarks on the
physical understanding of the space-like isometries leading to the
solution \eqref{iso1}-\eqref{iso2}.

\subsubsection*{Space-like Killing vectors}

Consider the $\ads_5 \times S^5$ solution \eqref{sol1}-\eqref{sol2}
with the $S^5$ parameterized by \eqref{thes5}.  Define the light-cone
coordinates $\tilde{x}^\pm = (t \pm \alpha)/2$.  The BMN Penrose limit
is now
\begin{equation}
R \rightarrow \infty \spa \tilde{x}^+ = \mu x^+ \spa \tilde{x}^- =
\frac{1}{\mu R^2} x^- \spa \rho = \frac{r}{R} \spa \theta =
\frac{\pi}{2} - \frac{\hat{r}}{R}~.
\end{equation}
This gives the pp-wave solution of
Ref.s~\cite{Blau:2001ne,Berenstein:2002jq}
\begin{subequations}\label{usualpp1-2}
\begin{align}
    ds^2 &= - 4 dx^+ dx^- - \mu^2 x^i x^i (dx^+)^2 + dx^i dx^i\spa \\
    F_{(5)} &= 2 \,\mu \,dx^+ \left(dx^1 dx^2 dx^3 dx^4 + dx^5 dx^6 dx^7
        dx^8 \right)\spa
\end{align}
\end{subequations}
with $i,j=1,...,8$.  This solution has 30 bosonic Killing vectors
\cite{Blau:2001ne,Michelson:2002wa}.  One can now make linear
combinations of these Killing vectors in order to obtain space-like
Killing vectors. Consider the following two Killing vectors
\begin{subequations}
\begin{align}
V_1& = \cos( \mu x^+ ) \frac{\partial}{\partial x^1} - \sin( \mu x^+ )
\frac{\partial}{\partial x^2} - \frac{\mu}{2} \left( x^1 \sin( \mu x^+
) + x^2 \cos( \mu x^+ ) \right) \frac{\partial}{\partial x^-} \spa\\
V_2 &= - \cos( \mu x^+ ) \frac{\partial}{\partial x^1} - \sin( \mu x^+
) \frac{\partial}{\partial x^2} + \frac{\mu}{2} \left( x^1 \sin( \mu
x^+ ) - x^2 \cos( \mu x^+ ) \right) \frac{\partial}{\partial x^-} ~.
\end{align}
\end{subequations}
Clearly $|V_1| = |V_2| = 1$ so both Killing vectors have norm 1.  Note
also that $[V_1,V_2] = 0$.  Obviously we can make similar space-like
Killing vectors by exploiting the $SO(4) \times SO(4)$ symmetry of the
pp-wave \cite{Michelson:2002wa}.

\subsubsection*{Coordinate transformations}

Since the Killing vectors $V_1$ and $V_2$ are purely space-like we can
transform to a coordinate system in which they correspond to manifest
space-like isometries.  Consider first a transformation that takes
$(x^-,x^1,x^2)$ into $(z^-,z^1,z^2)$ leaving all other coordinates
invariant, i.e. $z^+ = x^+$, $z^I=x^I$, $I=3,...,8$, with
\begin{equation}
\label{kill1}
\frac{\partial}{\partial z^1} = V_1~.
\end{equation}
It is straightforward to see that we can realize \eqref{kill1} with a
coordinate transformation where $(x^1,x^2)$ and $(z^1,z^2)$ are
connected by a rotation. The full coordinate transformation is then
\cite{Michelson:2002wa}
\begin{equation}
\begin{gathered}
\vecto{x^1}{x^2} = \matrto{\cos \mu z^+ }{\sin \mu z^+ }{-\sin \mu z^+
}{\cos \mu z^+ } \vecto{z^1}{z^2} \spa\\
x^- = z^- - \frac{\mu}{2} z^1 z^2 ~.
\end{gathered}
\end{equation}
This gives the eq.s~\eqref{iso1}-\eqref{iso2}.  Thus, we see that the
solution \eqref{iso1}-\eqref{iso2} indeed is connected to the solution
\eqref{usualpp1-2} of
Ref.s~\cite{Blau:2001ne,Berenstein:2002jq} by a coordinate
transformation.

We can also find a coordinate transformation that takes
$(x^-,x^1,x^2)$ into $(y^-,y^1,y^2)$ leaving all other coordinates
invariant, i.e. $y^+ = x^+$, $y^I=x^I$, $I=3,...,8$, with
\begin{equation}
\label{kill2}
\frac{\partial}{\partial y^1} = V_1 \spa \frac{\partial}{\partial y^2}
= V_2 ~.
\end{equation}
This is more restricted and we get the coordinate transformation
\cite{Michelson:2002wa}
\begin{equation}
\begin{gathered}
x^1 = \cos (\mu y^+) ( y^1 - y^2 ) \spa x^2 = - \sin (\mu y^+) ( y^1 +
y^2 ) \spa\\
x^- = y^- + \mu \sin ( 2 \mu y^+ ) y^1 y^2 ~.
\end{gathered}
\end{equation}
This gives the solution \eqref{ysol1}-\eqref{ysol2}, as promised.

\subsubsection*{Physical interpretation of $z$ coordinates}

We describe here the Newtonian physics of the motion in the
$(z^1,z^2)$ plane.  We can think of the motion in the $(x^1,x^2)$
coordinate system as a 2-dimensional Newtonian mechanical system where
a spring is attached between the point $(x^1,x^2)$ and the origin, and
with $t = x^+$ being the Newtonian time.  We thus have Hamiltonian $H
= (\vec{p}^2 /(2m)) + (k/2)\vec{x}^2$ and the Newton force law is $m
\vec{\ddot{x}} = - k \vec{x}$.

We now do the coordinate transformation
\begin{equation}
\vecto{z^1}{z^2} = \matrto{\cos \omega t }{-\sin \omega t }{\sin
\omega t }{\cos \omega t } \vecto{x^1}{x^2}~.
\end{equation}
To describe the motion in the $(z^1,z^2)$ coordinate system we need to
add fictional forces to the Newton force law.  We get
\begin{equation}
m \vec{\ddot{z}} = - k \vec{z} - 2 m \omega
\vecto{-\dot{z}^2}{\dot{z}^1} + m \omega^2 \vec{z} \spa
\end{equation}
corresponding to the Hamiltonian
\begin{equation}
H = \frac{\vec{p}^2}{2m} + \frac{1}{2} ( k - \omega^2 m ) \vec{z}^2 -
\omega ( z^1 p_2 - z^2 p_1 ) ~.
\end{equation}
Choosing the critical value $\omega = \sqrt{k/m}$ for the rotation
velocity we have
\begin{subequations}
\begin{align}
m \vec{\ddot{z}} &=  - 2 m \omega \vecto{-\dot{z}^2}{\dot{z}^1} \spa\\
H &= \frac{\vec{p}^2}{2m} - \omega ( z^1 p_2 - z^2 p_1 )~.
\end{align}
\end{subequations}
The motion is thus that of a free charged particle in a constant
magnetic field.%
\footnote{See also Ref.~\cite{Kiritsis:2002kz} for a discussion of
free trajectories in the pp-wave background.}

Obviously, we still do not have a free motion in the $z^1$ direction
at this point. But the point is now that the coordinate transformation
$x^- = z^- - (\mu/2) z^1 z^2$ induces an extra term in the Hamiltonian
of the form $C (z^1 p_2 + z^2 p_1 )$.  Choosing $C = \omega$ we
therefore get
\begin{subequations}
\begin{align}
H &= \frac{\vec{p}^2}{2m} + 2 \omega z^2 p_1 \spa\\
m \vec{\ddot{z}} &= - 2 m \omega \vecto{-\dot{z}^2}{\dot{z}^1} + 4 m
\omega^2 \vecto{0}{z^2}~.
\end{align}
\end{subequations}
The Hamiltonian is now independent of $z^1$ giving $\dot{p}_1 = 0$
which means that this is a free direction.  We can also see from
Newtons force law that it now is possible to have a motion with
constant velocity $\dot{z}^1$ if we choose $z^2 = \dot{z}^1 /
(2\omega)$.

%%%%%%%%%%%%%%%%%%%%%%%%%%%%%%%%%%%%%%%%%%%%%%%%%%%%%%%%%%%%%%%%%%%%
%%%%%%%%%%%%%%%%%%%%%%%%%%%%%%%%%%%%%%%%%%%%%%%%%%%%%%%%%%%%%%%%%%%%
\section{$\CN=2$ quiver gauge theory}
\label{appqgt}

We give here some details on the $\CN=2$ quiver gauge theory
(QGT) discussed in section \ref{seccirc}, namely the theory
 corresponding to
$N$ D3-branes on the orbifold $\C^2/\Z_M$.

By studying the low energy spectrum of open strings ending on the
stack of D3-branes  (for a recent review see
Ref.~\cite{Bertolini:2001gq}), it is easy to see that the
$\CN=2$ QGT has gauge group
\begin{equation}
U(N)^{(1)} \times U(N)^{(2)} \times \cdots \times U(N)^{(M)}\spa
\end{equation}
where the gauge coupling $\gqgt^2$ is the same for all group factors
and is expressed as $\gqgt^2 = 4\pi g_s M$ in terms of the string
coupling $g_s$. There are $M$ vector multiplets $(A_{\mu
I},\Phi_I,\psi_{\Phi, I},\psi_{I})$ ($I=1,...,M$ and $\Phi_I$ is a
complex scalar transforming in the adjoint representation of
$U(N)^{(I)}$) and  $M$ bi-fundamental hypermultiplets which in
${\mathcal N}=1$ notation  read $(A_I,B_I,\chi_{A, I},\chi_{B,
I})$. The complex scalar $A_I$ transforms in the $({ N}_I,{ \bar
N}_{I+1})$ and $B_I$ transforms in the $({ \bar N}_I,{ N}_{I+1})$
where ${ N}_{I}$ (${\bar N}_{I}$) represents the fundamental
(anti-fundamental) representation  of the gauge group $U(N)^{(I)}$
($U(N)^{(M+1)}$ is identified with $U(N)^{(1)}$). This theory has $16$
supercharges and is conformal \cite{Kachru:1998ys}.

\subsubsection*{Connection to $\CN=4$ SYM theory}

We now describe how the $\CN=2$ QGT discussed above can be
obtained as a projection of $\CN=4$ SYM theory with gauge
group $U(NM)$. This is useful since one can then   simply use an
$\CN=4$ formalism in performing gauge theory computations. As
discussed in
Ref.s~\cite{Douglas:1996sw,Johnson:1997py,Lawrence:1998ja}, the idea
is to consider $NM$ D3-branes on the covering space of $\C^2/\Z_M$ and
perform a $\Z_M$ projection on the worldvolume fields and the
Chan-Paton factors.

The Euclidean action for $\CN=4$ SYM is
\begin{equation}
\label{actn4}
S_E = \frac{1}{\gym^2} \int d^4 x \mbox{Tr} \left(  \frac{1}{4}
F_{\mu\nu} F_{\mu \nu}  + \frac{1}{2} D_\mu \phi^i D_\mu \phi^i +
\frac{1}{4} [\phi^i , \phi^j] [\phi^i , \phi^j] \right) + S_E^{\rm fer}\spa
\end{equation}
with $\phi^i$, $i=1,...,6$ being the six scalars,  $D_\mu =
\partial_\mu + [ A_\mu, \cdot ]$, $F_{\mu \nu} = D_{[\mu} A_{\nu]}\,$.
$S_E^{\rm fer}$ is the fermionic part of the action.  Note that all
fields are $NM \times NM$ matrices, so explicitly one should write the
scalar fields as $\phi^i_{ab}$ with $a$ belonging to the fundamental
representation of $U(NM)$ and $b$ belonging to the conjugate
representation. Define now the three complex scalars
\begin{equation}
\label{ws}
W_1 = \frac{1}{\sqrt{2}} ( \phi^1 + i \phi^2 ) \spa W_2 =
\frac{1}{\sqrt{2}} ( \phi^3 + i \phi^4 ) \spa W_3 = \frac{1}{\sqrt{2}}
( \phi^5 + i \phi^6 ) ~.
\end{equation}
Then we write the scalar part of the langrangian of $\CN=4$
SYM as
\begin{multline}
\frac{1}{\gym^2} \mbox{Tr} \left( \frac{1}{2} D_\mu \phi^i D_\mu
\phi^i + \frac{1}{4} [\phi^i , \phi^j] [\phi^i , \phi^j] \right)
\\  = \frac{1}{\gym^2} \mbox{Tr} \left( \frac{1}{2} D_\mu W_1 D_\mu
\bar{W_1} + \frac{1}{2} D_\mu W_2 D_\mu \bar{W_2} + \frac{1}{2} D_\mu
W_3 D_\mu \bar{W_3} \right) + \mathcal{L}_D + \mathcal{L}_F\spa
\end{multline}
with
\begin{subequations}
\begin{align}
\mathcal{L}_D &= - \frac{1}{\gym^2} \frac{1}{2} \mbox{Tr} \left(
[W_1,\bar{W_1}] + [W_2,\bar{W_2}] + [W_3,\bar{W_3}] \right)^2 \spa\\
\mathcal{L}_F &= \frac{1}{\gym^2} \mbox{Tr} \left( | [W_1,W_2] |^2 + |
[W_1,W_3] |^2 + | [W_2,W_3] |^2 \right)~.\label{Fterms}
\end{align}
\end{subequations}
One obtains the $\CN=2$ QGT by using the following relation
between the $\CN=4$ SYM fields and the fields of the
$\CN=2$ QGT
\begin{equation}\label{AmPAB}
\begin{aligned}
    A_{\mu} &= \left(
        \begin{array}{cccccc} A_{\mu 1} & &&& & \\ & A_{\mu 2} & &&& \\ & &\ddots&& &
        \\  & &&& & A_{\mu M}\end{array}\right)  \spa &
    W_3 \equiv \Phi &=
        \left(\begin{array}{cccccc} \Phi_1 & &&& & \\  & \Phi_2& &&& \\
        &&\ddots && & \\ & &&& & \Phi_M\end{array}\right)\spa\\
    W_1 \equiv A &= \left(
        \begin{array}{ccccccc} 0 & A_1 &&&& & \\ & 0 & A_2 &&&& \\ & &\ddots&&& & \\
        & &&&& 0 & A_{M-1}  \\ A_M & &&&& & 0 \end{array} \right) \spa &
    W_2 \equiv B &= \left(
        \begin{array}{ccccccc} 0 & &&&& & B_M \\ B_1 & 0 &&&&& \\ & &&\ddots && & \\
        & &&& & 0 & \\ & &&&& B_{M-1}& 0 \end{array} \right)~.
\end{aligned}
\end{equation}
These are the projected fields of $\CN=4$ SYM
corresponding to the orbifold projections acting on the Chan-Paton
factors of the  open strings ending on the $NM$ D3-branes in the
covering space of the orbifold space. Each non-vanishing entry of
the above matrices is an $N \times N$ matrix and corresponds to the
scalar fields of the ${\mathcal N}=2$ theory.

For future purposes, let us write down the normalization of the
scalar propagators which, given the identifications \eqref{AmPAB}, can be easily inferred from the action
\eqref{actn4} to be
\begin{equation}
\label{scor2}
\langle \Phi^{ab}_I (x) \bar \Phi^{cd}_J (0) \rangle = \langle
A^{ab}_I (x) \bar A^{cd}_J (0) \rangle = \langle B^{ab}_I (x) \bar
B^{cd}_J (0) \rangle = \delta^{ac}\delta^{bd} \frac{\gqgt^2}{8 \pi^2}
\frac{\delta_{IJ}}{[x^2]}\spa
\end{equation}
where $a,b,c,d$ are now $U(N)$ adjoint indices.

In Ref.s~\cite{Bershadsky:1998mb,Bershadsky:1998cb} it was shown both
using string theory and field theory arguments that the correlation
functions of the $\CN=2$ QGT theory are the same as that of
the corresponding $\CN=4$ SYM theory in the planar limit,
provided we identify $\gqgt^2 = \gym^2 M = 4 \pi g_s M$. This relation
also follows from identifying the action \eqref{actn4} with the action
for $\CN=2$ QGT.

\subsubsection*{Fermions from $\CN=4$ SYM theory}

We can also project the fermions fields in a similar fashion.  We
write the fermion part of the $\CN=4$ SYM theory action as
\begin{equation}
S_E^{\rm fer} = \frac{1}{\gym^2} \int d^4 x \mbox{Tr} \left(
\frac{1}{2} \bar{\Psi} \, \Gamma^\mu D_\mu \Psi + \frac{1}{2}
\bar{\Psi} \, \Gamma^i [ \phi^i , \Psi] \right)
\end{equation}
where $\Psi$ is a spinor in ten dimensions transforming in the adjoint of
$U(NM)$ and $\Gamma_\mu$ are the Gamma matrices of $SO(10)$.  The spinor
$\Psi$ is divided into four different four-dimensional spinors
$\chi_A$,$\chi_B$,$\psi_\Phi$ and $\psi$ so that $\chi_A$ is the
superpartner of the scalar field $A$, $\chi_B$ the superpartner of
$B$, $\psi_\Phi$ the superpartner of $\Phi$ and $\psi$ the gaugino
field.  The projections of the $\CN=4$ SYM spinors
$\chi_A$,$\chi_B$,$\psi_\Phi$ and $\psi$ to the spinors
$\chi_{A,I}$,$\chi_{B,I}$,$\psi_{\Phi,I}$ and $\psi_I$, $I=1,...,M$,
of $\CN=2$ QGT are given by
\begin{equation}\label{psichis}
\begin{aligned}
    \psi &= \left(
        \begin{array}{cccccc} \psi_{1} & &&& & \\
        & \psi_{2} & &&& \\ & &\ddots&& & \\  & &&& &
        \psi_{M}\end{array}\right)  \spa &
    \psi_\Phi &=
        \left(\begin{array}{cccccc} \psi_{\Phi,1} & &&& & \\  & \psi_{\Phi,2}
        & &&& \\ &&\ddots && & \\ & &&& & \psi_{\Phi,M} \end{array}\right)\spa\\
    \chi_A &= \left(
        \begin{array}{ccccccc} 0 & \chi_{A,1} &&&& & \\
        & 0 & \chi_{A,2} &&&& \\ & &\ddots&&& & \\ & &&&& 0 & \chi_{A,{M-1}}
        \\ \chi_{A,M} & &&&& & 0 \end{array} \right) \spa &
    \chi_B &= \left(
        \begin{array}{ccccccc} 0 & &&&& & \chi_{B,M} \\
        \chi_{B,1} & 0 &&&&& \\ & &&\ddots && & \\ & &&& & 0 & \\ & &&&&
        \chi_{B,{M-1}}& 0 \end{array} \right)~.
\end{aligned}
\end{equation}

\subsubsection*{Scalar Chiral Primaries in $\CN=4$ SYM and
$\CN=2$ QGT}

Consider the real scalars $\phi^i$, $i=1,...,6$, which transform in
the adjoint of $U(NM)$.  The single-trace scalar chiral primaries are
given by the symmetric traceless combination of $\mbox{Tr} (
\phi^{i_1} \cdots \phi^{i_n} )$. More explicitly, define the
symmetrized trace
\begin{equation}
V^{i_1 \cdots i_n} \equiv \mbox{Tr}  \left( \phi^{(i_1} \cdots
\phi^{i_n)} \right) ~.
\end{equation}
Then the chiral primaries $C^{i_1 \cdots i_n}$ are given by
\begin{equation}
C^{i_1 \cdots i_n} = V^{i_1 \cdots i_n}  - \frac{1}{6} \delta^{(i_1
i_2} V^{i_3 \cdots i_n ) jk} \delta_{jk} \cdots\spa
\end{equation}
where we only wrote the single contractions explicitly.  This
chiral primary is thus a tensor $C^{i_1 \cdots i_n}$ that  transform in the
irreducible $(n,0,0)$ representation (in terms of the
conventional Dykin labels) of $SO(6)$.

We can also consider the above chiral primaries in $SU(3)$ notation.
Our notation is here that $W^a$ for $a=1,2,3$ are defined by
\eqref{ws} whereas $W^a$ for $a=\bar{1},\bar{2},\bar{3}$ are defined
by $W^{\bar{a}} = \bar{W}^a$ for $a=1,2,3$. Define now the object
\begin{equation}
Y^{a_1 \cdots a_n} \equiv \mbox{Tr}  \left( W^{(a_1} \cdots W^{a_n)}
\right)\spa
\end{equation}
for $a_i = 1,2,3,\bar{1},\bar{2},\bar{3}$.  Then the scalar chiral
primaries are given by the tensor
\begin{equation}
\label{chpr}
P^{a_1 \cdots a_n} = Y^{a_1 \cdots a_n}  - \frac{1}{3} \eta^{(a_1 a_2}
Y^{a_3 \cdots a_n ) bc} \eta_{bc} \cdots\spa
\end{equation}
where
\begin{equation}
\eta_{1\bar{1} } = \eta_{2\bar{2} } = \eta_{3\bar{3} } =
\eta^{1\bar{1} } = \eta^{2\bar{2} } = \eta^{3\bar{3} } = 1 \spa \ \
\eta_{ab} = 0 \ \ \mbox{for all other entries}
\end{equation}
is the metric that we use in  tensor contractions.

If we consider $n=2$ as a special example we see that
\begin{equation}
P^{ab} = \mbox{Tr} ( W^{(a} W^{b)} )  - \frac{1}{3} \eta^{ab}
\eta_{cd} \mbox{Tr} ( W^{c} W^{d} ) ~.
\end{equation}
Since $P^{1\bar{1}} + P^{2\bar{2}} + P^{3\bar{3}} = 0$  we thus get
that the only chiral primary with $n=2$ that involves $A\bar{A}$ and
$B \bar{B}$ but not $\Phi \bar{\Phi}$ is $P^{1\bar{1}} - P^{2\bar{2}}
= \mbox{Tr} ( A\bar{A} - B \bar{B} )$.  This will be important in
section \ref{secgauge}.

The above describes the scalar chiral primaries of  $\CN = 4$
SYM. In $\CN = 2$ QGT the operators \eqref{chpr} define via
the truncation \eqref{AmPAB} scalar chiral primaries.  These
chiral primaries are the scalar chiral primaries of the {\it untwisted
sector}.  The scalar chiral primaries of the {\it twisted sectors} are
instead given by the traceless symmetric combination of
\begin{equation}
\mbox{Tr} \left( S^m W^{a_1} \cdots W^{a_n} \right) \spa \ m =
0,1,...,M-1\spa
\end{equation}
with
\begin{equation}
S \equiv \theta \left(
\begin{array}{cccccc} 1 & &&& & \\
& \theta & &&& \\ & &\ddots&& & \\  & &&& &
\theta^{M-1}\end{array}\right) \spa \theta \equiv \exp \left(
\frac{2\pi i}{M} \right)
\end{equation}
being the {\it twist matrix}.  We clearly have $M$ sectors, 1
untwisted and $M-1$ twisted, since $S^M = 1$.  These $M$ tensors are
the scalar chiral primaries of $\CN = 2$ QGT.

However, it is crucial to note that the $\Z_M$ projection above makes
the operator $\mbox{Tr} \left( S^m W^{a_1} \cdots W^{a_n} \right)$
vanish unless $W^{a_1} \cdots W^{a_n}$ has the right R-charge.  In
table \ref{tableAB} we listed the quantum numbers of $A$, $B$ and
$\Phi$ and their hermitian conjugates.  It is now possible to show
that $\mbox{Tr} \left( S^m W^{a_1} \cdots W^{a_n} \right)$ only is
non-zero if the total $J_L$ charge is a multiple of $M/2$.

\section{Details on the ${\cal N}=2$ operators}
\label{appn2}

In this appendix we supply the discussion of the gauge theory
operators in section~\ref{secgauge} with computational details and
proofs.

\subsubsection*{Level matching}

We first present the proof of the level matching conditions
\eqref{levmat1},  using the form of the ${\cal N}=2$ QGT operators in
terms of generating functions. In the following it is implicitly
assumed that $x=y=0$ in all expressions, after differentiation.

We start by focusing on the case with winding and momentum only for which,
up to normalization factors, the state is
\begin{equation}
\label{wina}
\CO_m \equiv \mbox{Tr} \left[ S^m \CG_{J_R+J_L,J_R-J_L}(\omega^m)
\right] \spa \omega = e^{2\pi i/(J_R M)}
\end{equation}
in terms of the generating function \eqref{genfun4}.  The latter can
be rewritten according to
\begin{multline}
\CG_{K,L} (\omega^m) = \partial_x^K
\partial_y^L \prod_{r=1}^{K+L} \left[
\omega^{\frac{m}{2}} \omega^{-\frac{mr}{2}} xA + \omega^{-
\frac{m}{2}} \omega^{\frac{mr}{2}} yB \right] \\
=\omega^{\frac{m(K-L)}{2}} \partial_x^K
\partial_y^L \prod_{r=1}^{K+L-1} \left[
\omega^{-\frac{mr}{2}} xA + \omega^{\frac{mr}{2}} yB \right] \left[
\omega^{-\frac{m(K+L)}{2}} xA + \omega^{\frac{m(K+L)}{2}} yB
\right]~.
\end{multline}
Then using the fact that the twist matrix \eqref{twistm} satisfies
\begin{equation}
\label{twma}
A S = \theta S A \spa BS = \theta^{-1}  SB \spa \theta = e^{2\pi i/M}\spa
\end{equation}
 we have
\begin{equation}
\begin{split}
    \mbox{Tr} &\left[ S^m \CG_{K,L} \left( \omega^m \right) \right] \\
    &= \omega^{\frac{m(K-L)}{2}} \partial_x^K
        \partial_y^L \mbox{Tr} \left[ \left(
        \omega^{-\frac{m(K+L)}{2}} xA + \omega^{\frac{m(K+L)}{2}} yB \right)
        S^m \right. \\
    &\hskip 3cm  \times \left.
        \prod_{r=1}^{K+L-1} \left( \omega^{-\frac{mr}{2}} xA +
        \omega^{\frac{mr}{2}} yB \right) \right] \\
    &= \omega^{\frac{m(K-L)}{2}} \partial_x^K
        \partial_y^L \mbox{Tr} \left[ S^m \left( \theta^m
        \omega^{-\frac{m(K+L)}{2}} xA + \theta^{-m} \omega^{\frac{m(K+L)}{2}}
        yB \right)\right.  \\
    &\hskip 3cm  \times \left.
        \prod_{r=1}^{K+L-1} \left( \omega^{-\frac{mr}{2}} xA +
        \omega^{\frac{mr}{2}} yB \right) \right] ~.
\end{split}
\end{equation}
Using now that $K = J_R + J_L$, $L = J_R - J_L$ along with
the values of $\omega$ and $\theta$ in \eqref{wina}, \eqref{twma}
one finds that
$\theta^m \omega^{-\frac{m(K+L)}{2}} = 1$ and hence
\begin{equation}
\mbox{Tr} \left[ S^m \CG_{K,L} \left( \omega \right) \right] =
\omega^{\frac{m(K-L)}{2}} \mbox{Tr} \left[ S^m \CG_{K,L} \left( \omega
\right) \right]\spa
\end{equation}
showing that, as desired, the state ${\cal{O}}_m$ only survives for
$J_L=0$ if we have $m\neq 0$.

Consider now the case with one insertion where we need
\begin{equation}
\CG_{K,L;l} (\omega) = \partial_x^K
\partial_y^L \prod_{r=0}^{l-1} \left[
\omega^{-\frac{r}{2}} xA +  \omega^{\frac{r}{2}} yB \right] \Phi
\prod_{r=l}^{K+L-1} \left[ \omega^{-\frac{r}{2}} xA +
\omega^{\frac{r}{2}} yB \right] ~.
\end{equation}
Then, by similar manipulations as done above we can move the $\Phi$
insertion $l$ spots forward using cyclicity of the trace, yielding
\begin{equation}
\label{thereli}
\mbox{Tr} \left[ S^m \CG_{K,L;l} (\omega^m) \right] = \omega^{-
\frac{1}{2} l m ( K-L)} \mbox{Tr} \left[ S^m \Phi \CG_{K,L} (\omega^m)
\right] = \omega^{- l m  J_L } \mbox{Tr} \left[ S^m \Phi \CG_{K,L}
(\omega^m) \right] ~.
\end{equation}
Now, according to \eqref{mow}, for a given level-number $n$, momentum $k
= 2J_L /M$ and winding number $m$ the appropriate state is
\begin{equation}
\CO_{n,k,m} \equiv \sum_{l=0}^{2J_R} \mbox{Tr} \left[ S^m
\CG_{K,L;l} (\omega^m) \right] \beta^{nl} \spa \beta = e^{2\pi i/(2J_R)}\spa
\end{equation}
which using \eqref{thereli} is equal to
\begin{equation}
\CO_{n,k,m} = \mbox{Tr} \left[ S^m \Phi \CG_{K,L} (\omega^m)\right]
\sum_{l=0}^{2J_R} ( \beta^{n} \omega^{-m J_L } )^l ~.
\end{equation}
Hence in order for this to be non-zero we need $\beta^n =
\omega^{mJ_L}$ which precisely reduces to the level matching $n = mk$
after using the explicit forms of $\beta$, $\omega$ and $J_L$.

\subsubsection*{${\cal{N}}=2$ QGT operators and their normalization}

In section \ref{secgauge} we presented the gauge theory operators both
in terms of generating functions as well as words in ${\cal{N}}=4$
notation. Here we present the corresponding explicit forms in terms of
${\cal{N}}=2$ fields after substitution of \eqref{AmPAB}.  The
resulting expressions are useful in order to derive the normalization
of the operators, their anomalous dimension, while also providing an
alternative proof of level matching.

We start with the ground state \eqref{gs} with non-zero momentum $k$.
Using \eqref{genfun1}, \eqref{wordsum} and substituting the
${\cal{N}}=2$ form \eqref{AmPAB} one arrives at
\begin{subequations}
\begin{gather}
\label{wordsuma}
| k, m=0 \rangle \map {\cal{O}}_k =  {\cal{C}} \sum_{\si } \sum_{I=1}^M {\rm Tr}[ {\cal{W}}_{\si,I} ]\spa\\
\label{wordn2}
 {\cal{W}}_{\si,I} \equiv U_{\si(1)}^I U_{\si(2)}^{I + \frac{1}{2}
[\si(1) +\si(2)]} U_{\si(3)}^{I + \frac{1}{2}[\si(1) + 2 \si(2) +
\si(3)]} \ldots U_{\si(2J_R)}^{I + \frac{1}{2} [ \si (1) + 2
\sum_{r=0}^{2J_R-1} \si (i) + \si(2J_R)]}\spa\\
\si (i) = \pm 1 \spa U_1^I = A_I \spa  U_{-1}^I = B_I\spa\\
\label{norm1}
{\cal{C}} \equiv  \frac{1}{\sqrt{w_{J_R+J_L,J_R-J_L}}}
\frac{1}{N^{J_R} \sqrt{2J_R M}}\spa
\end{gather}
\end{subequations}
where all upper indices in the ${\cal{N}}=2$ words \eqref{wordn2} are
meant to be the pullback (modulo $M$) in the fundamental range
$I=1\ldots M$. For brevity, we have omitted that the sum over
$\si$ is over $(J_R+J_L,J_R-J_L)$-type words only, while
we also recall that the definition of $w_{K,L}$ is given
in \eqref{wdef}.
To check that the last letter in the word indeed
correctly contracts with the first letter, note that the superscript
of the last letter in the trace is
\begin{multline}
\label{Ired}
I + \frac{1}{2} [ \si (1) + 2 \sum_{r=0}^{2J_R-1} \si (i) + \si(2J_R)]
= I - \frac{1}{2} [ \si (1) + \si(2J_R) ] + {\cal{I}}(\si) \\
= I - \frac{1}{2} [ \si (1) + \si(2J_R) ] + k M \simeq I - \frac{1}{2}
[ \si (1) + \si(2J_R) ]\spa
\end{multline}
where we used the definition of the index in \eqref{index}  along with
its value  ${\cal{I}}(\si)=2J_L = k M$ and the last step uses that the
superscripts are periodic in $M$.  Then indeed we see on inspection
that the last word matches the first, since $ U_{\si (2J_R)}^{I
-\frac{1}{2} [ \si (1) + \si(2J_R) ]} U_{\si(1)}^I$ has the correct
structure to move from one $U(N)$ factor to the next one.

Let us next verify the normalization factor \eqref{norm1} by computing
in the planar limit the free two-point function of the operator
\eqref{wordsuma}. Using eq.~\eqref{scor2} we compute the free two-point
function
\begin{subequations}
\begin{align}
\langle {\cal{O}}_k (x) \bar  {\cal{O}}_{k'} (0) \rangle &= {\cal{C}}^2
\sum_{\si, \si' }  \sum_{I,J=1}^M \langle {\rm Tr}[{\cal{W}}_{\si,I} (x)]
{\rm Tr}[ \bar {\cal{W}}_{\si',I'} (0)] \rangle \nn\\
\label{normk1}
&=  {\cal{C}}^2 \de_{k,k'} 2J_R  \sum_\si \sum_{I=1}^M \langle {\rm
Tr}[{\cal{W}}_{\si,I} (x)] {\rm Tr}[ \bar {\cal{W}}_{\si,I} (0)]
\rangle \\
\label{normk2}
& =    {\cal{C}}^2 \de_{k,k'} 2J_R \bino{2J_R}{J_R+J_L} M \langle {\rm Tr}[W_{\si,I}
] (x) {\rm Tr}[ \bar {\cal{W}}_{\si,I} (0)] \rangle \\
\label{normk3}
&=  {\cal{C}}^2 \de_{k,k'} 2J_R  \bino{2J_R}{J_R+J_L} M N^{2J_R} \left(\frac{g_{\rm
QGT}^2}{8 \pi^2}\frac{1}{|x^2|}\right)^{2J_R}~.
\end{align}
\end{subequations}
Here the second step is a crucial simplification, analogues of
which hold for all our ${\cal{N}}=2$ gauge theory operators in the
planar limit.  To see this, we note that since the scalar
propagators are diagonal in $I$-space, it follows by comparison
with \eqref{wordn2} that we need $ I = I'$ and $\si (i) = \si'
(i)$, $\forall \;\,i = 1 \ldots 2J_R $ up to possible
identifications due to cyclicity of the trace. This means
for the case at hand first
of all that we need that both $\si$ and $\si'$ are of the
same type, and hence $k=k'$. The overall factor of $2J_R$
is furthermore obtained as follows.
Suppress for
a moment the $I$-dependence and consider a string of $A$'s and
$B$'s, with $J_R+J_L$ $A$'s and $J_R-J_L$ $B$'s. There are ${\tiny
\bino{2J_R}{J_R+J_L}}$ such combinations but since the trace is
cyclic not all combinations are independent. There are some words
that have a symmetry, eg. ${\rm Tr}[ (AB)^{J_R}]$, which is
invariant under a cyclic shift of order two. However, these
special words make up a very small fraction of all the words, that
vanishes rapidly as $J_R$ increases. We can safely neglect all
these symmetric words. Then if we compute the two-point function,
at leading order, each word can be contracted with $2J_R$ other
words, accounting for the prefactor in \eqref{normk1}. The result
\eqref{normk2} is then obtained by replacing the sum over $\si$ by
the number of words, the $I$-sum by the number $M$ of $U(N)$
factors, leaving just a two-point function of a representative
${\cal{N}}=2$ word (since all of them give the same) for which use
the scalar propagators in \eqref{scor2}. This then yields
\eqref{normk3} where the factor $N^{2J_R}$ arises from the
contraction of the $U(N)$ indices in the planar limit.  Finally,
substituting our normalization factor \eqref{norm1},   we arrive
at the resulting two-point function
\begin{equation}
\label{2ptgs}
\langle {\cal{O}}_k (x) \bar  {\cal{O}}_{k'} (0) \rangle =\de_{k,k'}
\left(\frac{g_{\rm QGT}^2}{8 \pi^2}\frac{1}{|x^2|}\right)^{2J_R}\spa
\end{equation}
which concludes our derivation of ${\cal{C}}$ in this case.

The derivation of all other normalization factors in the paper
proceeds along the same lines. We leave those in \eqref{1osc}
as an exercise for the reader, turning immediately to
the oscillator states in \eqref{twoins} for which we define
\begin{subequations}
\begin{gather}
(a^{\Phi}_{n})^\dagger (a^{\Phi}_{-n})^\dagger |k = 0, m=0
\rangle \map {\cal{O}}_{n}^{(\rm o)} = {\cal{C}}_{(\rm o)}
\sum_{l=0}^{2J_R} \sum_{\si \in \si(J_R,J_R)} \sum_{I=1}^M {\rm
Tr}[\Phi_I {\cal{W}}_{\si,I; l} ] \beta^{nl}\spa\\
{\cal{W}}_{\si,I;l} \equiv U_{\si(1)}^I  \ldots U_{\si(l)}^{I +
 \ldots}  \Phi^{I + \ldots} U_{\si (l+1)}^{I + \ldots} \ldots
 U_{\si(2J_R)}^{I - \frac{1}{2} (\si(1) + \si(2J_R))}\spa\\
\label{norm2}
{\cal{C}}_{(\rm o)} =\frac{1}{\sqrt{w_{J_R,J_R}}} \frac{1}{N^{J_R+1}
\sqrt{2J_R M}}~.
\end{gather}
\end{subequations}
Then the free two-point function is calculated to be
\begin{equation}\label{step1}
\begin{split}
\langle {\cal{O}}&_n^{(\rm o)} (x) \bar  {\cal{O}}_{n'}^{(\rm o)} (0)
\rangle \\
&=   {\cal{C}}_{(\rm o)}^2 \sum_{\si, \si' }
\sum_{I,I'=1}^M
\sum_{l,l' =0}^{2J_R} \langle  {\rm Tr}[ \Phi_I (x) {\cal{W}}_{\si,I; l} (x)]
{\rm Tr} [\bar \Phi_{I'} (0)  \bar {\cal{W}}_{\si',I'; l'} (0)] \rangle \beta^{nl -n'l'}\\
& =  {\cal{C}}_{(\rm o)}^2 \sum_{l=0}^{2J_R} \sum_\si
\sum_{I=1}^M \langle
{\rm Tr}[\Phi_I (x){\cal{W}}_{\si,I;l} (x)] {\rm Tr}[\bar \Phi_{I} (0)
 \bar {\cal{W}}_{\si,I;l}
(0)] \rangle \beta^{(n -n')l} \\
& = {\cal{C}}_{(\rm o)}^2   \bino{2J_R}{J_R} M \sum_{l=0}^{2J_R}
\langle {\rm Tr}[\Phi_{I} (x)   {\cal{W}}_{\si,I;l}  (x)]
{\rm Tr}[\bar \Phi_{I} (0)   \bar
{\cal{W}}_{\si,I;l} (0)] \rangle \beta^{(n -n')l} \\
& = {\cal{C}}_{(\rm o)}^2  \bino{2J_R}{J_R} M  2J_R \delta_{n,n'}
N^{2(J_R+1)} \left(\frac{g_{\rm QGT}^2}{8
\pi^2}\frac{1}{|x^2|}\right)^{2(J_R+1)}~.
\end{split}
\end{equation}
Then using the normalization factor \eqref{norm2} we record for the
free two-point function in the planar limit the result
\begin{equation}
\label{2pt2ins}
\langle {\cal{O}}_n^{(\rm o)} (x) \bar  {\cal{O}}_{n'}^{(\rm o)} (0)
\rangle_{(\rm free)} =\delta_{n,n'} \left(\frac{g_{\rm QGT}^2}{8
\pi^2}\frac{1}{|x^2|}\right)^{2(J_R+1)}~.
\end{equation}

Finally we discuss the operators \eqref{wingau} with winding in the
present ${\cal{N}}=2$ notation. For the general state with non-zero
momentum and winding one finds
\begin{equation}
\label{wins}
| k, m \rangle \map {\cal{O}}_{k,m} = {\cal{C}} \sum_{\si } \sum_{I=1}^M (\om^m)^{{\cal{N}}(\si)}
(\theta^m)^{I + \frac{1}{2} [ 3 - \si(1)]} {\rm Tr}[ {\cal{W}}_{\si,I}]\spa
\end{equation}
where ${\cal{C}}$ is as in \eqref{norm1},
the ${\cal{N}}=2$ words ${\cal{W}}_{\si,I}$ are defined in
\eqref{wordn2} and the weight ${\cal{N}}(\si)$ in \eqref{weight}.
To calculate the free two-point function of these operators
follow the steps
\begin{equation}\label{step1w}
\begin{split}
\langle {\cal{O}}&_{k,m} (x) \bar  {\cal{O}}_{k',m'} (0)
\rangle \\
&= {\cal{C}}^2 \sum_{\si, \si' }
\sum_{I,I'=1}^M
 \om^{m {\cal{N}}(\si) - m' {\cal{N}}(\si')}
\theta^{m(I + \frac{1}{2} [ 3 - \si(1)]) -m'
(I' + \frac{1}{2} [ 3 - \si'(1)])} \\
& \hskip 5cm \times\langle {\rm Tr}[ {\cal{W}}_{\si,I} (x)]
{\rm Tr}[ \bar {\cal{W}}_{\si',I'} (0)] \rangle \\
& = {\cal{C}}^2 \de_{k,k'}
 2J_R \sum_{\si }
\sum_{I=1}^M
 \om^{(m-m') {\cal{N}}(\si) }
\theta^{(m-m')I}
\langle {\rm Tr}[ {\cal{W}}_{\si,I} (x)]
{\rm Tr}[ \bar {\cal{W}}_{\si,I} (0)] \rangle \\
& = {\cal{C}}^2 \de_{k,k'} 2J_R \bino{2J_R}{J_R+J_L} M \de_{m,m'}
\left(\frac{g_{\rm QGT}^2}{8
\pi^2}\frac{1}{|x^2|}\right)^{2J_R}\spa
\end{split}
\end{equation}
showing that the orthogonality of winding states
with different $m$ is intimately connected to the twist matrix.
Using ${\cal{C}}$ in \eqref{norm1}, we
thus record the final result
\begin{equation}
\label{2ptgsw}
\langle {\cal{O}}_{k,m} (x) \bar  {\cal{O}}_{k',m'} (0) \rangle =
\de_{k,k'} \de_{m,m'}
\left(\frac{g_{\rm QGT}^2}{8 \pi^2}\frac{1}{|x^2|}\right)^{2J_R}\spa
\end{equation}
which includes the $m=0$ result \eqref{2ptgs} as a special case.

In this form the level matching is easily checked by using cyclicity of
trace to move the last letter in ${\cal{W}}_{\si,I}$ to the first spot
in the word. Then after redefining the two summation variables $\si $
and $I$ in \eqref{wins} as
\begin{equation}
\label{siIshift}
\tilde \si (i) = \si (i-1) \spa \tilde \si (1) = \si (2J_R) \spa
\tilde I = I - \frac{1}{2} [ \si (1) + \si (2J_R)]\spa
\end{equation}
and using the identity
\begin{equation}
{\cal{N}}(\si) = N(\tilde \si )+\frac{1}{2} {\cal{I}}(\tilde \si)+ J_R
\tilde \si (1) \spa {\cal{I}}(\tilde \si) = {\cal{I}}(\si) = 2 J_L\spa
\end{equation}
one finds for each word in the sum the phase factor
\begin{equation}
\label{phf1}
\om^{m[J_L +J_R \si(1)] } \theta^{- m \si(1)} = \om^{m J_L}\spa
\end{equation}
where we used \eqref{omega}, \eqref{twistm} to obtain the second
form. Then the  level matching $J_L=0$ for $m \neq 0$ follows
immediately.  More generally for the operator with one insertion
\begin{subequations}
\begin{gather}
(a^{\Phi}_{n})^\dagger |k, m\rangle \map {\cal{O}}_{n,k,m}
 = {\cal{C}}_{(\rm o,w)}
 \sum_{\si } \sum_{I=1}^M
{\rm Tr}[ {\cal{W}}_{\si,I; l} ]\\
{\cal{C}}_{(\rm o,w)} =\frac{1}{\sqrt{w_{J_R+J_L,J_R-J_L}}}
\frac{1}{N^{J_R+\frac{1}{2}} \sqrt{2J_R M}}
\end{gather}
\end{subequations}
we move, as in \eqref{wins},  the last letter to the first spot.
Since this increases the location of the insertion of $\Phi_I$ by one
unit this means that  we need to accompany \eqref{siIshift}
by the shift $\tilde l  = l + 1$ inducing an extra phase $\beta^{-n}$
in \eqref{phf1}. To cancel the overall shift we thus need $\beta^n =
\omega^{m J_L}$, which implies the level matching $n=km$.

%%%%%%%%%%%%%%%%%%%%%%%%%%%%%%%%%%%%%%%%%%%%%%%%%%%%%%%%%%%%%%%%%%
\subsubsection*{Derivation of anomalous dimension formula}

The anomalous dimension of an operator ${\cal O}$ can be evaluated by
computing the correlator $\langle {\cal O}(x) {\cal O}(0)\rangle$ at
one loop
\begin{multline}\label{corf0}
\langle {\cal O}(x) \bar{\cal O}(0)\rangle_{(\rm free)} =
\frac{{\cal N}}{|x|^{2\Delta_0}}  \\
\longrightarrow\langle
{\cal O}(x) \bar{\cal O}(0)\rangle_{({\rm free + 1-loop})}
=\frac{{\cal N}}{|x|^{2(\Delta_0 + \delta\Delta)}} = \frac{{\cal
N}}{|x|^{2\Delta_0}} \left(1 - 2 \delta\Delta \ln |x|\Lambda \right)\spa
\end{multline}
where ${\cal N}$ is a normalization factor, $\Delta_0$ is the
conformal dimension of the operator in the free theory, $\Lambda$ a
regulator and $\delta\Delta$ the anomalous dimension. The one-loop
corrected dimension is then
\begin{equation}
\label{anomgen}
\Delta = \Delta_0 + \delta\Delta ~.
\end{equation}
This shows that we are interested in the ratio \eqref{ratio} between
the one-loop contribution and the free part.

Here we derive the formula \eqref{ratioF} for the ratio between the
free and one-loop two-point function for the oscillator and winding
states, and discuss some of the other steps that lead to the result
\eqref{andimg} for the one-loop correction to the anomalous dimension.
As argued in section \ref{andsec} we only need to focus on one-loop
contributions coming from the F-terms given in \eqref{Fterms}.

We first consider the operator \eqref{Oo}, for which the free
two-point function is given in \eqref{2pt2ins}.  To compute the
one-loop contribution to this two-point function we may start with the
expression in \eqref{step1} and recall that the one-loop F-term
contribution to this will induce an interchange of $\Phi$
with a nearest neighbor $U_{\si (l \pm 1)}$.  This gives rise to a factor $[\delta_{l,l'-1}
+ \delta_{l,l'+1}]/2$ as compared to the $\delta_{l,l'}$ factor in the
free case.  Moreover, there are in all four interchanges possible,
accounting for a multiplicity factor $\mathfrak{m}=4$.  As we are
working in the planar limit, it is not difficult to see that again one
only gets contributions when $\si = \si'$ and $I= I'$.  Furthermore,
as a result of the one-loop interaction 2 out of the $2J_R+2$ scalar
propagators are replaced by the one-loop term involving the F-term
interaction, so that
\begin{equation}
\label{FtermPU}
\langle \Phi (x) \bar \Phi (0) \rangle \langle U_{\pm 1} (x) \bar
U_{\pm 1} (0) \rangle  \longrightarrow \langle \Phi (x)U_{\pm 1} (x)
\bar \Phi (0) \bar U_{\pm 1} (0) \rangle\spa
\end{equation}
where we recall $U_{1} = A =W_2$, $U_{-1}= B = W_3$ and
the corresponding one-loop diagram is given in figure \ref{01}.
Finally, we have to take into account the Fourier transform
\begin{equation}
\label{sumrelo}
\frac{1}{2J_R} \sum_{l,l'=0}^{2J_R} \frac{1}{2}[\delta_{l,l'-1} + \delta_{l,l'+1}]
\beta^{n l - n'l'} = \delta_{n,n'} \frac{\beta^n + \beta^{-n}}{2}
\end{equation}
and putting it all together leads to \eqref{ratioF} with the
particular substitutions given  in \eqref{oscv}.

To compute the one-loop correction for the winding state \eqref{Ow}
we start from the expression \eqref{step1w}.
 In this case the one-loop F-term will induce interchanges of $A_I$
with a nearest neighbor $B_I$ and vice versa. In analogy with
\eqref{FtermPU} the result of the F-term interaction is the replacement
\begin{equation}
\label{ABflip}
 \langle A (x) \bar A (0) \rangle \langle B (x) \bar
B (0) \rangle \longrightarrow \langle A (x) B (x) \bar A (0) \bar
B (0) \rangle\spa
\end{equation}
which corresponds again to the diagram in figure \ref{01}.
Moreover, it follows that  (modulo cyclic permutations)
$\delta_{\si,\si'}$ in the free case is replaced by $\delta_{\si
\circ P,\si'}$  where $\si \circ P $ stands for the word that
results from interchanging a nearest neighbor pair $AB$ or $BA$.
This shows that we can reduce again the double sum in
\eqref{step1w} to a single $\si$-sum since $\si'$ is fixed given
each choice of $\si$. This will also fix $I'$ given $I$, and the
resulting sum involves
\begin{equation}
\label{sumrelw}  \sum_{\si'} \omega^{m {\cal{N}}(\si) -
m'{\cal{N}}(\si')} \delta_{\si \circ P,\si'}  \sum_{I=1}^M
\theta^{(m-m') I} = \de_{m,m'} \frac{\omega^m+\omega^{-m}}{2}  ~.
\end{equation}
Here we  used the definition of the weight ${\cal{N}}(\si)$ in
\eqref{weight}, which implies for the $A \leftrightarrow B$
interchange defined above that
  ${\cal{N}}(\si) - {\cal{N}}(\si \circ P) = -1$ for
$ AB \rightarrow BA$  and ${\cal{N}}(\si) - {\cal{N}}(\si \circ P)
= 1$ for $BA \rightarrow AB$. Here, each of these two situations
occurs an equal number of times, so that we have taken the average
to obtain \eqref{sumrelw} (see below for the multiplicity factor).
 The relation \eqref{sumrelw} can be seen as the
analogue of \eqref{sumrelo}.

Finally, we need to take into account the multiplicity factor counting
the number of interchanges. For this we need to count the number of
words where $A$ and $B$ sit next to each other at $2r$ places.
(Since the number of $A$'s and $B$'s is the same, it is
clear that on $r$ places there is an $AB$, while on the other
$r$ places there is a $BA$).
The
number $r$ satisfies $1\leq r \leq J_R$.  If $r=1$ the word is ${\rm
tr}[A^{J_R} B^{J_R}]$ or a cyclic permutation, if $r=J_R$ the word is
${\rm tr}[(AB)^{J_R}]$ or a cyclic permutation. The number of words as
a function of $r$ is
\begin{equation}
N_r = 2 \bino{J_R-1}{r-1} \bino{J_R}{r}
\end{equation}
and indeed $\sum_{r=1}^{J_R} N_r = {\tiny \bino{2J_R}{J_R}}$
reproduces the total  number of words. As a consequence the
multiplicity factor is computed as
\begin{equation}
\mathfrak{m} = 2 \frac{\sum_{r=1}^{2J_R} r N_r}{\sum_{r=1}^{2J_R} N_r}
= \frac{2 J_R^2}{2J_R -1} \sim  J_R~.
\end{equation}
Taking all this together then leads to \eqref{ratioF} with the
particular substitutions given  in \eqref{winv}.

%%%%%%%%%%%%%%%%%%%%%%%%%%%%%%%%%%%%%%%%%%%%%%%%%%%%%%%%%%%%%%%%%%%%%
\subsubsection*{Operators corresponding to oscillators in the isometric
direction}

We conclude by considering
the operators \eqref{wordz1} corresponding to the oscillator modes
in the isometric $z_1$ direction. For simplicity we focus on
the $\CN =4$ counterpart of these operators in the $U(N)$ theory.
This means we take $M=1$ in the states \eqref{wordz1}, but we stress that
our results also hold for their $\CN =2$ counterparts, as one may
check that in the planar limit the contributions to the two-point
functions below are diagonal in the $U(N)^M$ product space. This
enables us then at the end to compute the anomalous dimension of
the $\CN=2$ operator in \eqref{OoJL}.

As a further simplification, we may first look at the gauge non-invariant
operator
\begin{equation}
\label{Pn}
{\cal{P}}_n = \sum_{\si \in \si (J_R,J_R)} d_\si^n {\cal{W}}_\si \spa
d_\si^n = \frac{1}{2} \sum_{l=0}^{2J_R-1} \si (l+1) \beta^{nl}
\end{equation}
relevant to the case of a single oscillator in the $z_1$ direction. Though
this operator vanishes after taking the trace, we may still use the
non-traced version \eqref{Pn} in order to compute the anomalous dimension
of the (level-matched) double insertion by simply multiplying the result
by two.

We start by computing the free two-point function
\begin{multline}
   \langle {\cal{P}}_{n} (x)   \bar  {\cal{P}}_{n'} (0) \rangle
= \sum_{\si, \si' } d_\si^n \bar d_{\si'}^{n'}
\langle  {\cal{W}}_{\si} (x) \bar {\cal{W}}_{\si'} (0) \rangle
 =  \sum_{\si } d_\si^n \bar d_{\si}^{n'}
\langle  {\cal{W}}_{\si} (x) \bar {\cal{W}}_{\si} (0) \rangle\\
=  \frac{1}{4} \sum_\si \left[
\sum_l \si (l+1)^2 \beta^{(n-n')l} + \sum_{l \neq l'}
 \si (l+1) \si (l'+1)  \beta^{nl-n'l'} \right]
\langle  {\cal{W}}_{\si} (x) \bar {\cal{W}}_{\si} (0) \rangle\spa
\label{st1}
\end{multline}
%\begin{eqnarray}
%& &    \langle {\cal{P}}_{n} (x)   \bar  {\cal{P}}_{n'} (0) \rangle
%= \sum_{\si, \si' } d_\si^n \bar d_{\si'}^{n'}
%\langle  {\cal{W}}_{\si} (x) \bar {\cal{W}}_{\si'} (0) \rangle
% =  \sum_{\si } d_\si^n \bar d_{\si}^{n'}
%\langle  {\cal{W}}_{\si} (x) \bar {\cal{W}}_{\si} (0) \rangle \nn \\
%& &=  \frac{1}{4} \sum_\si \left[
%\sum_l \si (l+1)^2 \beta^{(n-n')l} + \sum_{l \neq l'}
% \si (l+1) \si (l'+1)  \beta^{nl-n'l'} \right]
%\langle  {\cal{W}}_{\si} (x) \bar {\cal{W}}_{\si} (0) \rangle
%\label{st1}
%\hskip .5cm
%\end{eqnarray}
where the first step uses the fact that in the planar limit the
contributions are diagonal in word space.
Now use that the free 2-point function
$\langle  {\cal{W}}_{\si} (x) \bar {\cal{W}}_{\si} (0) \rangle$
is $\si$-independent as well as the result
\begin{eqnarray}
& & \sum_{\si \in \si (J_R,J_R)} \si (l+1) \si (l'+1)
 =  2 \bino{2J_R-2}{J_R} - 2 \bino{2J_R-2}{J_R-1}  \nn \\
& = &
 - \frac{1}{2J_R-1} \bino{2J_R}{J_R}
=- \frac{1}{2J_R-1} \sum_{\si \in \si (J_R,J_R)} \spa l \neq l' \ .
\end{eqnarray}
Then, after also using $\si(l+1)^2=1$ in the first term,
we can perform the $l,l'$ sums in \eqref{st1} to obtain
\begin{eqnarray}
   & & \langle {\cal{P}}_{n} (x) \bar  {\cal{P}}_{n'} (0) \rangle
\nn \\
&&=\frac{1}{4}  \left[ 2J_R \delta_{n,n'}
- \frac{1}{2J_R-1} \Big((2J_R)^2\delta_{n,0} \delta_{n',0} -
 2J_R \delta_{n,n'} \Big) \right] \sum_\si
\langle  {\cal{W}}_{\si} (x) \bar {\cal{W}}_{\si} (0) \rangle
\nn \\
& & =
\frac{(2J_R)^2}{4(2J_R-1)} \delta_{n,n'} (1- \delta_{n,0}) \sum_\si
\langle  {\cal{W}}_{\si} (x) \bar {\cal{W}}_{\si} (0) \rangle \ .
\end{eqnarray}
As a check note that indeed the two-point function of ${\cal{P}}_0
= 0$ vanishes. The way this works out is that the second term in
\eqref{st1} precisely cancels that first term in \eqref{st1}, as
one could have noted immediately from $\sum_l \si (l+1) = 0 $
(since $J_L=0$) in that case. For $n\neq 0$ on the other hand, the
computation above shows that the second term in \eqref{st1} is
negligible compared to the first term in the large $J_R$ limit. In
summary, we record
\begin{equation}
\langle {\cal{P}}_{n} (x) \bar  {\cal{P}}_{n'} (0) \rangle
=
\frac{2J_R}{4} \delta_{n,n'}(1- \delta_{n,0})
\langle {\cal{P}}_g   (x) \bar {\cal{P}}_g  (0)  \rangle\spa
\label{Pnfree}
\end{equation}
where we have taken the large $J_R$ limit and introduced the notation
$\langle {\cal{P}}_g   (x) \bar {\cal{P}}_g  (0)  \rangle $ for
the  free-ground state (without trace) two-point function.
In the following we restrict our discussion to the non-trivial
states ${\cal{P}}_{n \neq 0}$.

We now consider the one-loop planar corrections to the two-point
function $\langle {\cal{P}}_{n} (x)   \bar  {\cal{P}}_{n'} (0) \rangle$.
We first consider the diagrams that act diagonally in word space,
i.e. which only have planar contributions between equal words.
These are the D-terms plus radiative corrections to the propagators,
as well as some of the F-terms. 
Since they are diagonal in word space, we can use exactly the
same argument as above, and we get
\begin{equation}
\langle {\cal{P}}_{n} (x) \bar  {\cal{P}}_{n'} (0) \rangle_{\rm 1-loop, diag.}
=
\frac{2J_R}{4} \delta_{n,n'}(1- \delta_{n,0})
\langle {\cal{P}}_g   (x) \bar {\cal{P}}_g  (0)  \rangle_{\rm 1-loop, diag.}
\spa
\label{PnDterm}
\end{equation}
just as in \eqref{Pnfree}. 

Turning instead to the one-loop planar diagrams that do not act diagonally
in word space,
we note that the only diagrams contributing are the F-term diagrams
that interchange an $A$ with a $B$ as in \eqref{ABflip}.
It will be useful to
employ a similar notation as for the windings, introducing now
$P_q$ to be the permutation of the nearest neighbor pair
$U_{\si(q+1)} U_{\si (q+2)}$
 occurring after the $q$th spot in the word $\si$. 
More explicitly, if for example
${\cal{W}}_\si  = U_{\si(1)} .. U_{\si (q)} AB U_{\si (q+3)}
.. U_{\si(2J_R)} $ then ${\cal{W}}_{\si \circ P_q}  =
U_{\si(1)} .. U_{\si (q)} BA U_{\si (q+3)} .. 
U_{\si(2J_R)} $. 
Define now for a given $q$ the set $M_q$ as the set 
of words in $\si (J_R,J_R)$ that have $U_{\si(q+1)}U_{\si(q+2)}=AB$
or $U_{\si(q+1)}U_{\si(q+2)}=BA$.
The F-term contribution to the two-point function is then
\begin{multline}
 \langle {\cal{P}}_{n} (x) \bar  {\cal{P}}_{n'} (0) \rangle\vert_{\rm F}
= \sum_q \sum_{\si, \si' \in M_q} d_\si^n \bar d_{\si'}^{n'}
\ \delta_{\si \circ P_q,\si'} \langle  {\cal{W}}_{\si} (x) \bar {\cal{W}}_{\si'} (0) \rangle\\
 =\sum_q \sum_{\si \in M_q} \sum_{l,l'}
 \si (l+1) (\si\circ P_q) (l'+1)
 \beta^{nl- n'l'}
\langle  {\cal{W}}_{\si} (x) \bar {\cal{W}}_{\si \circ P_q } (0)
\rangle\spa
\end{multline}
where the correlator, which incorporates \eqref{ABflip}, is
independent of the choice of $\si$ and $P_q$. In parallel with the
free two-point function above, the leading contributions that are
also $n$-dependent, will now arise when $(\si \circ P_q)(l'+1) =
\si (l+1) $, so that $l'=l+1$ and $q=l$ or $l=l'+1$ and $q=l'$.
Substituting this and performing the $l$-sum then yields the ratio
\begin{equation}
\label{fracP} \frac{\langle {\cal{P}}_{n} (x) \bar  {\cal{P}}_{n'}
(0) \rangle\vert_{\rm F}} {\langle {\cal{P}}_{n} (x) \bar
{\cal{P}}_{n'} (0) \rangle\vert_{\rm free}} = \delta_{n,n'} 
\left( \beta^n + \beta^{-n} \right) 
\frac{\langle A (x) B (x) \bar A (0) \bar B (0)
\rangle}{
 \langle A (x) \bar A (0) \rangle \langle B (x) \bar B (0) \rangle
 } \spa n,n' \neq 0
\end{equation}
in terms of the free result in \eqref{Pnfree}.
We now observe that if we take \eqref{fracP} and formally
set $\beta=1$ but keeping $n \neq 0$ then all one-loop contributions
from \eqref{PnDterm} and \eqref{fracP} vanish, since they
reduce to a factor times the one-loop contribution to
$\langle {\cal{P}}_g   (x) \bar {\cal{P}}_g  (0)  \rangle$
which of course is zero.%
\footnote{Note that we cannot use the usual argument of
comparing $n\neq 0$ to $n=0$ since \eqref{Pnfree} for $n=n'$ has
an $n$-dependence in the form of $\delta_{n,0}$.}
Thus, with this in mind, we see that for $\beta \neq 1$ we get
a factor $\beta^n + \beta^{-n} - 2$ in front of the total one-loop
planar contribution to $\langle {\cal{P}}_{n} (x) \bar  {\cal{P}}_{n}
(0) \rangle$, and the resulting anomalous dimension indeed becomes
$\delta \Delta = \frac{\gqgt^2 N n^2}{2(2J_R)^2}$.

The corresponding result for the operator \eqref{OoJL} with
two $J_L$ insertions, as summarized in \eqref{ratioF}, \eqref{oscv1},
then immediately follows from \eqref{fracP} by multiplying by two.

\subsubsection*{Useful formulas}

We collect here some formulas which are useful for the gauge theory
computations carried on in section \ref{secgauge}.

Let us first fix our convention for Fourier transform. The Fourier
transform of a function $g(p)$ in $2\om$ dimension is a function
$f(x)$ defined as
\begin{equation}
f(x) = \int \frac{d^{2\om}p}{(2\pi)^{2\om}} e^{ipx} g(p) \spa g(p) =
\int d^{2\om}p e^{-ipx} f(x) \spa \delta(p) = \int
\frac{d^{2\om}x}{(2\pi)^{2\om}} e^{ipx}~.
\end{equation}
The Green function for the Laplacian in  $2\omega$ dimensions is
\begin{equation}
\label{fscor}
\Delta(x) = \int \frac{d^{2\om}p}{(2 \pi)^{2\om}} \frac{e^{ipx}}{p^2}
= \frac{\Ga(\om -1)}{4 \pi^\om [x^2]^{\om -1}}\spa
\end{equation}
while more generally,
\begin{equation}
\int \frac{d^{2\om}p}{(2 \pi)^{2\om}} \frac{e^{ipx}}{[p^2]^s} =
\frac{\Ga(\om -s)}{4^s \pi^\om \Ga(s) [x^2]^{\om -s}}~.
\end{equation}

In momentum space, the relevant integral for the one-loop diagram in
 figure \ref{01} is
\begin{equation}
\label{intxy}
\int \frac{d^{2\om}q}{q^2 (p-q)^2}\int \frac{d^{2\om}l}{l^2 (m-l)^2}
\delta_{p,m} = \left(\int \frac{d^{2\om}q}{q^2 (p-q)^2}\right)^2 =
\frac{\Ga(2-\om)^2 \Ga(\om-1)^4}{\Ga(2\om -2)^2}
\frac{1}{[p^2]^{2(2-\om)}}\spa
\end{equation}
where $\de_{p,m}$ follows from momentum conservation on the four-point
vertex. The above integral seems to have a quadratic divergence  in
four dimensions, i.e. for $\om=2$. However, this is not the case when
transforming back to coordinate space. Indeed the Fourier transform of
eq.~\eqref{intxy} is
\begin{equation}
\begin{split}
\frac{\Ga(2-\om)^2 \Ga(\om-1)^4}{\Ga(2\om -2)^2} & \int
\frac{d^{2\om}p}{(2 \pi)^{2\om}} \frac{e^{ipx}}{[p^2]^{2(2-\om)}} \\
&=\frac{\Ga(2-\om)^2 \Ga(\om-1)^4 \Ga(3\om -4)}{4^{4-2\om} \pi^\om
\Ga(2\om -2)^2 \Ga(4 - 2\om)} \frac{1}{[x^2]^{3\om -4}}  \\
&=\frac{2}{2-\om}\frac{1}{\pi^\om}\frac{1}{[x^2]^{3\om -4}}  =  32
\pi^2 \Delta(x)^2 \ln(|x|\Lambda)^2~.
\end{split}
\end{equation}
Here  $\ln\Lambda^2 \equiv 1/(2-\om)$ and we have evaluated at $\om=2$
all factors giving finite contribution. The above computation implies
that
\begin{equation}
\label{onelform}
\int d^{4}y \,\Delta(y)^2\Delta(x-y)^2 = \frac{1}{4 \pi^2} \Delta(x)^2
\ln(|x|\Lambda)\spa
\end{equation}
which is used in section \ref{andsec} to compute the one-loop
anomalous dimensions.

\section{$\CN=1$ quiver gauge theory}
\label{appqgt1}

Here we give some details on the $\CN=1$ quiver gauge theory
arising by considering $N$ D3-branes at the orbifold $\C^3 / (\Z_{M_1}
\times \Z_{M_2})$  discussed in section \ref{secDLCQ}. The gauge group
is a product of $M_1M_2$ factors
\begin{equation}
\prod_{I,J} U(N)^{(IJ)} \spa \mbox{where} \quad
I=1,\dots,M_1\;,\;J=1,\dots,M_2 ~.
\end{equation}
Contrary to the ${\mathcal N}=2$ case, we are using a double
index notation since in this case the orbifold group has two
generators, $\omega_1$ and $\omega_2$, defined in
eq.~\eqref{defn1}. The gauge coupling is the same for all group
factors, $\gqgt^2 = 4\pi g_s M_1 M_2$ in terms of the string coupling
$g_s$. The field content of the gauge theory consists of $M_1 M_2$
vector multiplets $(A_{\mu,IJ},\psi_{IJ})$ and $3 M_1 M_2$
bifundamental chiral multiplets ${\bf W}_{IJ}  \equiv
(W_{IJ},\chi_{IJ})$. Similar to the ${\mathcal N}=2$ case the chiral
multiplets can be organized in three different $M_1M_2 \times M_1M_2$
matrices ${\bf W}_i$ (every entry being an $N\times N$ matrix)
associated to the three complex planes $a_i$ transverse to the
D3-branes. Given eq.~\eqref{defn1} one can show \cite{Hanany:1998it}
that the complex fields ${\bf W}_i$ decompose in the following
representations of the constituent gauge factors
\begin{subequations}
\begin{align}
{\bf W}_1 &\rightarrow   \oplus_{I,J} \left( N_{I,J}, \bar
N_{I+1,J}\right)\spa \\ {\bf W}_2 &\rightarrow   \oplus_{I,J} \left(
N_{I,J}, \bar N_{I-1,J-1}\right)\spa \\ {\bf W}_3 &\rightarrow
\oplus_{I,J} \left( N_{I,J}, \bar N_{I,J+1}\right)\spa
\end{align}
\end{subequations}
where $N_{I,J}$ ($\bar N_{I,J}$) represents the fundamental
(anti-fundamental) representation of the gauge group $U(N)^{(IJ)}$.
The above structure automatically relates the ${\mathcal N}=1$ theory
to the parent ${\mathcal N}=4$ theory defined on the covering
space. The latter is a suitable truncation of the gauge theory
obtained form $N M_1M_2$ D3-branes on $\C^3$. The three ${\mathcal
N}=1$ chiral multiplets ${\bf W}_i$ sum-up into the 3 complex fields
entering the ${\mathcal N}=4$ multiplet. The connection between
${\mathcal N}=1$ and ${\mathcal N}=4$ field then proceeds along the  same
lines as for the ${\mathcal N}=2$ QGT discussed previously:  each $N
\times N$ non-vanishing entry of the $M_1 M_2 \times M_1 M_2$ matrices
${\bf W}_i$ is a given chiral field ${\bf W}_{IJ}$ of the ${\mathcal
N}=1$ QGT. In all we have three $N M_1M_2 \times N M_1M_2$ matrices
corresponding to the (truncated) three chiral fields transforming in
the adjoint representation of the $U(N M_1 M_2)$ ${\mathcal N}=4$
theory.

\begin{figure}[ht]
\begin{center}
{\scalebox{1}{\includegraphics{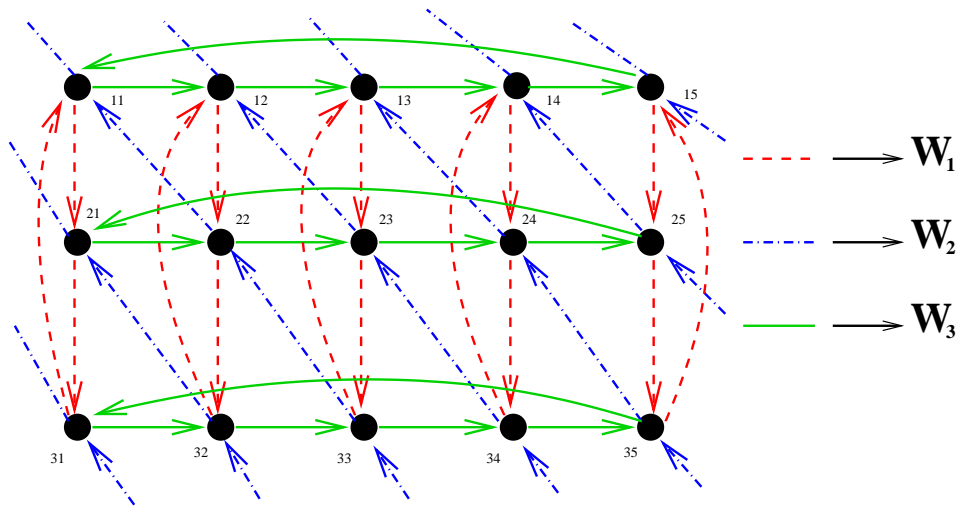}}}
\caption{\small The quiver diagram for the orbifold $\C^3/(\Z_5 \times
\Z_3)$. Each dot represents a $U(N)$ gauge factor, in double index
notation. The red dashed lines correspond to chiral multiplets
belonging to ${\bf W}_1$ , the blue lines to chiral multiplets
belonging to ${\bf W}_2$ and green ones to chiral multiplets belonging
to ${\bf W}_3$. Arrows go from fundamental to anti-fundamental
representations.}
\label{quiv2}
\end{center}
\end{figure}

As for the ${\mathcal N}=2$ theory case, the field content of the
${\mathcal N}=1$ at hand can be efficiently summarized into quiver
diagrams, which will be of increasing complexity.
 As an explicit example, the quiver diagram for
$M_1=5,M_2=3$ is depicted in figure \ref{quiv2}.

\end{appendix}

\addcontentsline{toc}{section}{References}

%The following two lines is for bibtex only:
%\bibliographystyle{utphys}
%\bibliography{bibrot,biblioniels}
%\bibliographystyle{../INPUT/utphys}
%\bibliography{../BIB/bibrot,../BIB/biblioniels}

\providecommand{\href}[2]{#2}\begingroup\raggedright\endgroup

\end{document}